\documentclass[11pt,aps,preprint,nofootinbib]{revtex4-1}

\usepackage{graphicx}
\usepackage{caption}
\usepackage{subcaption}

\usepackage{ragged2e}

\usepackage{comment} 
\usepackage{dcolumn}
\usepackage{bm}%
\usepackage{float}	
\usepackage{amssymb}
\usepackage[T1]{fontenc} 
\usepackage{color}
\usepackage[dvipsnames]{xcolor}
\usepackage{mdframed}
\usepackage{amsmath}
\usepackage{textcomp}
\usepackage{hyperref}
\usepackage{arydshln}
\usepackage{enumerate}
\usepackage{mathrsfs}
\usepackage{verbatim}
\usepackage{slashed}
\usepackage{epsfig}
\usepackage{enumerate}
\usepackage{color}
\usepackage{wrapfig}
\usepackage{amsfonts}
\usepackage{multirow}
\usepackage{tikz}
\usepackage{amsmath}
\usepackage[normalem]{ulem}

 \usepackage[normalem]{ulem}

\expandafter\ifx\csname package@font\endcsname\relax\else
 \expandafter\expandafter
 \expandafter\usepackage
 \expandafter\expandafter
 \expandafter{\csname package@font\endcsname}%
\fi
\definecolor{darkcyan}{cmyk}{1,0,0,0.4}

\newcommand{\beq}{\begin{equation}}
\newcommand{\eeq}{\end{equation}}
\newcommand{\bsp}{\begin{split}}
\newcommand{\esp}{\end{split}}
\newcommand{\bit}{\begin{itemize}}
\newcommand{\eit}{\end{itemize}}

\def\barr{\begin{array}}
\def\earr{\end{array}}

\newcommand{\ba}{\begin{array}}
\newcommand{\ea}{\end{array}}
\newcommand{\bd}{\begin{displaymath}}
\newcommand{\ed}{\end{displaymath}}
\newcommand{\bsube}{\begin{subequation}}
\newcommand{\esube}{\end{subequation}}
\newcommand{\bea}{\begin{eqnarray}}
\newcommand{\eea}{\end{eqnarray}}
\newcommand{\bal}{\begin{align}}
\newcommand{\ealign}{\end{align}}
\newcommand{\eal}{\end{align}}
\newcommand{\ben}{\begin{enumerate}}
\newcommand{\een}{\end{enumerate}}
\newcommand{\nn}{\nonumber}

\newcommand{\Slash}[1]{{\ooalign{\hfil/\hfil\crcr$#1$}}}

\graphicspath{ {figures/} }

\begin{document}

\title{Limiting the Heavy-quark and Gluon -philic Real Dark Matter}
\author{Sukanta Dutta}
\email{Sukanta.Dutta@gmail.com}
\affiliation{Delhi School of Analytics, Institute of Eminence, University of Delhi, Delhi.}
\affiliation{SGTB Khalsa College, University of Delhi, Delhi,}
\author{Lalit Kumar Saini}
\email{sainikrlalit@gmail.com}
\affiliation{Department of Physics \& Astrophysics, University of Delhi}

\date{\today}
\begin{abstract}
We investigate the phenomenological viability of real spin half, zero and one dark matter candidates, which interact predominantly with third generation heavy quarks and gluons via the twenty-eight gauge invariant higher dimensional effective operators. The   corresponding Wilson coefficients are constrained one at a time from  the relic density  $\Omega^{\rm DM} h^2$ $\approx$ 0.1198. Their contributions to the thermal averaged annihilation cross-sections are shown to be consistent with the FermiLAT and H.E.S.S. experiments' projected upper bound on the annihilation cross-section in the $b\,\bar b$ mode.
\par The tree-level gluon-philic and one-loop  induced heavy-quark-philic DM-nucleon direct-detection cross-sections  are analysed. The non-observation  of any excess over expected background in the case of  recoiled Xe-nucleus events for spin-independent DM-nucleus scattering in XENON-1T sets the upper limits on the eighteen Wilson coefficients. Our analysis validates the real DM candidates for the large range of  accessible mass spectrum below 2 TeV for all but one  interaction induced by the said operators.

\end{abstract}

\maketitle

\section{Introduction}\label{intro}
\par Regardless of the unequivocal  astrophysical evidences from rotation velocity curves~\cite{Rubin1970} via mass-to-luminosity ratio, Bullet Cluster \cite{Clowe:2006eq}, and precision measurements of the Cosmic Microwave Background 
(CMB) from WMAP \cite{Komatsu:2014ioa} and PLANCK \cite{Aghanim:2018eyx} satellites {\it etc.},  we have no clue about the fundamental nature of dark-matter (DM), which forms the  dominant matter component of the Universe. Dedicated experiments for the direct-detection of DM  such as LUX~\cite{Akerib:2012ys}, XENON-1T~\cite{XENON:2018voc}, DarkSide50~\cite{DarkSide:2018bpj}, PandaX-4T ~\cite{PandaX-4T:2021bab} and CRESST-III~\cite{CRESST:2019jnq}, PICO-60~\cite{PICO:2019vsc}, PICASSO \cite{Behnke:2016lsk} are designed to measure the momentum of the recoiled atom and/ or nucleus  due to the  scattering of DM particles off  the sub-atomic constituents of the detector material \cite{Schumann:2019eaa, Bauer:2017qwy, Billard:2021uyg}. We haven't seen any significant signal excess over the expected background yet. PandaX-4T \cite{PandaX-4T:2021bab} and PICO-60~\cite{PICO:2019vsc} have recently lowered the upper-limits of the measured sensitivities  at 90\% C.L. corresponding to  (a) spin-independent cross-sections at $3.3 \times 10^{-47}$ cm$^2$ for 30 GeV DM   and (b) spin-dependent cross-sections at $2.5 \times 10^{-41}$ cm$^2$  for 25 GeV DM respectively.  There are efforts being made to understand the nature of DM interactions by indirectly detecting DM resulting from DM pair annihilations  to SM particles using space based facilities such as Fermi-LAT~\cite{Ackermann:2015zua}, PLANCK data~\cite{Aghanim:2018eyx}, MAGIC~\cite{Acciari:2020pno} and some ground based large neutrino detectors such as  HESS~\cite{Abdallah:2016ygi}, Ice Cube~\cite{IceCube:2020wxa}, ANTARES~\cite{ANTARES:2019svn}, Super-Kamiokande~\cite{Super-Kamiokande:2011lwo} {\it etc}.  
\par Several experiments, past and ongoing, have constrained a plethora of viable UV complete dark matter models formulated by writing renormalizable Lagrangians with heavy non-SM mediators (spin 0$^\pm$, 1/2, 1, 2) facilitating interactions between DM and SM particles \cite{Bell:2016ekl, Albert:2016osu,Ko:2016zxg,Englert:2016joy,Dutta:2017jfj,Bauer:2017fsw,Bauer:2017ota,Baek:2017vzd,Carrillo-Monteverde:2018phy,Kraml:2017atm,LHCDarkMatterWorkingGroup:2018ufk}. The models with Higgs bosons as mediators have been excluded by the recent collider experiments \cite{ATLAS:2021pdg}. Analogous analysis has also been performed in the domain of the effective field theory (EFT) for the EW-Boson-philic DM operators \cite{Cotta:2012nj,Chen:2013gya,Crivellin:2015wva}. Recently, the GAMBIT collaboration performed a global analysis for signatures of SM gauge singlet Dirac fermion DM at LHC in the EFT set-up with simultaneous activation of fourteen effective operators constructed in association with light-quarks, gluons and photons  (up-to mass dimension seven) where the authors have disfavored  the exclusive contribution of   DM  $\le $ 100 GeV \cite{GAMBIT:2021rlp}. The possibility of sterile neutrinos as DM candidates interacting with the third generation fermions has been recently analysed in the EFT approach in reference \cite{Bischer:2020sop}.  The collider signatures of the effective Lepto-philic DM operators have been studied for the proposed ILC  through $e^+e^-\to \gamma + \slash\!\!\!\!E_{\rm T}$ \cite{Barman:2021hhg,Bharadwaj:2020aal,Bharadwaj:2019zdp,Chen:2015tia,Dreiner:2012xm,Chae:2012bq,Fox:2011pm} and $e^+e^-\to Z^0 + \slash\!\!\!\!E_{\rm T}$ \cite{Bell:2012rg, Rawat:2017fak} channels.  Sensitivity analysis for DM-quark effective interactions at LHC have been performed \cite{Kahlhoefer:2017dnp,Mitsou:2014wta,Boveia:2018yeb,DeSimone:2016fbz,Bhattacherjee:2012ch} in a model-independent way for the dominant (a) mono-jet + $\slash\!\!\!\!E_{\rm T}$, (b) mono-$b$ jet + $\slash\!\!\!\!E_{\rm T}$ and (c) mono-$t$ jet + $\slash\!\!\!\!E_{\rm T}$ processes.

\par Various top-philic DM inspired models have been studied and  constrained \cite{Baek:2017ykw,Gomez:2014lva, Kilic:2015vka,Colucci:2018vxz,Garny:2018icg,Fuks:2019ziv,Zhang:2012da} by the cosmological relic density criteria, direct-detection and indirect-detection  experiments. The interactions of heavy-quark-philic DM are also constrained by the ongoing collider experiments. The CMS \cite{CMS:2019zzl,CMS:2018ysw} and ATLAS \cite{ATLAS:2020yzc} collaborations  investigated the dominant scalar-mediated production of $t$-quark-philic DM particles in association with a single top quark or a pair of $t\, \bar t$. CMS collaboration recently conducted an exclusive search analysis for the $b$-philic DM in reference 
  \cite{CMS:2016uxr}. Many authors in the literature have explored mono-jet + $\slash \!\!\!E_T$ mode at the LHC for the  viable signatures of the scalar current induced heavy-quark-philic DM interactions \cite{Lin:2013sca,Haisch:2012kf,Bhattacherjee:2012ch, Goodman:2010ku}.  A comprehensive search analysis for heavy scalar mediated top-philic DM models  for mono-jets+ $\slash \!\!\!E_T$, mono-$Z$, mono-$h$ and $t\,\bar t$ pair productions at LHC can be found in \cite{Arina:2016cqj,Mattelaer:2015haa}.    

\par The phenomenology of  spin 1/2, 0 and 1 real DM deserves a special mention in the context of their contributions to the DM-nucleon scattering events in the direct-detection experiments. In the absence of the contributions from vanishing vector operators for Majorana and real vector DM, the spin-independent DM-nucleon scattering cross-sections are found to be dominated by the respective scalar and dimension-8 twist operators \cite{Drees:1993bu,Hisano:2010ct, Hisano:2010yh, Hisano:2011cs, Hisano:2011um, Hisano:2012wm, Hisano:2015bma, Hisano:2015rsa}. The pseudo-scalar current contribution to the DM - nucleon scattering cross-section is found to be spin-dependent and velocity suppressed.   Similar studies have also been undertaken for lepto-philic real DM candidates \cite{Bharadwaj:2020aal,Bharadwaj:2019zdp}. 

\par In this context, it is worthwhile to investigate the WIMP DM phenomenology induced by heavy-quark-philic and gluon-philic scalar, pseudo-scalar, axial-vector, and twist-2 real-DM operators and explore whether they satisfy the relic density criteria and other experimental constraints for a DM mass  between 10 GeV and 2 TeV. We introduce the effective Lagrangian for real spin 1/2, 0 and 1 DM particles in section \ref{model}. In section \ref{dmpheno}, we discuss the phenomenology of the dark-matter. We investigate the cosmological constraints on the DM  in sub-section \ref{relic}. In sub-section \ref{indirect}, we predict and analyse the thermally averaged DM pair annihilation cross-section. Sub-section \ref{direct} details the computation of  DM-nucleon scattering cross-sections and expected recoil nucleus event(s)  for XENON-1T \cite{XENON:2018voc} set up. Section \ref{summary}  summarises our study and observations.


\section{Heavy-Quark and Gluon-philic DM Effective Operators}
\label{model}
It is a fact that in a beyond-standard model (BSM) renormalizable gauge theory, as long as the energy range of DM-SM interaction is much below the mediator mass, the study of DM interactions can be restricted by the DM and SM degrees of freedom and their symmetries.

The higher dimensional effective operators are obtained from the BSM Lagrangian by writing the operator product expansion (OPE) of currents in the limit 
$p^2/\, m_{\rm Med.}^2 <<$ 1, where $p^\mu$ represents the four momentum of the virtual mediator of mass $m_{\rm Med.}$.
This facilitates the effective contact interaction between DM and any SM third generation heavy quark/ gluon,  assuming that the mediator  mass scale $m_{\rm Med.}$  is of the order of the effective theory cut-off $\left(\sim \Lambda_{\rm eff}\right)$, which is much heavier than the masses of the SM and  DM  fields in general. 
\par For example, in renormalizable models, the interaction between a Majorana DM $\chi$ and a third generation quark $\psi$ can be written  as
\bea
{\cal L}^{\rm Dim.\,4}&=& \bar{\psi} \left(a + b\ \gamma_5\right) \chi\, \eta  + \bar{\psi}\,\gamma^\mu \left(c + d\ \gamma_5\right) \chi \,\zeta_\mu+\, {\rm h.c.} \label{dim4lag}
\eea
where $\eta$ and $\zeta_\mu$ are  electrically charged scalar and vector fields respectively. This interaction Lagrangian  facilitates the Majorana DM - quark interaction via $t$-channel exchange of the heavy $\eta$ and/ or $\zeta_\mu$. Thus, expanding the propagator in powers of $p^2/m_{\rm Med}^2$, yields the higher-dimensional effective four-fermion interaction for Majorana DM and heavy-quarks. The scalar, pseudo-scalar, axial-vector, and twist-2 operators are all induced by this expansion \cite{Drees:1993bu}. When compared to a spin-2 graviton mediated model, the twist operator corresponds to the contribution from traceless part of the energy-momentum tensor T$^{\mu\nu}$~\cite{Carrillo-Monteverde:2018jcs}. In various WIMP-inspired renormalizable electroweak models, such effective interactions between the SM and Majorana DM particles are also realised by $s$-channel processes, where the interactions are mediated by non-SM heavy scalar/pseudo-scalar/axial-vector or spin-2 tensor particles ~\cite{Morgante:2018tiq,Lee:2014caa}. 
\par Since the DM-gluon interactions can be naturally realised via one-loop interactions of the DM particles either with SM heavy quarks or BSM non-singlet coloured spin 1/2, 0 and 1 exotics at the next order in the strong coupling constant, it becomes all the more necessary to include the study of effective operators constructed independently with a Majorana DM bilinear and a pair of gluons at the leading order of $\sim \alpha_{\rm s}/\,\pi$.
\par The  phenomenological effective Lagrangian for the heavy-quark-philic and gluon-philic Majorana DM, $\chi$, is written  as:
	\begin{eqnarray}
	{\cal  L}_{\rm eff}^{\chi} &=& \frac{C^q_{\chi_S}}{\Lambda^3}\, \mathcal{O}^q_{\chi_S} +\frac{C^q_{\chi_{\rm PS}}}{\Lambda^3}\, \mathcal{O}^q_{\chi_{\rm PS}} + \frac{C^g_{\chi_S}}{\Lambda^4}\, \mathcal{O}^g_{\chi_S} +\frac{C^g_{\chi_{\rm PS}}}{\Lambda^4}\, \mathcal{O}^g_{\chi_{\rm PS}} \nn\\
	&&\,\,\,\,\,\, + \frac{C^q_{\chi_{\rm AV}}}{\Lambda^2}\, \mathcal{O}^q_{\chi_{\rm AV}}+  \frac{C^p_{\chi_{T_{1}}}}{\Lambda^4}\, \mathcal{O}^p_{\chi_{T_{1}}}+\frac{C^p_{\chi_{T_{2}}}}{\Lambda^5}\, \mathcal{O}^p_{\chi_{T_{2}}} \,\,\, \label{MDMLagaux} 
	\end{eqnarray}
The $\mathcal{O}^q_{\chi_S}, \mathcal{O}^q_{\chi_{\rm PS}}, \mathcal{O}^q_{\chi_{\rm AV}}$ and $\mathcal{O}^q_{\chi_{T_{i}}}$ representing the third generation quark-philic scalar, pseudo-scalar, axial-vector and twist-2 type-1 and type 2 operators  respectively  along with $\mathcal{O}^g_{\chi_S}, \mathcal{O}^g_{\chi_{\rm PS}}$ and $\mathcal{O}^g_{\chi_{T_{i}}}$ representing the gluon-philic scalar, pseudo-scalar and twist-2 type-1 and type-2 operators respectively are defined as
\begin{eqnarray}
&&\begin{array}{cclccl}\mathcal{O}^q_{\chi_S} &=& m_q\, \left(\bar{\chi}\,\chi\right)\,\,\left( \bar{q}\, q\right);&\hskip 2 cm  & \mathcal{O}^g_{\chi_S} =& \frac{\alpha_s}{\pi}\, m_\chi\, \left(\bar{\chi}\,\chi\right)\, \, G^A_{\mu\nu}\,\, {G^A}^{\mu\nu};\\
	\mathcal{O}^q_{\chi_{\rm PS}} &=& m_q\, \left(\bar{\chi}\,\gamma_5\,\chi\right)\,\,\left(\bar{q}\,\gamma_5\,q\right);& \hskip 2 cm   &\mathcal{O}^g_{\chi_{\rm PS}} =& \frac{\alpha_s}{\pi}\, m_\chi\, \left(i\,\bar{\chi}\,\gamma_5\,\chi\right) \, \,G^A_{\mu\nu}\,\,\widetilde{{G^A}^{\mu\nu} };\\
	\mathcal{O}^q_{\chi_{\rm AV}} &=& \left(\bar{\chi}\,\gamma_\mu \,\gamma_5\, \chi\right)\,\left( \bar{q} \,\gamma^\mu\, \gamma_5\, q\right); &\hskip 2 cm& & \\
\mathcal{O}^q_{\chi_{T_{1}}} &=&\left( \bar{\chi} \,i \partial^\mu \,\gamma^\nu \,\chi\right)\,\, \,\mathcal{O}^q_{\mu\nu};& \hskip 2 cm   &  \mathcal{O}^g_{\chi_{T_{1}}} =&\left( \bar{\chi} \,i \partial^\mu \,\gamma^\nu \,\chi\right)\,\, \,\mathcal{O}^g_{\mu\nu};\\
\mathcal{O}^q_{\chi_{T_{2}}} &=&\left( \bar{\chi}\, i \partial^\mu\, i\partial^\nu\, \chi\right)\, \,\,\mathcal{O}^q_{\mu\nu};& \hskip 2 cm & \mathcal{O}^g_{\chi_{T_{2}}} =&\left( \bar{\chi}\, i \partial^\mu\, i\partial^\nu\, \chi\right)\, \,\,\mathcal{O}^g_{\mu\nu};
	\end{array}
		 \label{MDMLag}
\end{eqnarray}
The second rank twist tensor currents ${\cal O}^{q}_{\mu\nu}$ and ${\cal O}^{g}_{\mu\nu}$ for the heavy-quarks and gluons respectively are given as
\begin{subequations} 
\begin{eqnarray}\label{twist2op}
 {\cal O}^{q}_{\mu\nu}&\equiv &i\,\frac{1}{2}\overline{q_L}\biggl(
{D_\mu^{}}_L\gamma_\nu^{} +{D_\nu^{}}_L\gamma_\mu^{}-\frac{1}{2}g_{\mu\nu}^{}
{\Slash{D}}_L\biggr)q_L+\ i\,\frac{1}{2}\overline{q_R}\biggl(
{D_\mu^{}}_R\gamma_\nu^{} +{D_\nu^{}}_R\gamma_\mu^{}-\frac{1}{2}g_{\mu\nu}^{}
{\Slash{D}}_R\biggr)q_R\nn\\
\\
{\cal O}^g_{\mu\nu}&\equiv& 
\left[\left(G^A\right)^{~\rho}_{\mu} \left(G^A\right)_{\nu\rho}-\frac{1}{4}\,g_{\mu\nu}\,\,
\left(G^A\right)_{\rho\sigma}\,\,\left(G^A\right)^{\rho\sigma}\right],
\end{eqnarray}
\end{subequations}
where ${D_\mu}_L$ and ${D_\mu}_R$ are the covariant derivatives for left and right handed quarks respectively in SM. The contribution from the vector operator vanishes for the real particles. We exclude the contribution of the dimension-9 twist-2 type-2 operators $\mathcal{O}^g_{\chi_{T_{2}}}$ in our analysis because we are only interested in the  effective operators up to mass dimension-8. 

\par We extend the domain of our analysis to include the effective contact  interactions of  real scalar $\phi^0$ and vector $V^0_\mu$ DM candidates  with  SM third generation heavy  quarks  and gluons.  The effective scalar  DM Lagrangian are given as 
\begin{eqnarray}
\mathcal{L}_{eff}^{\phi^0} &=& \frac{C^q_{\phi^0_S}}{\Lambda^2}\, \mathcal{O}^q_{\phi^0_S} + \frac{C_{\phi^0_S}^g}{\Lambda^2}\, \mathcal{O}_{\phi^0_S}^g + \frac{C^p_{\phi^0_{T_{2}}}}{\Lambda^4}\, \mathcal{O}^p_{\phi^0_{T_{2}}}\,\,\, \end{eqnarray}
where
\begin{eqnarray}
\begin{array}{cclcccc}
\mathcal{O}^q_{\phi^0_S} &=& \left(\phi^0\, \phi^0\right)\, m_q\, \left( \bar{q}\,q\right); &\hskip 2cm & \mathcal{O}^g_{\phi^0_S} &=& \frac{\alpha_s}{\pi}\,\left(\phi^0\, \phi^0\right)\, G^A_{\mu\nu}\,\, G^{A\mu\nu} ;\\
\mathcal{O}^q_{\phi^0_{T_{2}}} &=& \left(\phi^0\, i \partial^\mu\, i\partial^\nu\, \phi^0\right)\, \,\,\mathcal{O}^q_{\mu\nu}; &\hskip 2cm & \mathcal{O}^g_{\phi^0_{T_{2}}} &=& \left(\phi^0\, i \partial^\mu\, i\partial^\nu\, \phi^0\right)\, \,\,\mathcal{O}^g_{\mu\nu};\\
\end{array}\label{SDMLag}
\end{eqnarray}
The $\mathcal{O}^{q/g}_{\phi^0_S}$ and $\mathcal{O}^{q/g}_{\phi^0_{T_{2}}}$ are the scalar and second rank twist-2 type-2 operators respectively. There are no contributions from pseudo-scalar and axial vector currents for the scalar DM operators. 
\par The effective vector DM Lagrangian is given as
\begin{eqnarray}
\mathcal{L}_{eff}^{V^0} &=& \frac{C^q_{V^0_S}}{\Lambda^2}\, \mathcal{O}^q_{V^0_S} +
\frac{C^q_{V^0_{\rm PS}}}{\Lambda^4}\, \mathcal{O}^q_{V^0_{\rm PS}} +
 \frac{C_{V^0_S}^g}{\Lambda^2}\, \mathcal{O}_{V^0_S}^g + \frac{C_{V^0_{\rm PS}}^g}{\Lambda^4}\, \mathcal{O}_{V^0_{\rm {\rm PS}}}^g  + \frac{C^q_{V^0_{\rm AV}}}{\Lambda^2}\, \mathcal{O}^q_{V^0_{\rm AV}}+ \frac{C^p_{V^0_{T_{2}}}}{\Lambda^4}\, \mathcal{O}^p_{V^0_{T_{2}}}
 \label{VDMLag}\end{eqnarray}
The heavy-quark-philic  $\mathcal{O}^q_{V^0_S}$, $\mathcal{O}^q_{V^0_{\rm PS}}$, $\mathcal{O}^q_{V^0_{\rm AV}}$, $\mathcal{O}^q_{V^0_{T_{2}}}$ representing the third generation quark-philic scalar, pseudo-scalar, axial-vector and twist-2 type-2 operators respectively and gluon-philic $\mathcal{O}^g_{V^0_S}$, $\mathcal{O}^g_{V^0_{\rm PS}}$ and $\mathcal{O}^g_{V^0_{T_{2}}}$ operators corresponding to the scalar, pseudo-scalar and second rank twist-2 type-2 interactions respectively are defined as
\begin{eqnarray}
\begin{array}{cclcccl}
\mathcal{O}^q_{V^0_S} &=& \left(V^0\right)^\rho \,\left(V^0\right)_\rho\, m_q\,\left(  \bar{q}\,q\right); &\hskip 3 cm & \mathcal{O}^g_{V^0_S} &=& \frac{\alpha_s}{\pi} \,\left(V^0\right)^\rho \,\left(V^0\right)_\rho\, G^A_{\mu\nu}\,\, {G^A}^{\mu\nu};\\
	\mathcal{O}^q_{V^0_{\rm PS}} &=& \left(V^0\right)^{\rho\sigma} \,\,\widetilde{\left(V^0\right)_{\rho\sigma}}\, m_q\,\left(i\,\bar{q}\,\gamma_5\,q\right)  ;&\hskip  3 cm & \mathcal{O}^g_{V^0_{\rm PS}} &=& \frac{\alpha_s}{\pi} \,\left(V^0\right)^{\rho\sigma} \,\widetilde{\left(V^0\right)_{\rho\sigma}}\, G^A_{\mu\nu}\,\widetilde{{G^A}^{\mu\nu}};  \\
	\mathcal{O}^q_{V^0_{\rm AV}} &=& i \epsilon_{\mu\nu\rho\sigma} \left(V^0\right)^\mu i \partial^\nu \left(V^0\right)^\rho \bar{q} \gamma^\sigma \gamma_5 q\,;&\hskip  3 cm & & & \\
\mathcal{O}^q_{V^0_{T_{2}}} &=& \left(V^\rho\right) i \partial^\mu i\partial^\nu \left(V_\rho\right)\, \mathcal{O}^q_{\mu\nu} ;
&\hskip  3 cm & \mathcal{O}^g_{V^0_{T_{2}}} &=& \left(V^\rho\right) i \partial^\mu i\partial^\nu \left(V_\rho\right)\, \mathcal{O}^g_{\mu\nu};
\\
\end{array}
\label{VDMLag}
\end{eqnarray}
where $\epsilon^{\mu\nu\rho\sigma}$ is the totally antisymmetric tensor with $\epsilon^{0123}$ = +1.
\par In this study, each operator's phenomenology is investigated independently, assuming that the unique interaction contributes to the total relic density for DM.


\section{Phenomenological Constraints on Real DM}
\label{dmpheno}
\subsection{Contribution of DM to the relic density}\label{relic}
The present day relic abundance of the DM species $n_{\rm DM}(t)$  can be calculated by solving the Boltzmann equation 
\begin{eqnarray}
	\frac{dn_{\rm DM}}{dt} + 3 \,H_0\, n_{\rm DM} &=& -\left\langle \sigma_{\rm ann} \left\vert \vec v_{\rm DM}\right \vert \right\rangle \bigg( (n_{\rm DM})^2 - (n_{\rm DM}^{\rm eq})^2 \bigg);\,\,\, 
	 \label{BE}
\end{eqnarray}
where $n_{\rm DM}^{\rm eq}$ is  the DM number density at thermal equilibrium, $H_0$ is the Hubble constant, $\vert \vec v_{\rm DM}\vert$ is relative velocity of the DM pair and $\left\langle \sigma_{\rm ann} \left\vert \vec v_{\rm DM}\right \vert \right\rangle$ is the thermal average of the annihilation cross-section~\cite{Gondolo:1990dk}. It is customary to parametrise $\rho_{\rm DM} \equiv \Omega_{\rm DM} h^2\,\rho_c$, where $\rho_ c \equiv 1.05373 \times 10^{-5}\, h^2 /c^2 \,\,{\rm GeV\, cm^{-3}}$ is the critical density of the universe and dimensionless $h$ is the current Hubble constant in units of 100 km/s/Mpc. Solving the Boltzmann equation \cite{Kolb:1990vq} we then get
\begin{eqnarray}
	{\Omega}^{\rm DM} h^2 =\frac{\rho^{\rm DM}}{\rho_{\rm critical}}\,h^2&\approx& 0.12 \left( \frac{2.2 \times 10^{-26}\ {\rm cm^3/s}}{\left\langle \sigma_{\rm ann} \left\vert\vec v_{\rm DM}\right\vert\right\rangle}\right) \left( \frac{80}{g_{\rm eff}}\right)^{1/2} \left( \frac{m_{DM}/T_F}{23}\right)  \label{RelicForm}
\end{eqnarray}
where parameter $x_F \equiv m_{\rm DM}/T_F$ is  a function of degrees of freedom of the DM $g$, the effective massless degrees-of-freedom $g_{\rm eff}$ ($\sim$ 106.75 and 86.75 above $m_t$ and $m_b$  mass-thresholds respectively) at freeze-out temperature $T_F$: 
\begin{eqnarray}
	x_F = \ln \Bigg[ a(a+2) \sqrt{\frac{45}{8}}\ \frac{g\, M_{pl}}{2 \pi^3}\ \frac{m_{\rm DM}\langle\sigma_{\rm ann}\left\vert \vec v_{\rm DM} \right\vert\rangle}{\sqrt{x_F} \, g_{\rm eff}(x_F)} \Bigg];\hskip 0.2cm a\sim {\cal O}(1)\,\,\,{\rm and}\,\,\, M_{pl}=1.22 \times 10^{19}\,{\rm GeV}.\nn\\ \label{xF}
\end{eqnarray}

\par  The predicted  DM relic density  $\Omega^{\rm DM} h^2$ of  $0.1138\pm 0.0045$ and $0.1198\pm 0.0012$  by WMAP   \cite{Komatsu:2014ioa} and Planck   \cite{Aghanim:2018eyx} collaborations respectively restrict  the thermal averaged  DM annihilation cross-section $\left\langle \sigma \left\vert \vec v_{\rm DM}\right\vert \right\rangle\ge 2.2 \times 10^{-26}$ ${\rm cm^3\ s^{-1}}$ as a smaller thermally averaged cross-section would render large DM abundance which will over-close the universe. We have analytically calculated the DM pair annihilation cross-sections  to a pair of third generation heavy quarks and  gluons in the Appendix \ref{annihilationCS}. 
\par We investigate the DM  relic density contributions propelled by the thermally averaged annihilation cross-sections given in the appendix \ref{TherAvgCS} corresponding to the scalar, pseudo-scalar,  axial-vector and twist-2 currents of the heavy quarks for a given Majorana/\, scalar/\, vector DM.  The  annihilation channels induced by the scalar, pseudo-scalar and twist-2 currents of gluons  also contribute to the relic abundance of Majorana, scalar and vector DM.

\par Throughout our investigation, we assumed that the operators would be actuated one at a time. As a result, the constraints on all twenty-eight Wilson coefficients $  \left\vert C^{q,\,g}_{{\rm DM}_i\,O_j}/\, \Lambda^n\right\vert$ in TeV$^{-n}$ (${\rm DM}_i\equiv \chi,\,\phi^0,\,V^0$ and $O_j\equiv S/\,{\rm PS}/\, {\rm AV}/\, T_1/\, T_2$ are derived by triggering  the lone contribution from the associated operator to fulfill the relic density $\Omega^{\rm DM} h^2\approx 0.1198\pm 0.0012$ \cite{Aghanim:2018eyx}. Due to its only action, these constraints can be regarded as a cosmologically acceptable lower-limits on the corresponding Wilson Coefficients, which translate to upper bounds on the respective cutoffs $  \left\vert C^{q,\,g}_{{\rm DM}_i\,O_j}\right\vert^{-1/n}\,\Lambda$ in units of TeV for a given DM mass.
When the Wilson coefficient is greater than its lower-limit for a given DM mass, it partially meets the relic density.

\begin{figure}[h!]
	\centering
\hspace{-0.5cm}
  \begin{subfigure}{0.5\textwidth}
      \centering
	\includegraphics[scale=0.5]{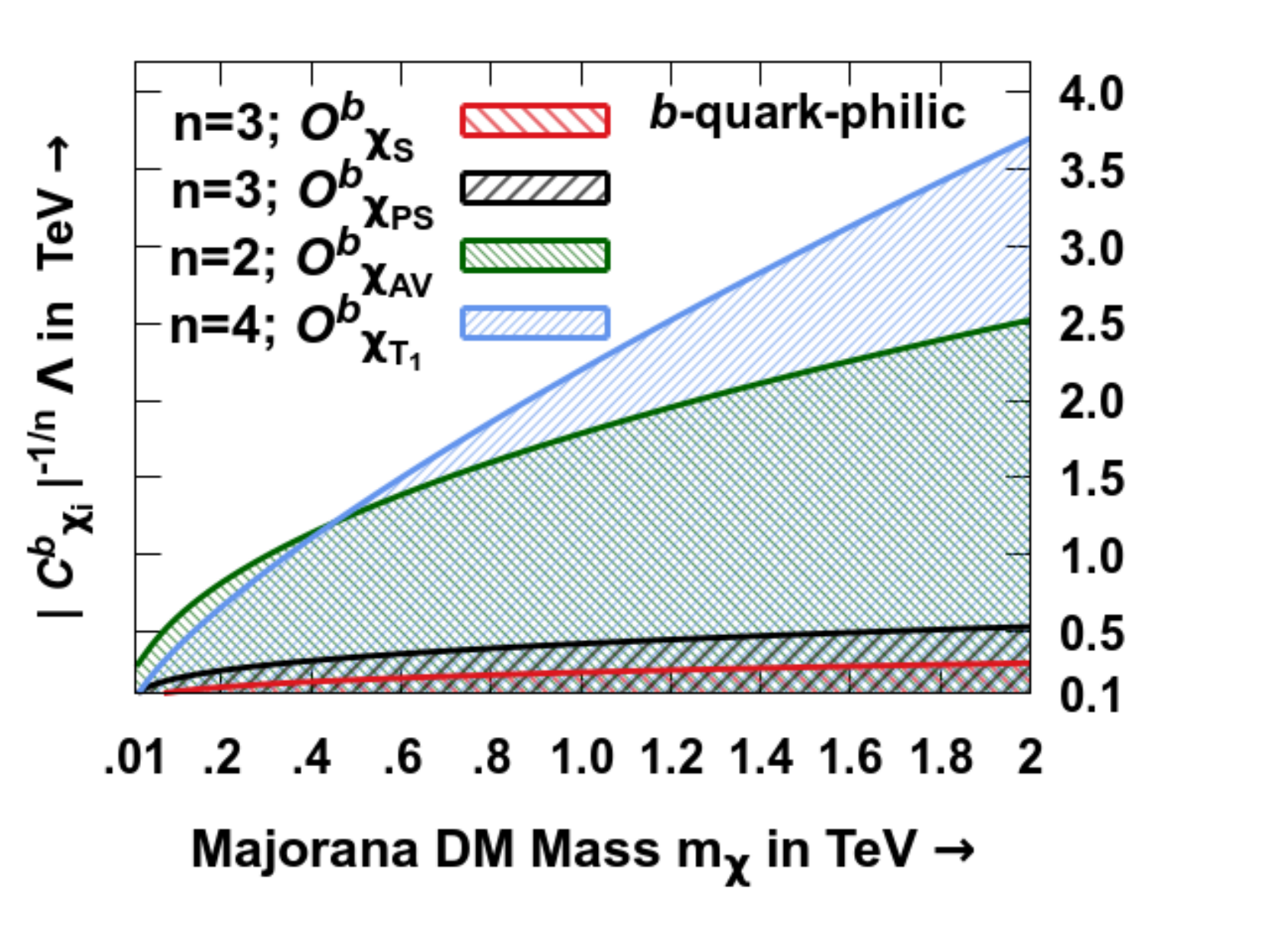}
      \caption{\em{}}	\label{fig:RelicFDMB}
  \end{subfigure}%
\hspace{0.2cm}
  \begin{subfigure}{0.5\textwidth}
      \centering
	\includegraphics[scale=0.5]{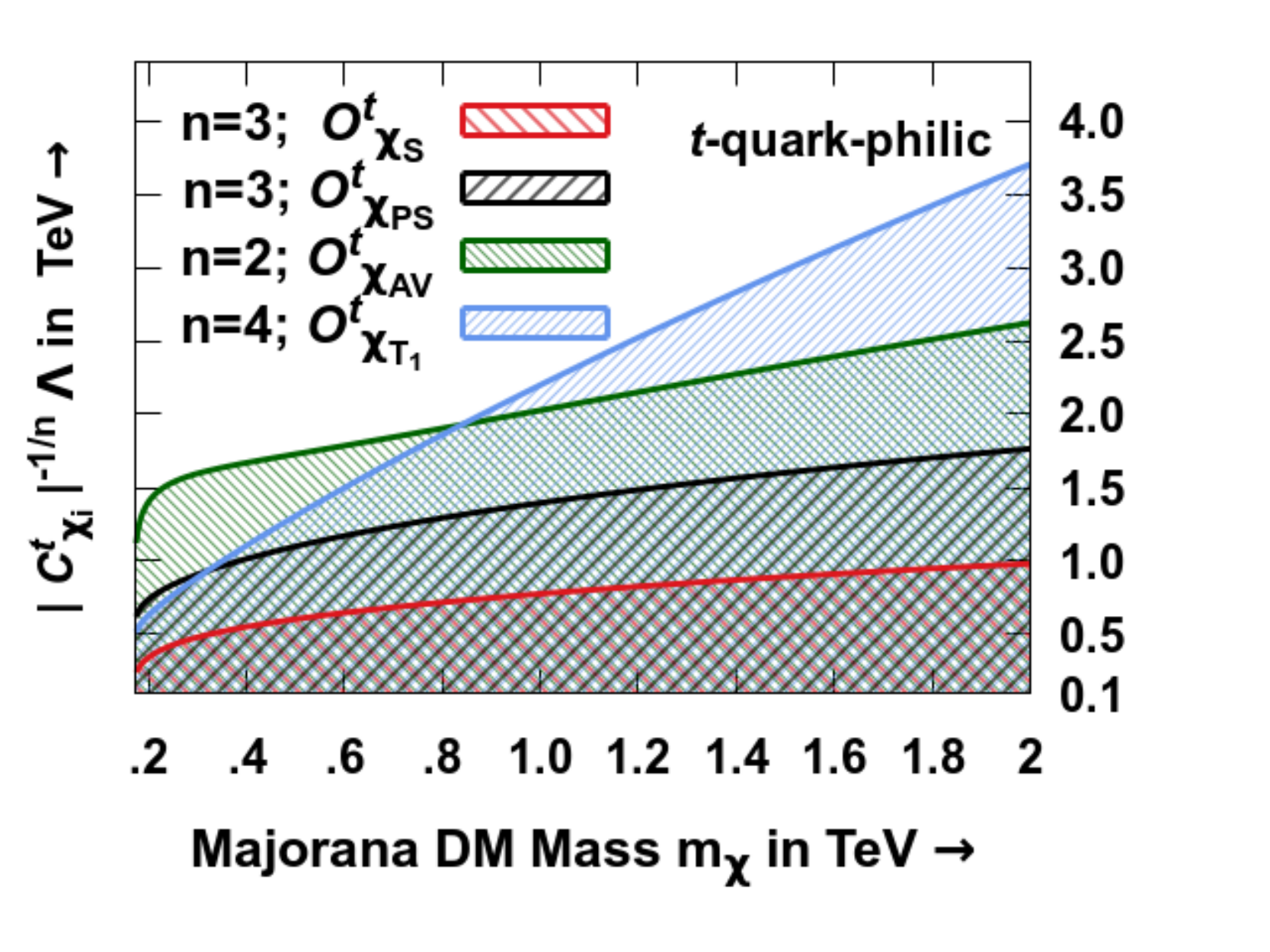}
      \caption{\em{}}	\label{fig:RelicFDMT}
  \end{subfigure}%
  \vskip 0.1cm
  \begin{subfigure}{0.5\textwidth}
      \centering
	\includegraphics[scale=0.5]{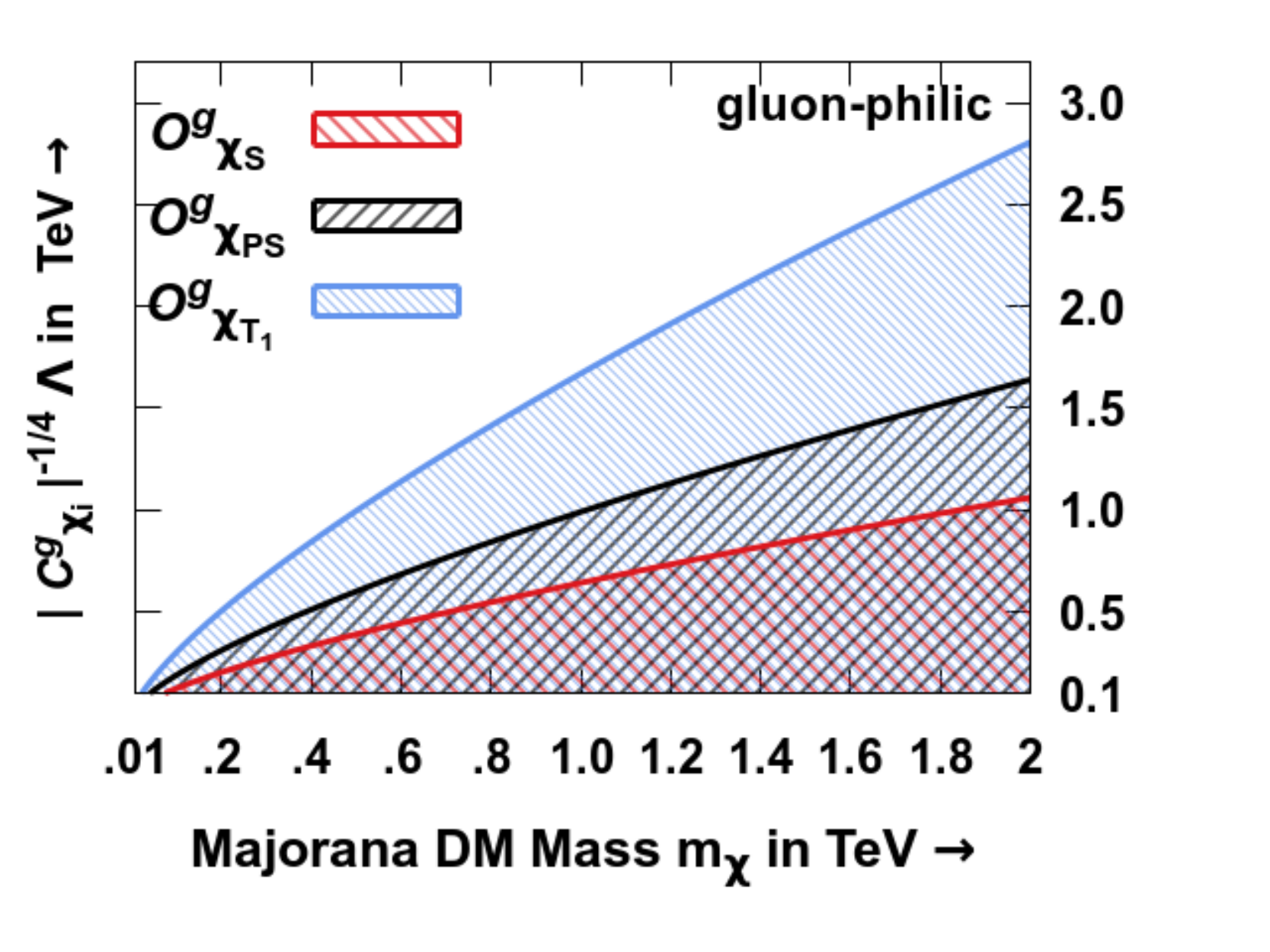}
      \caption{\em{}}	\label{fig:RelicFDMG}
  \end{subfigure}%
	\caption[nooneline]{\justifying \em {We depict relic density contours satisfying $\Omega^{\chi} h^2$ = 0.1198~\cite{Aghanim:2018eyx} and shaded cosmologically allowed regions in the plane defined by Majorana DM mass $m_\chi$ and $\vert C^{q/g}_{\chi_{{\rm S},\,{\rm PS},\,{\rm AV},\, T_1}}\vert^{-1/n}\Lambda$. The $b$-quark-philic and $t$-quark-philic Majorana DM contours in figures\ref{fig:RelicFDMB} and \ref{fig:RelicFDMT} are drawn corresponding to  scalar, pseudo-scalar,  axial-vector  and twist-2 type-1 operators. The  gluon-philic Majorana DM contours in figure \ref{fig:RelicFDMG} are drawn for scalar, pseudo-scalar and twist-2 type-1 operators.}} \label{fig:RelicFermDM}  
\end{figure}

\begin{figure}[h!]
	\centering
\hspace{-0.5cm}
  \begin{subfigure}{0.5\textwidth}
      \centering
	\includegraphics[scale=0.5]{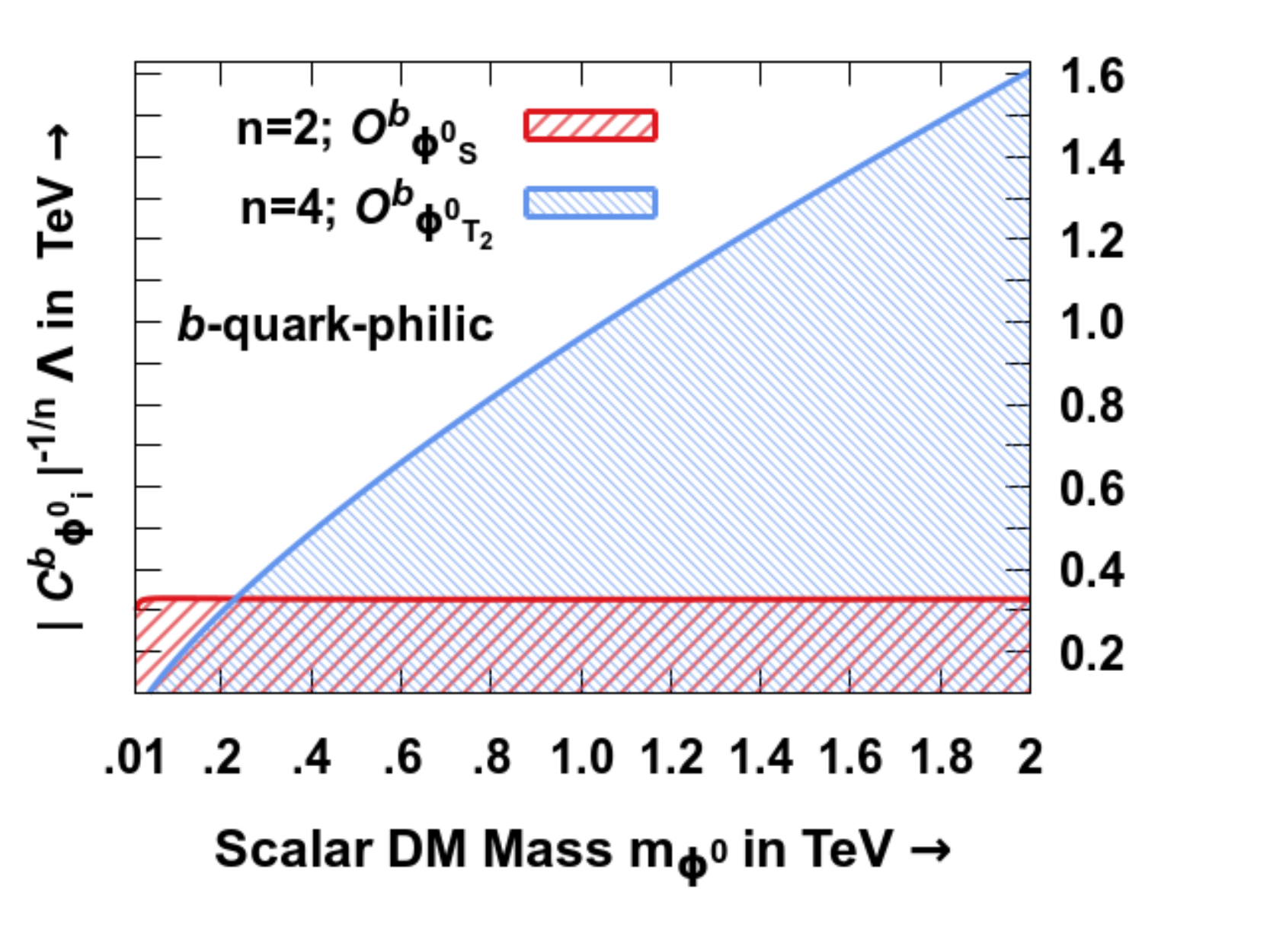}
      \caption{}	\label{fig:RelicSDMB}
  \end{subfigure}%
\hspace{0.2cm}
  \begin{subfigure}{0.5\textwidth}
      \centering
	\includegraphics[scale=0.5]{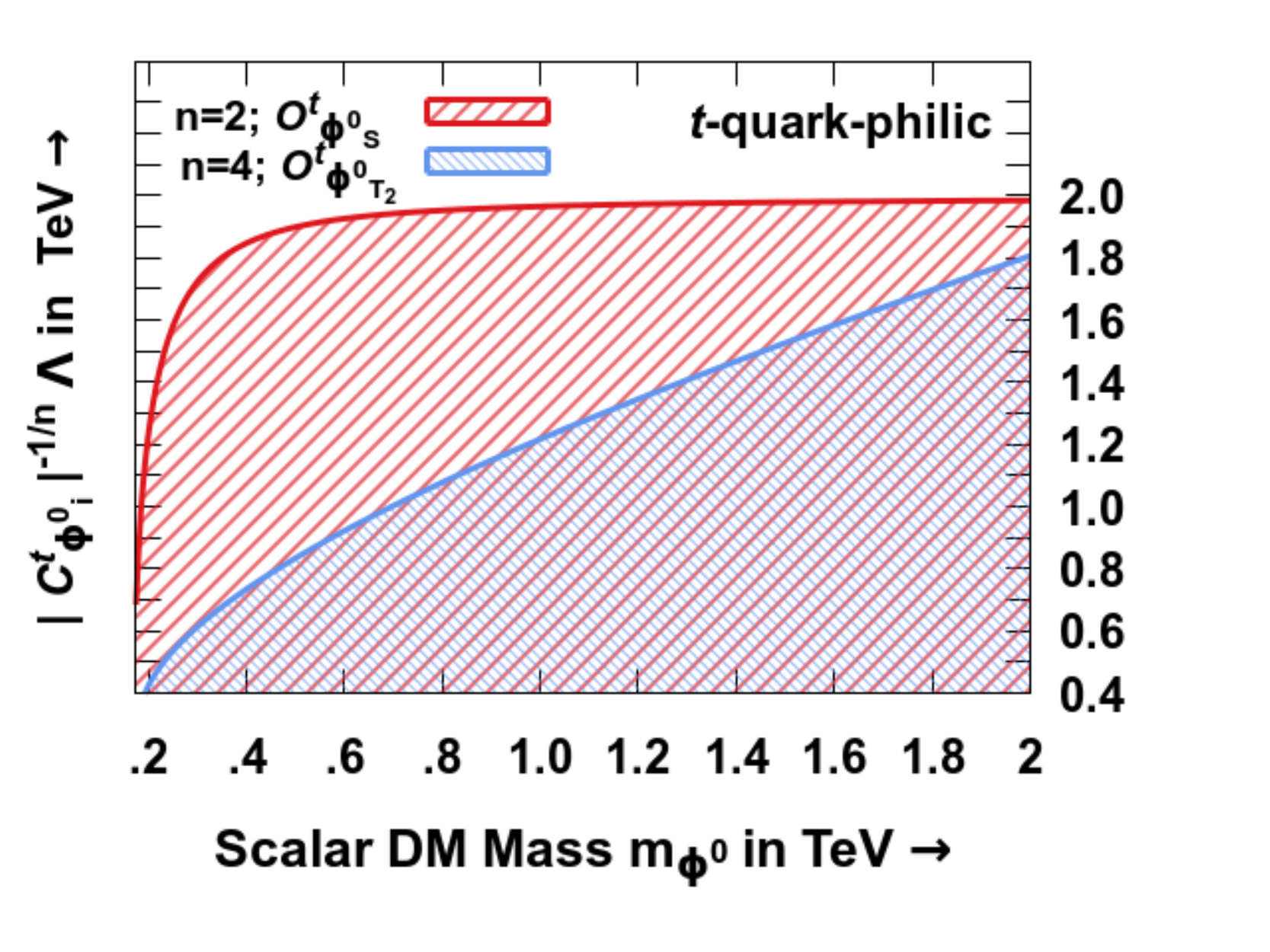}
      \caption{}	\label{fig:RelicSDMT}
  \end{subfigure}%
    \vskip 0.1cm
  \begin{subfigure}{0.5\textwidth}
      \centering
	\includegraphics[scale=0.5]{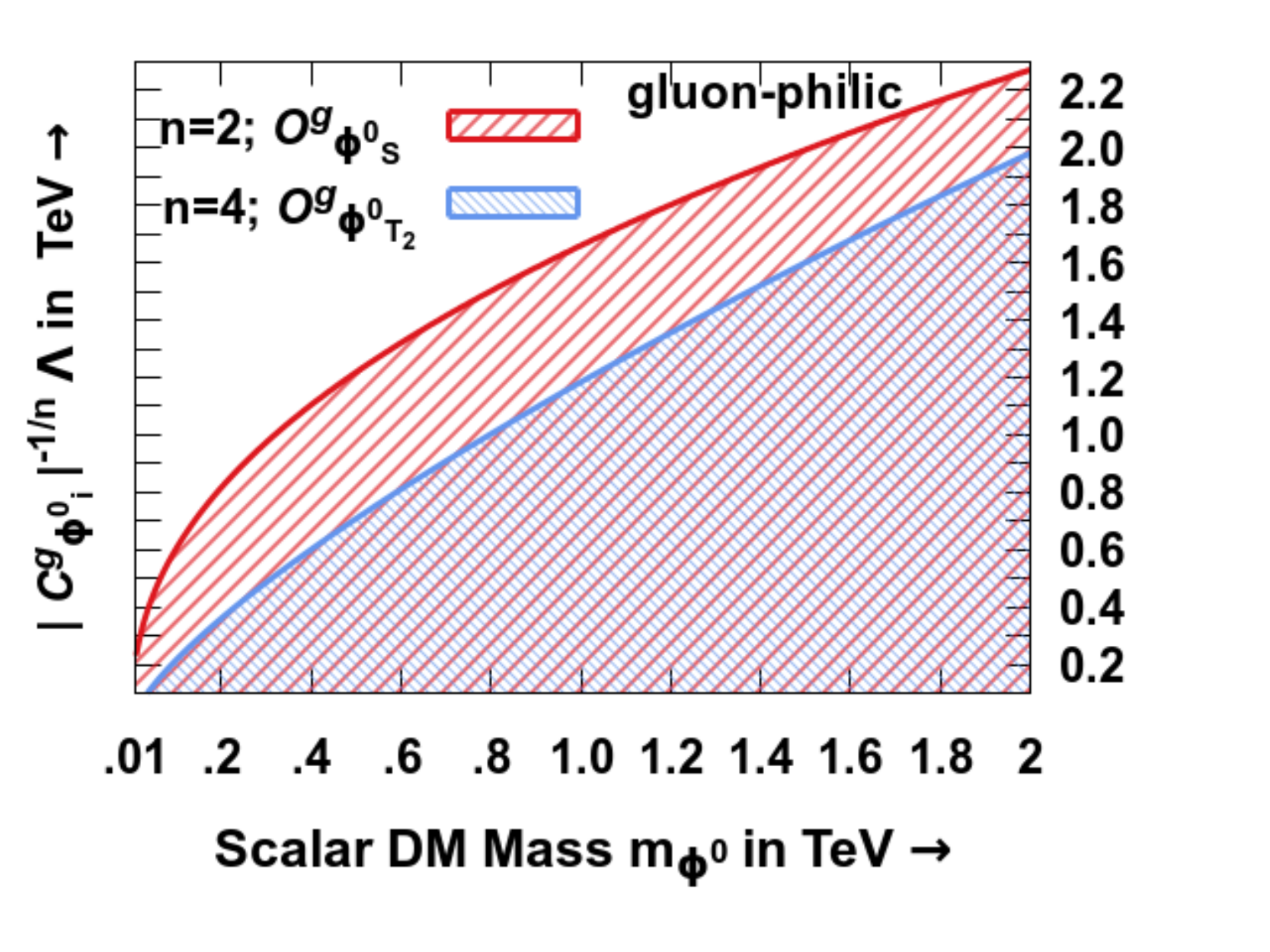}
      \caption{}	\label{fig:RelicSDMG}
  \end{subfigure}%
	\caption[nooneline]{\justifying \em{We depict relic density contours satisfying $\Omega^{\phi^0} h^2$ = 0.1198~\cite{Aghanim:2018eyx} and shaded cosmologically allowed regions  in the plane defined by Majorana DM mass $m_{\phi^0}$ and $\vert C^{q/g}_{\phi^0_{{\rm S},\, T_1}}\vert^{-1/n}\Lambda$. The $b$-quark-philic, $t$-quark-philic and gluon-philic scalar DM contours in figures \ref{fig:RelicSDMB},  \ref{fig:RelicSDMT} and \ref{fig:RelicSDMG} are drawn corresponding to scalar and twist-2 type-2 operators respectively.}} \label{fig:RelicScalDM}  
\end{figure}

\par It is important to note that because we can raise the annihilation cross-section by turning on more Wilson coefficients, we can easily imagine a scenario in which they all reside below their lower-limit without over-closing the universe with DM. To put it another way, if we set two distinct Wilson coefficients to their lower limits for a given DM mass, we plainly under-produce DM because the overall annihilation cross-section is larger. Therefore, in order to be consistent with a multiple species of DM model satisfying the  relic density constraint, this lower-bound on the specific Wilson coefficient will further reduce  when more than one coupling are triggered simultaneously. However, the rest of our analysis is focused on the scenario where the sole operator contributes to the relic density.

\par
\begin{figure}[h!]
	\centering
\hspace{-0.5cm}
  \begin{subfigure}{0.5\textwidth}
      \centering
	\includegraphics[scale=0.5]{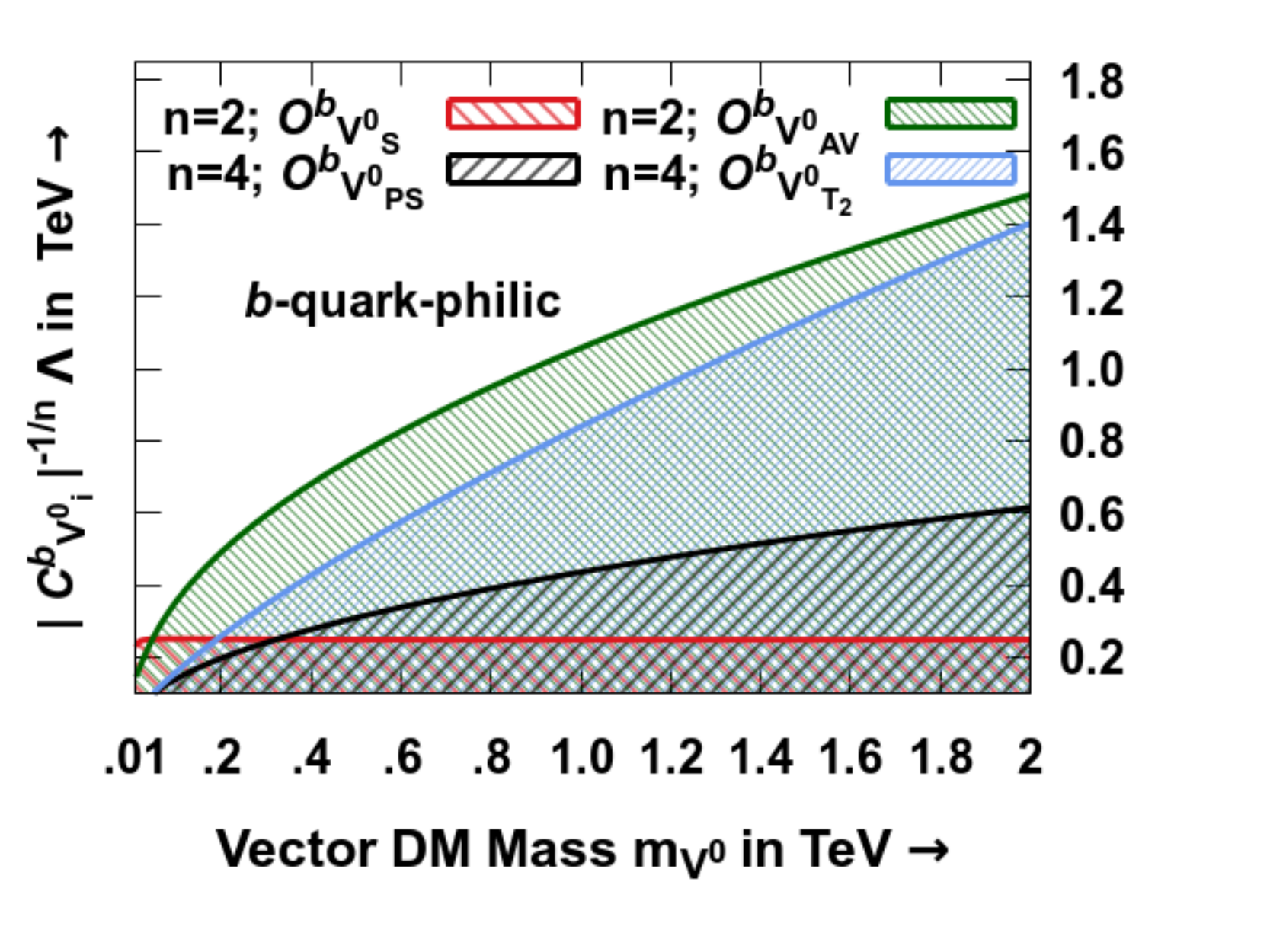}
      \caption{}	\label{fig:RelicVDMB}
  \end{subfigure}%
\hspace{0.2cm}
  \begin{subfigure}{0.5\textwidth}
      \centering
	\includegraphics[scale=0.5]{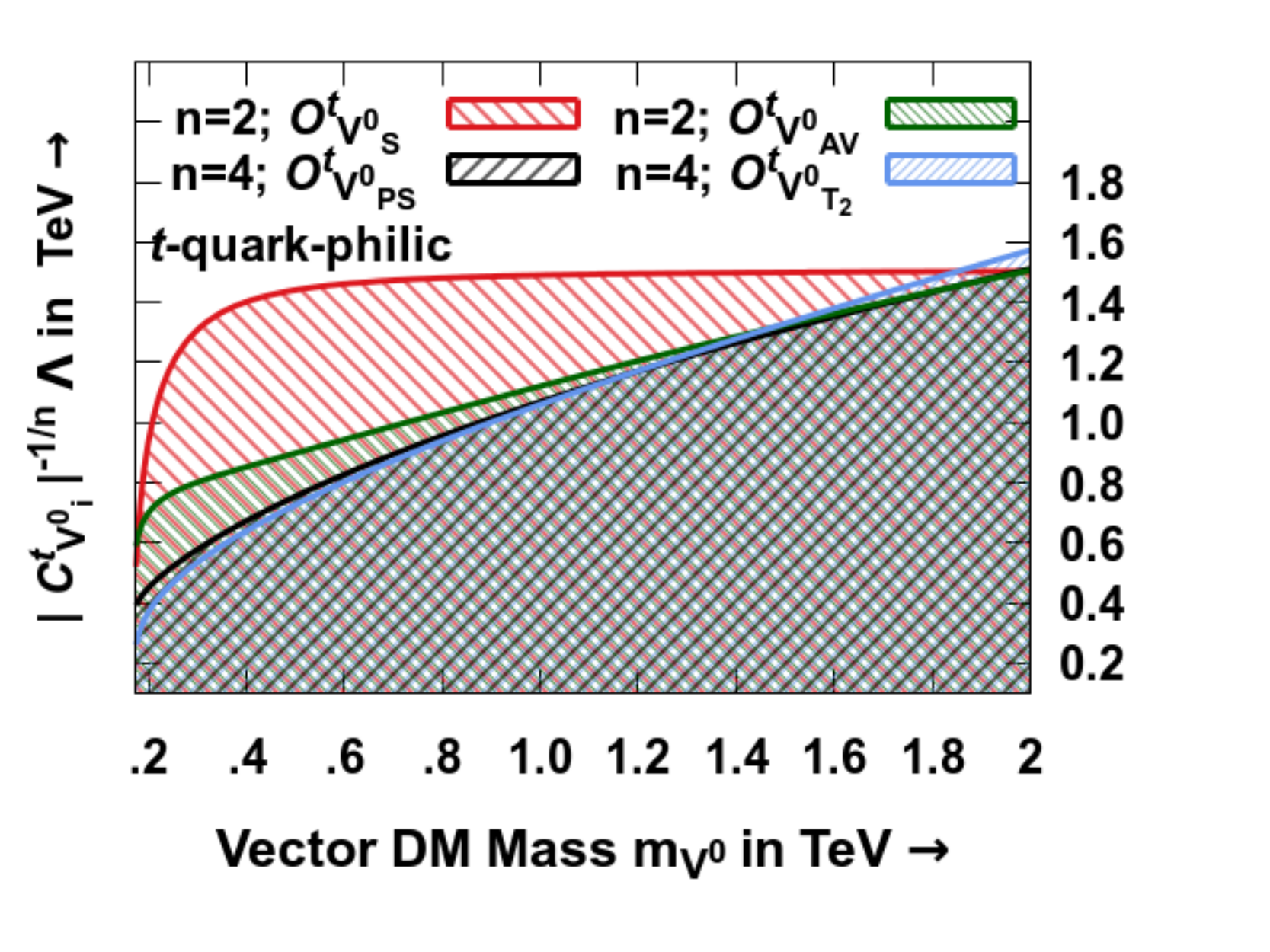}
      \caption{}	\label{fig:RelicVDMT}
  \end{subfigure}%
    \vskip 0.1cm
  \begin{subfigure}{0.5\textwidth}
      \centering
	\includegraphics[scale=0.5]{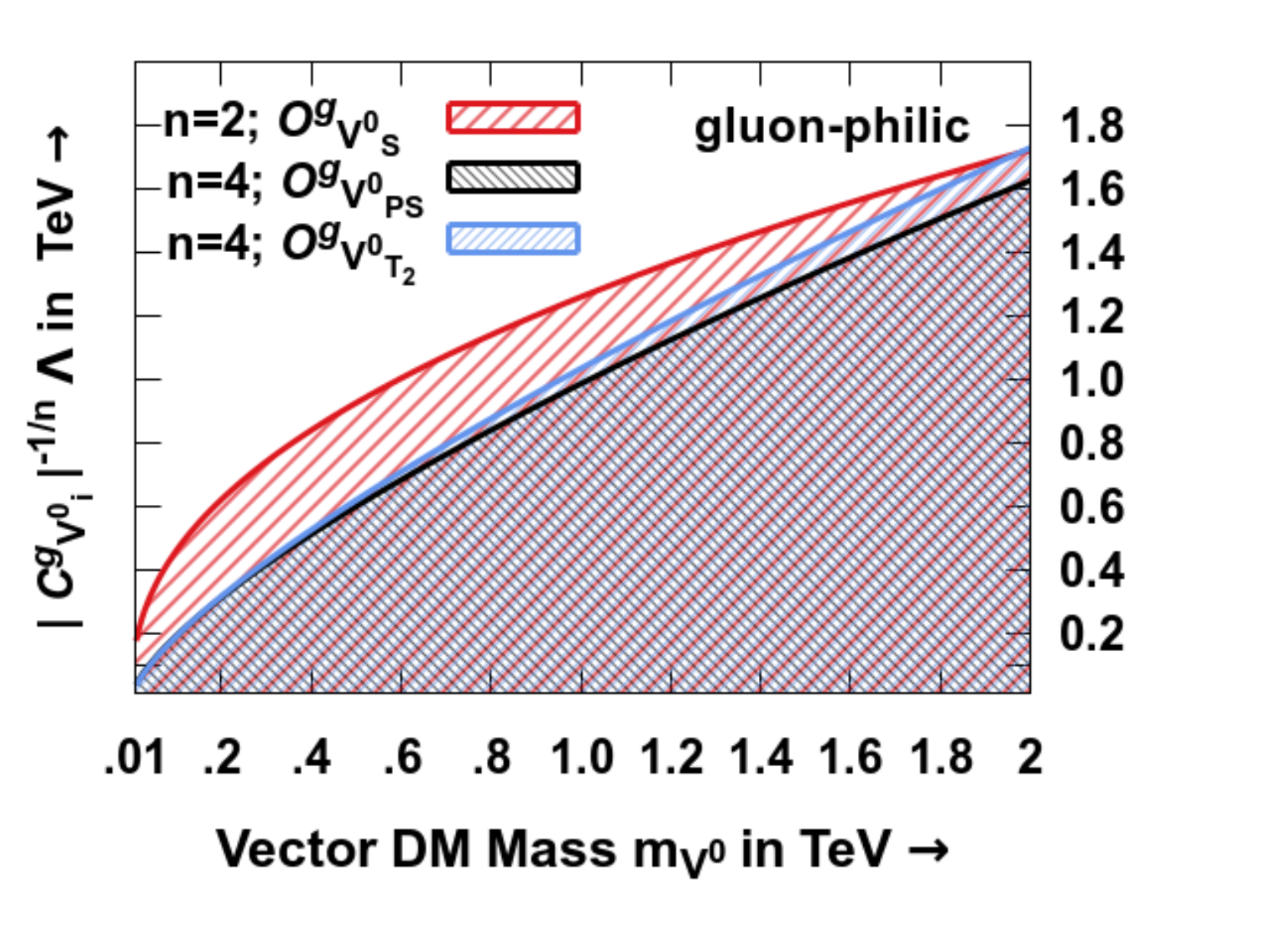}
      \caption{}	\label{fig:RelicVDMG}
  \end{subfigure}%
	\caption[nooneline]{\justifying \em{We depict relic density contours satisfying $\Omega^{V^0} h^2$ = 0.1198~\cite{Aghanim:2018eyx} and shaded cosmologically allowed regions  in the plane defined by Majorana DM mass $m_{V^0}$ and $\vert C^{q/g}_{V^0_{{\rm S},\,{\rm PS},\,{\rm AV},\, T_1}}\vert^{-1/n}\Lambda$. The $b$-quark-philic and $t$-quark-philic  vector DM contours in figures \ref{fig:RelicVDMB} and \ref{fig:RelicVDMT} are drawn corresponding to  scalar,  pseudo-scalar, axial-vector and twist-2 type-2 operators.  The gluon-philic vector DM contours for figure \ref{fig:RelicVDMG} are drawn corresponding to  scalar,  pseudo-scalar, and twist-2 type-2 operators.}} \label{fig:RelicVectDM}  
\end{figure}

 For numerical computation of the DM relic density, we have used MadDM \cite{Ambrogi:2018jqj, Backovic:2013dpa}, which implements the exact and closed expression for thermal averaged annihilation cross-section as discussed by the authors in the reference \cite{Gondolo:1990dk}. The input model file required by  the MadDM is generated using FeynRules \cite{Christensen:2008py, Alloul:2013bka}, which calculates all the required couplings and Feynman rules by using the full Lagrangian given in equations \eqref{MDMLag}, \eqref{SDMLag} and \eqref{VDMLag} corresponding to Majorana, scalar and vector DM particles respectively.  We have further verified and validated our numerical results from MicrOMEGAs~\cite{Belanger:2010pz}. The constant relic density contours are depicted in figures \ref{fig:RelicFermDM}, \ref{fig:RelicScalDM}, and \ref{fig:RelicVectDM} corresponding to Majorana, real scalar and real vector DM candidates respectively in the  plane defined by  $ m_{{\rm DM}_i} -\left\vert C^{q,\,g}_{{\rm DM}_i\,O_j}\right\vert^{-1/n}\,\Lambda$. 

\par The thermal averaged DM pair annihilation cross-section  is found to be $\sim 2\times 10^{-26}$ cm$^3$ s$^{-1}$ corresponding to a relic density of 0.1198, which essentially provides the functional dependence and shape-profile of the Wilson coefficients with the varying DM mass corresponding to each four-point effective operator. In order to understand the shape profile of  the relic density contours, $\left\langle\sigma\,v\right\rangle$ may be expanded analytically in power-series of $\left\vert v_{\rm DM}\right\vert^2$ and then the contribution of the leading terms may be investigated. The velocity of the DM at freeze-out is considered to be $0.3\, c$. We observe that
\begin{itemize}
 \item due to the constrained thermal averaged cross-section, the cut-off $\Lambda$ for Majorana DM scalar and pseudo-scalar operators in figures  \ref{fig:RelicFDMB} and  \ref{fig:RelicFDMT} varies as $\left(m_f\,m_\chi\right)^{1/3}$ and therefore shows a steep rise for top-philic DM in comparison to the bottom-philic case. However, the pseudo-scalar contribution dominates over its scalar counterpart due to its contribution from the leading velocity independent term in equation \eqref{FDMIDQPS}. The cut-offs for leading contributions from the quark-philic twist operator and velocity suppressed gluon-philic scalar, pseudo-scalar, and twist operators are found to increase monotonously with increasing DM mass as $m_\chi^{3/4}$.

\item the $s$-wave contribution in the thermal averaged annihilation cross-section $\left\langle \sigma_{\rm AV} \left\vert\vec v_{\chi}\right\vert  \right\rangle$   given in equation \eqref{FDMIDQAV} is proportional to $ m_f^2$    which follows from the chirality conserving property of axial-vector operator~\cite{Kumar:2013iva}. The variation of  $\left\langle \sigma_{\rm AV} \left\vert\vec v_{\chi}\right\vert  \right\rangle$  with the increasing DM mass is found to be largely determined  by the terms proportional to $m_{\chi}^2$  in  the converging power-series expansion  as explicitly shown in equation \eqref{FDMIDQAV}.    As a consequence, to satisfy the relic density constraint, the cut-off varies with respect to DM mass as $(m_\chi\, \vert \overrightarrow{v}_\chi\vert)^{1/2}$, i.e, faster than scalar and pseudo-scalar but slower than the twist operator, as shown in figures~\ref{fig:RelicFDMB} and~\ref{fig:RelicFDMT}.

  \item the cut-offs for the scalar and twist operators induced by heavy-quark-philic scalar and vector DM candidates are depicted in figures \ref{fig:RelicSDMB}, \ref{fig:RelicSDMT} and \ref{fig:RelicVDMB}, \ref{fig:RelicVDMT} respectively. In case of scalar interactions, the cuts-offs are found to be independent of the DM masses, whereas the cut-off for the heavy-quark-philic vector DM pseudo-scalar operator is found to be monotonously increasing as $(m_{V^0})^{1/2}$. In case of twist operators, the cut-off variation goes as $(m_{\phi^0})^{1/2}$ for scalar DM, while for vector DM scenario, it is d-wave suppressed and goes as  $\left\vert\vec v_{V^0}\right\vert^{1/2} \left(m_{V^0}\right)^{3/4}$.
  
 \item the gluon-philic scalar, pseudo-scalar, and twist interactions corresponding to Majorana DM are $p$-wave suppressed and the cut-off dependences are depicted in figure~\ref{fig:RelicFDMG}. The cut-offs corresponding to gluon-philic scalar, pseudo-scalar, and twist operators for vector DM candidates are constrained to vary as $\left(m_{V^0}\right)^{1/2}$, $\left\vert\vec v_{V^0}\right\vert^{1/4} \left(m_{V^0}\right)^{3/4}$  and  $\left\vert\vec v_{V^0}\right\vert^{1/2} \left(m_{V^0}\right)^{3/4}$ respectively, as shown in figure \ref{fig:RelicVDMG}. The observed relative suppression of the constrained cut-off for the twist operator is due to the velocity dependence. The same suit follows for the scalar DM candidates corresponding to the gluon-philic scalar and twist operators in figure \ref{fig:RelicSDMG}.
\end{itemize}


\subsection{Indirect detection of DM pairs}\label{indirect}
\par Today, the WIMP Dark Matter in the universe is expected to be trapped in large gravitational potential wells, which further enhances the number density of DM in the region, resulting in frequent collisions among themselves. This facilitates DM-DM pair annihilation into a pair of SM particles (photons, leptons, hadronic jets, etc) at the Galactic centre, in Dwarf Spheroidal galaxies (dSphs), Galaxy clusters, and Galactic halos.  The dwarf spheroidal satellite galaxies of the Milky Way are especially promising targets for DM indirect detection due to their large dark matter content, low diffuse Galactic $\gamma$-ray foregrounds as they travel the galactic distance, and lack of conventional astrophysical $\gamma$-ray production mechanisms.  Their flux is observed by the satellite-based $\gamma$-ray observatory Fermi-LAT~\cite{Ackermann:2015lka}, PLANCK ~\cite{Aghanim:2018eyx}, primary cosmic rays measurements by AMS-02 \cite{Aguilar:2014mma,Aguilar:2016kjl} on International Space Station and the ground-based Cherenkov telescope H.E.S.S.~\cite{Abramowski:2011hc, Abramowski:2013ax, Abdallah:2016ygi, Abdallah:2020sas}, HAWC \cite{Vargas:2016hcp, Albert:2017vtb, Lundeen:2020pnm}, MAGIC \cite{Acciari:2020pno}.

\par $t$-quark-philic DM annihilation yields a pair of $t\bar t$,  where the $t\, \, (\bar t)$  decays into $b\,\, (\bar b)$  associated with $W^+$ ($W^-$) with 100 \% branching fraction. The charged gauge Bosons then decays in either  leptonic or semi-leptonic or hadronic modes. The gluons and/or pair of $b\,\bar b$ produced by DM pair annihilation hadronize partially to neutral pions, which  then decay to photon pairs ($\pi^0 \rightarrow \gamma\gamma$) with a $99\%$ branching fraction, yielding a broad spectrum of photons. In order to compare the photon spectra observed in FermiLAT ~\cite{Ackermann:2015zua} and H.E.S.S~\cite{Abdallah:2016ygi} experiments, the photon flux resulting from the  DM pair annihilations is realised by interfacing the MadDM algorithm \cite{Ambrogi:2018jqj} with the showering and hadronization simulated using PYTHIA 8.0 code \cite{Sjostrand:2014zea}. In this subsection, we compare the thermally averaged DM annihilation cross-sections  corresponding to the $b\bar{b}$, $t\bar{t}$ and $gg$ channels in the dSphs environment with the upper bounds calculated from the respective recasted experimental limits of the observed photon spectra.

\par The analytic expressions of thermally averaged DM pair annihilation cross-sections $\left\langle \sigma \left\vert\vec v_{\rm DM}\right\vert\right \rangle$ for Majorana, scalar, and vector DM are given in the Appendix \ref{TherAvgCS} and agree with what is known for other fermions  in the literature  ~\cite{Zheng:2010js, Yu:2011by, Rawat:2017fak, Bharadwaj:2020aal}. These are derived from the cross-sections  in the Appendix \ref{annihilationCS} in which the center of mass energy squared $s$ is expanded as   $\approx$ $4 \, m^2_{\rm DM} + m^2_{\rm DM}\, \left\vert \vec v_{\rm DM} \right\vert^2+ \frac{3}{4}\, m_{\rm DM}^2 \left\vert\vec v_{\rm DM}\right\vert^4$ + $\mathcal{O}\left(\left\vert\vec v_{\rm DM}\right\vert^6\right)$. It is to be noted that  DM velocity  $\left\vert\vec v_{\rm DM}\right\vert$ is roughly of the order of $\sim 10^{-3}\, c$ at the center of Galaxy and  $10^{-5}\, c$ at dSphs in contrast to that of $10^{-1}\,c$ at freeze-out. In comparison to all chiral blind $s$-dominated processes, the leading term contributions from the $p$-wave and $d$-wave channels in $\left\langle \sigma \left\vert \vec v_{\rm DM}\right\vert \right\rangle$ are proportional to $\left\vert\vec v_{\rm DM}\right\vert^2$ and $\left\vert \vec v_{\rm DM}\right\vert^4$, respectively, and thus suppressed. This renders the leading $p$-wave suppressed thermal averaged DM pair annihilation cross-section to be of the order of $10^{-32}\, {\rm cm^{3}\, s^{-1}}$  in the Galactic centre and  $10^{-36}\, {\rm cm^{3}\, s^{-1}}$    at the dSphs respectively. Henceforth, our the numerical analysis of the leading order contribution to the velocity dependent/ independent thermal averaged cross-sections are performed for dSphs and only those $\left\langle \sigma \left\vert \vec v_{\rm DM}\right\vert \right\rangle$ are depicted in the figures which are larger than $10^{-29}\, {\rm cm^{3}\, s^{-1}}$.

\begin{figure}[h!]
	\centering
\hspace{-0.5cm}
  \begin{subfigure}{0.5\textwidth}
      \centering
	\includegraphics[scale=0.5]{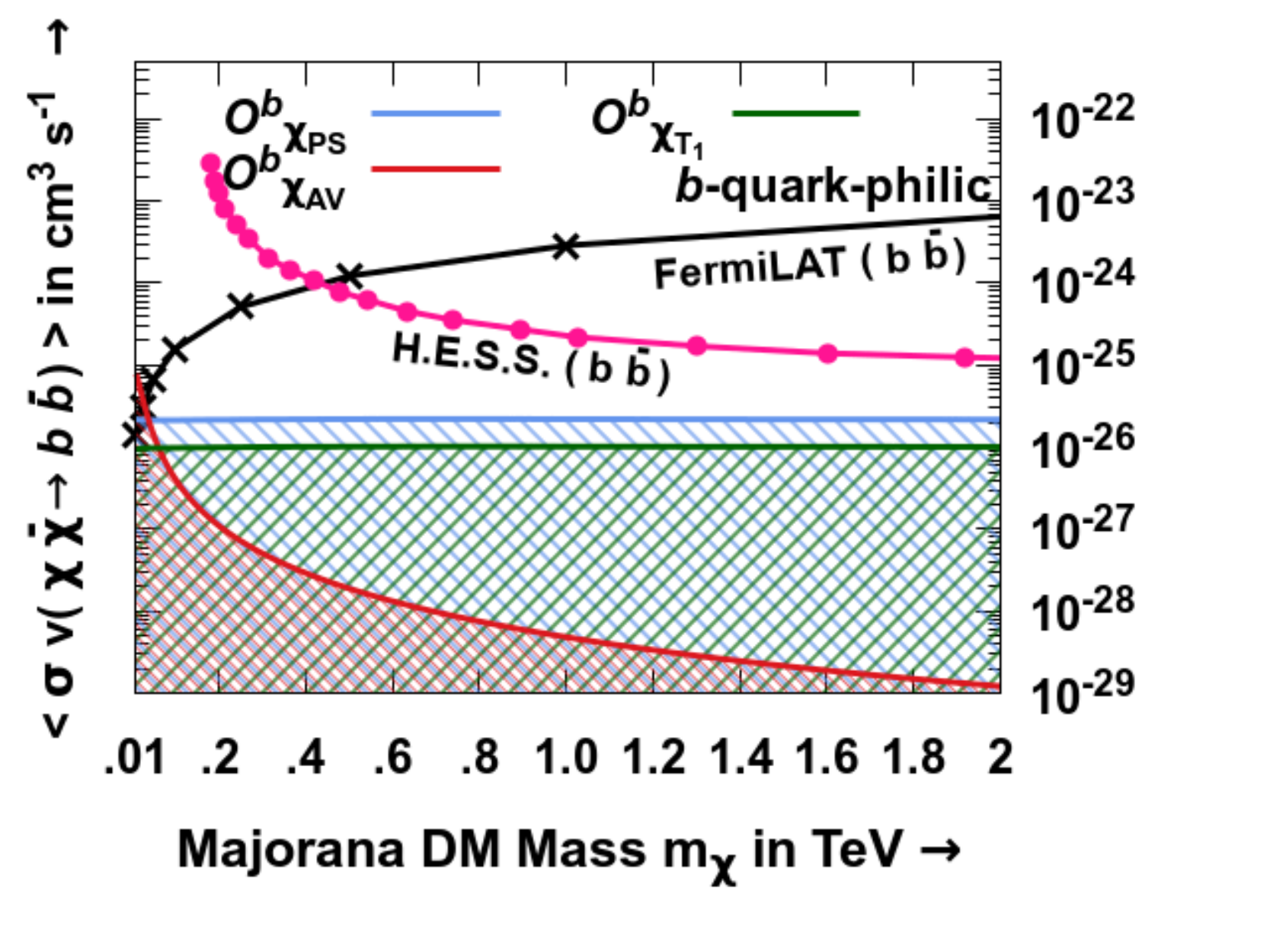}
      \caption{}	\label{fig:IDFb}
  \end{subfigure}%
\hspace{0.1cm}
  \begin{subfigure}{0.5\textwidth}
      \centering
	\includegraphics[scale=0.5]{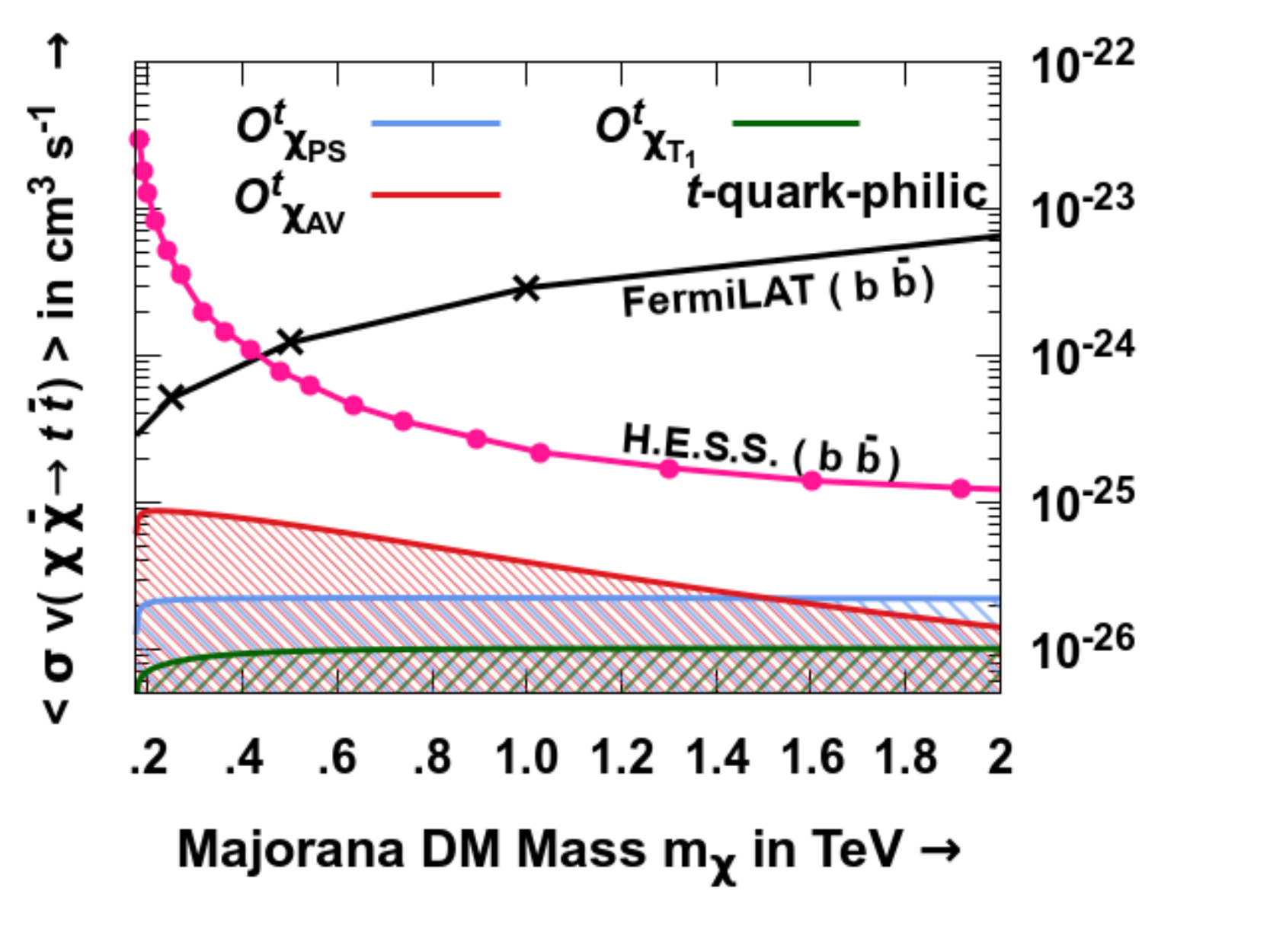}
      \caption{}	\label{fig:IDFt}
  \end{subfigure}%
	\caption[nooneline]{\justifying \em{ Figures \ref{fig:IDFb} and \ref{fig:IDFt} depict the thermally averaged cross-sections for Majorana DM pair annihilation into  $b\bar{b}$ and $t\bar{t}$ pairs, respectively. The contributions from pseudo-scalar, axial-vector and twist-2 type-1 operators   in both the panels are evaluated using values for $  \left\vert C^{q}_{\chi_{{\rm PS},\,{\rm AV},\, T_1}}/\, \Lambda^n\right\vert $ satisfying  $\Omega^{\chi } h^2$ = $0.1198\pm 0.0012$  \cite{Aghanim:2018eyx} as shown in figure \ref{fig:RelicFermDM} and hence, the unshaded regions above the respective curves are cosmologically allowed.   Regions above the recasted experimental limits obtained from  FermiLAT~\cite{Ackermann:2015zua} as well as H.E.S.S.~\cite{Abdallah:2016ygi} are excluded.}}
	\label{fig:IDFermDM}  
\end{figure}

\begin{figure}[h!]
	\centering
\hspace{-0.5cm}
  \begin{subfigure}{0.5\textwidth}
      \centering
	\includegraphics[scale=0.5]{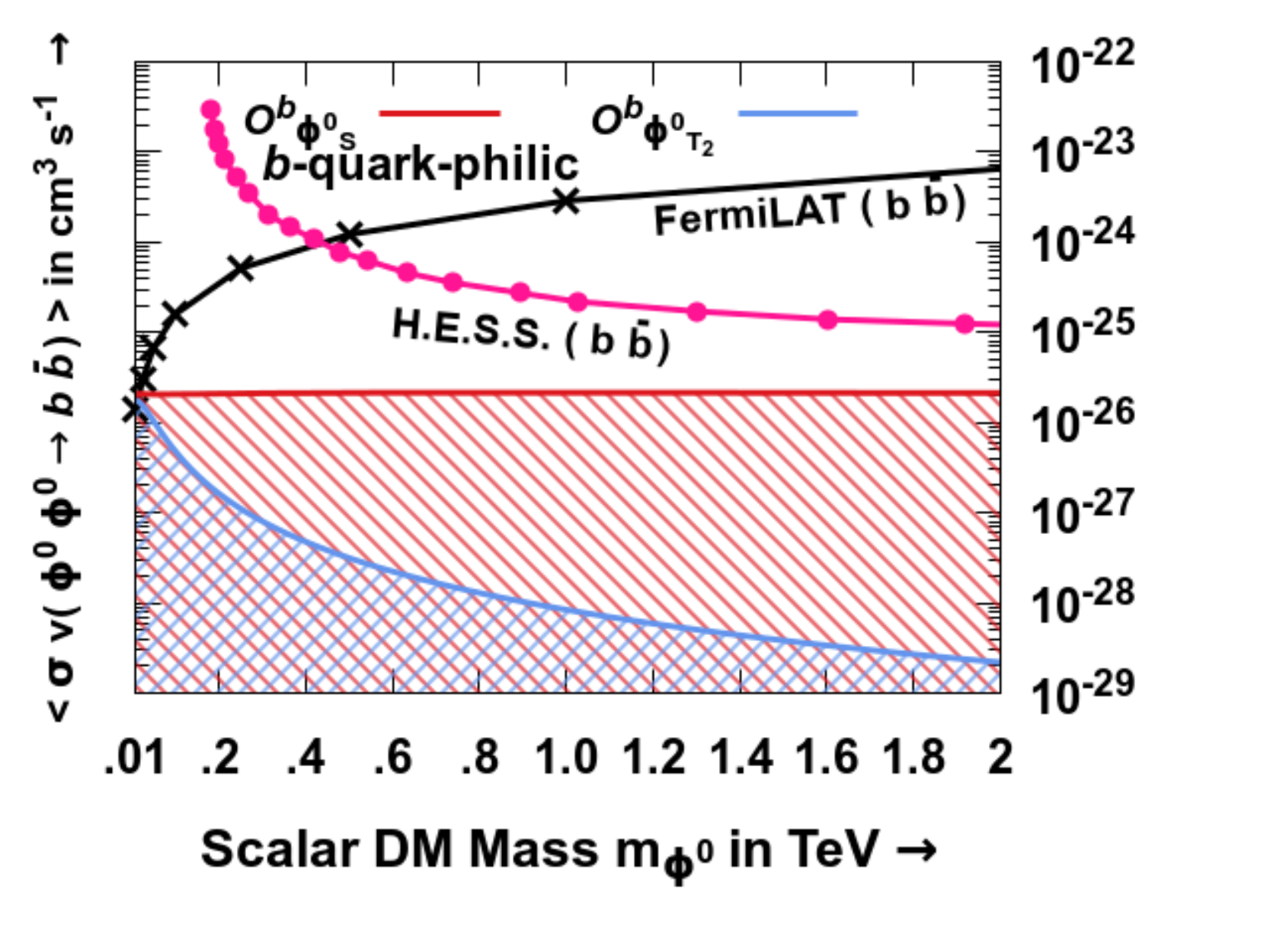}
      \caption{}	\label{fig:IDSb}
  \end{subfigure}%
\hspace{0.1cm}
  \begin{subfigure}{0.5\textwidth}
      \centering
	\includegraphics[scale=0.5]{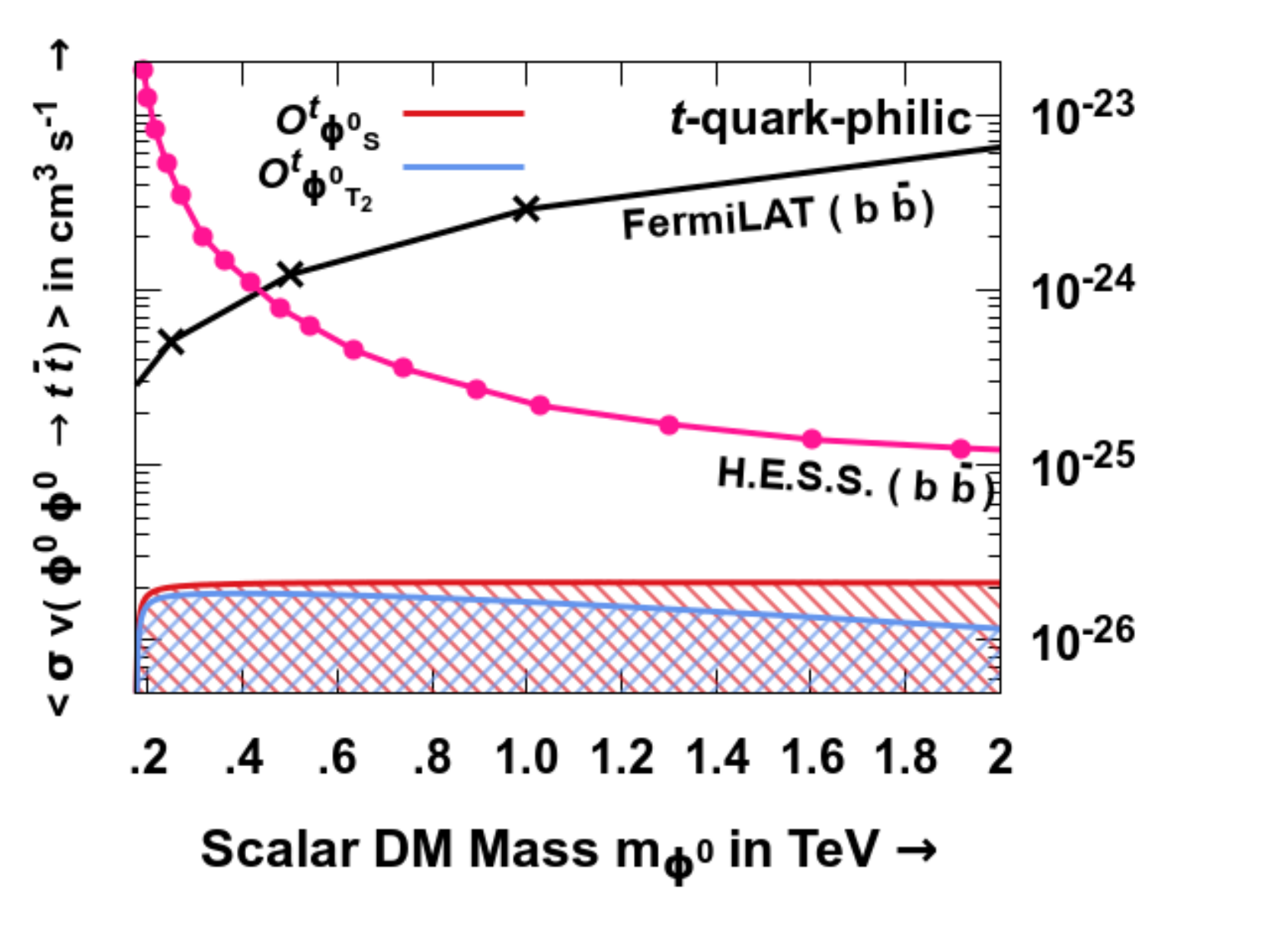}
      \caption{}	\label{fig:IDSt}
      \end{subfigure}%
      \\
        \begin{subfigure}{0.5\textwidth}
      \centering
	\includegraphics[scale=0.5]{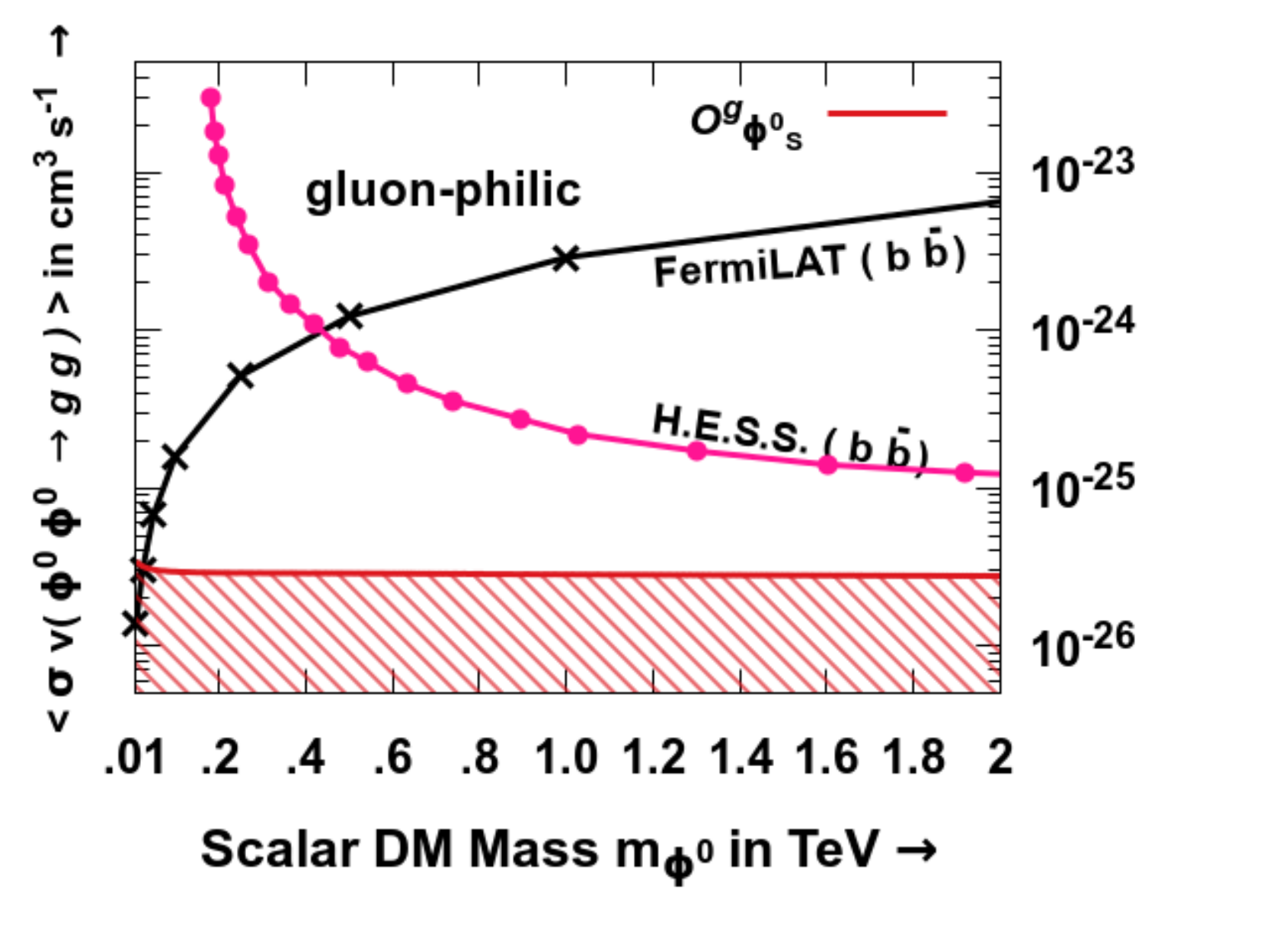}
      \caption{}	\label{fig:IDSg}
  \end{subfigure}%

	\caption[nooneline]{\justifying \em{ Figures \ref{fig:IDSb}, \ref{fig:IDSt}  and \ref{fig:IDSg} depict the thermally averaged cross-sections for real scalar DM pair annihilation into  $b\bar{b}$, $t\bar{t}$ and $g\,g$ pairs, respectively. The   contributions from scalar and twist-2 type-2 operators in the panels are evaluated using their respective  values for $  \left\vert C^{q,\,g}_{\phi^0_{S,\, T_2}}/\, \Lambda^n\right\vert$ satisfying  $\Omega^{\phi^0} h^2$ = $0.1198\pm 0.0012$  \cite{Aghanim:2018eyx} as shown in figure \ref{fig:RelicScalDM}  and hence, the unshaded regions above the respective curves are cosmologically allowed.  Regions above the  recasted experimental limits obtained from  FermiLAT~\cite{Ackermann:2015zua} as well as H.E.S.S.~\cite{Abdallah:2016ygi} are excluded.}}
	\label{fig:IDScalDM}  
\end{figure}

\par  For  varying $m_{\rm DM}$,  we investigate the contributions to the  thermally averaged  DM pair annihilation cross-sections  into   $b\,\bar b$, $t\,\bar t$ and $g\,g$ pairs. Using  the respective  Wilson coefficients  satisfying relic density constraint for a given $m_\chi$ as shown in figure  \ref{fig:RelicFermDM}, we calculate and depict the variation of cosmological  bound on the thermally averaged Majorana DM pair annihilation cross-sections with Majorana DM mass for $\chi\,\bar{\chi} \rightarrow b\,\bar{b}$ and $\chi\,\bar{\chi} \rightarrow t\,\bar{t}$ in figures \ref{fig:IDFb} and \ref{fig:IDFt} respectively. 
\begin{figure}[h!]
	\centering 
\hspace{-0.5cm}
  \begin{subfigure}{0.5\textwidth}
      \centering
	\includegraphics[scale=0.5]{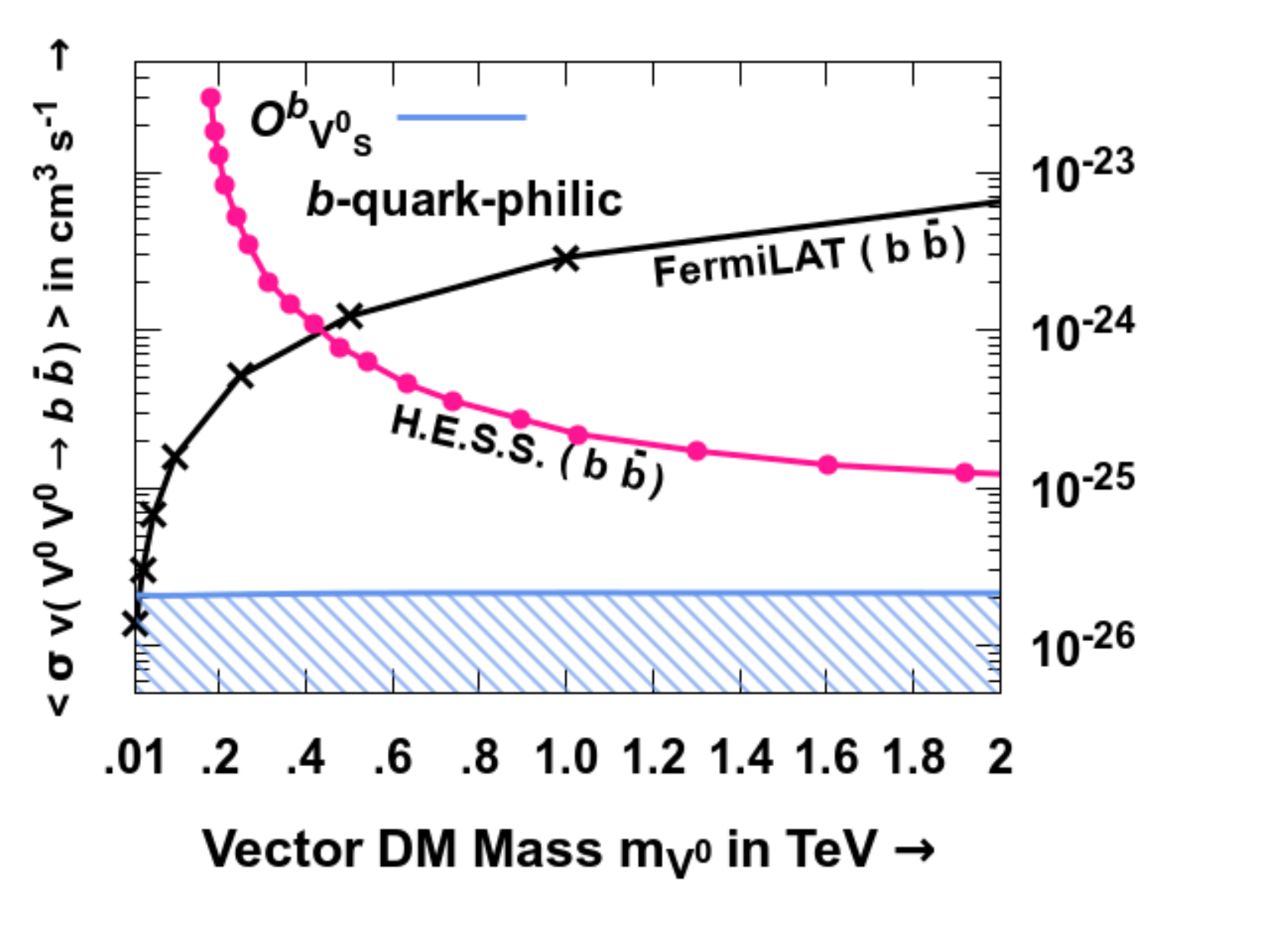}
      \caption{}	\label{fig:IDVb}
  \end{subfigure}%
\hspace{0.1cm}
  \begin{subfigure}{0.5\textwidth}
      \centering
	\includegraphics[scale=0.5]{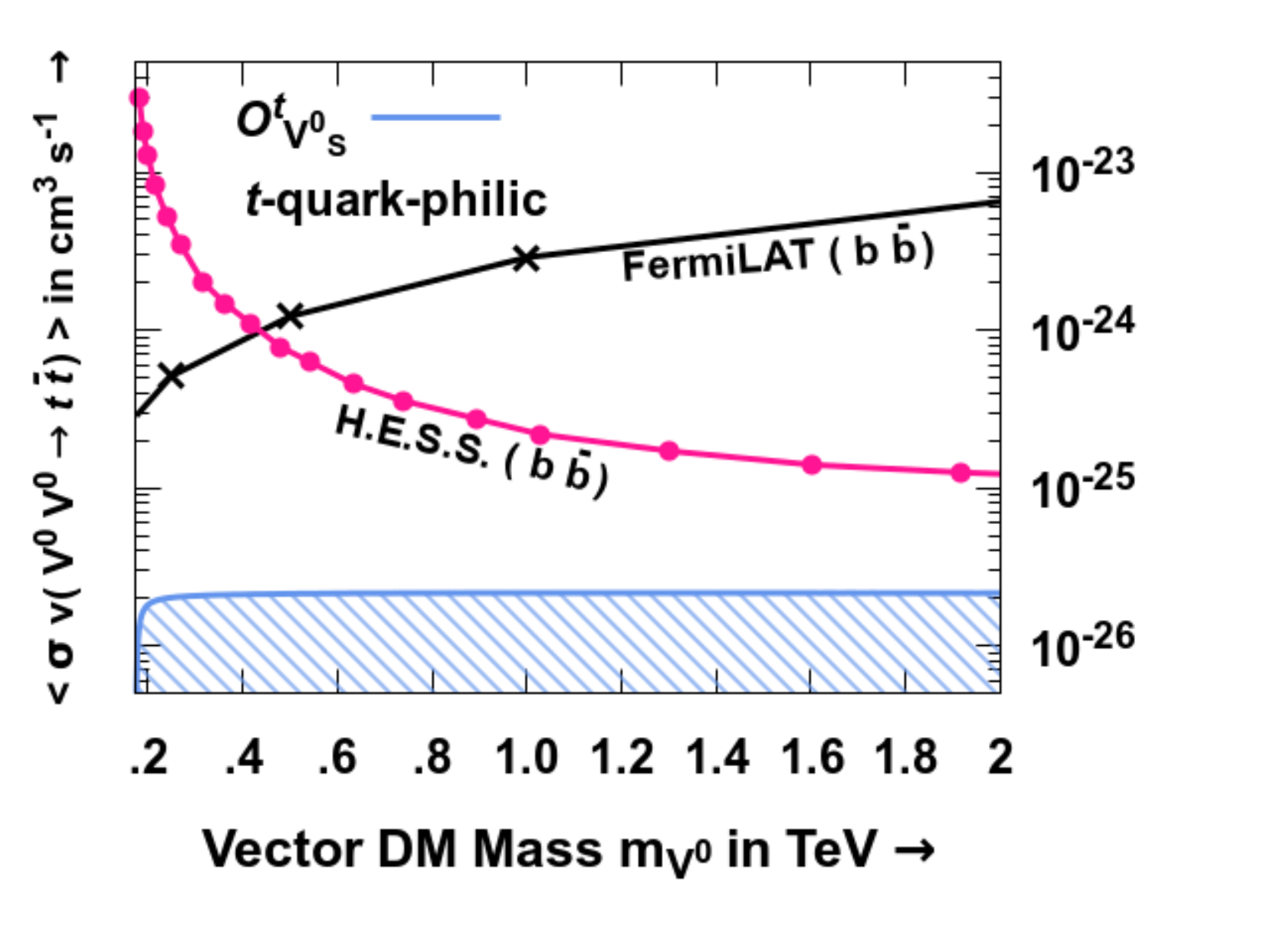}
      \caption{}	\label{fig:IDVt}
  \end{subfigure}%
  \\
  	\centering 
  \begin{subfigure}{0.5\textwidth}
      \centering
	\includegraphics[scale=0.5]{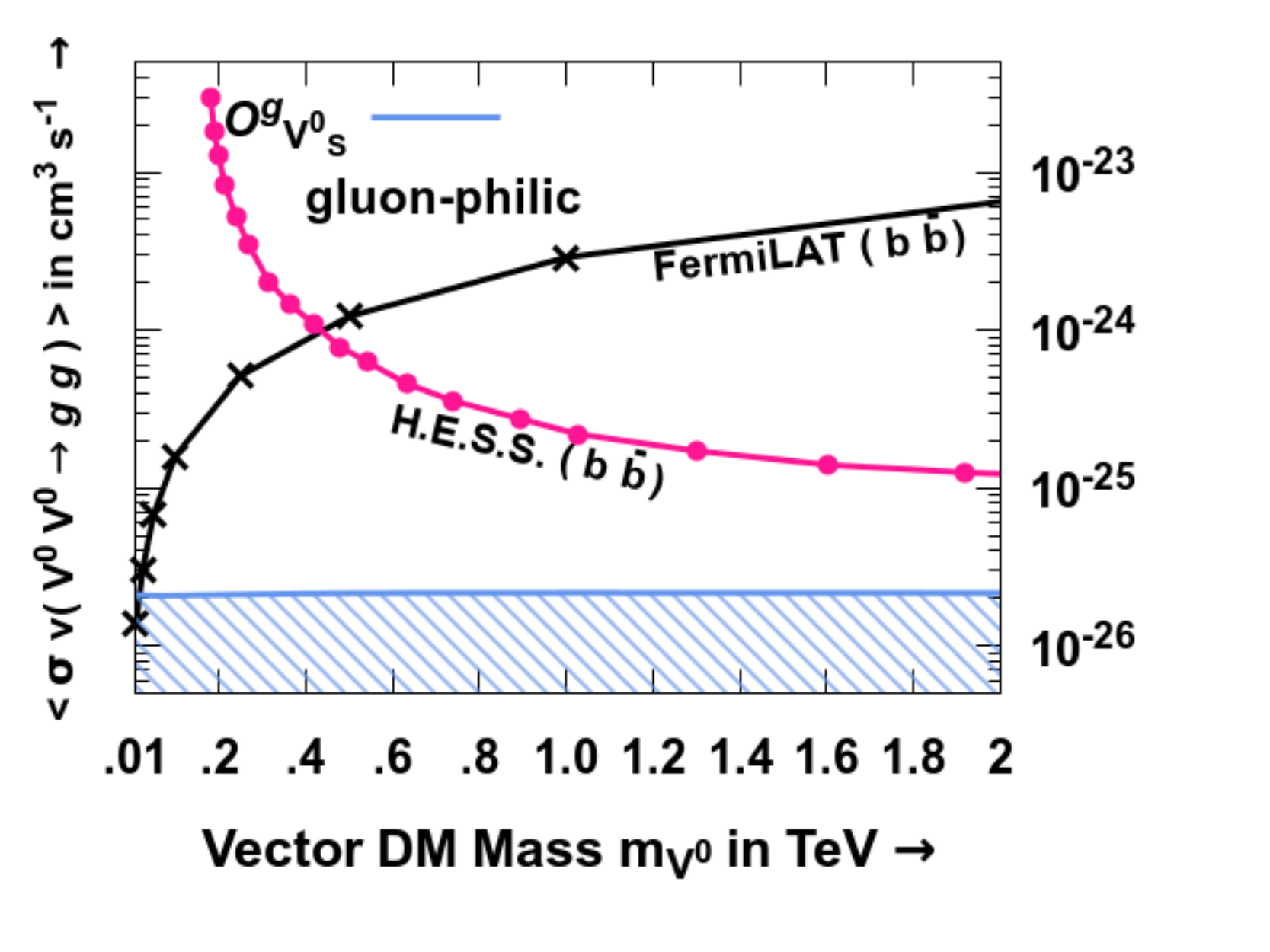}
      \caption{}	\label{fig:IDVg}
  \end{subfigure}%

	\caption[nooneline]{\justifying \em{Figures \ref{fig:IDVb}, \ref{fig:IDVt} and \ref{fig:IDVg} depict the thermally averaged cross-sections for real vector DM pair annihilation into  $b\bar{b}$, $t\bar{t}$ and $g\,g$ pairs, respectively. The contributions from scalar and pseudo-scalar operators in the panels are evaluated using their respective  values for $  \left\vert C^{q,\,g}_{V^0_{S/\,PS}}/\, \Lambda^n\right\vert$ satisfying  $\Omega^{V^0} h^2$ = $0.1198\pm 0.0012$  \cite{Aghanim:2018eyx} as shown in figure \ref{fig:RelicVectDM}  and hence, the unshaded regions above the respective curves are cosmologically allowed.  Regions above the  the experimental limits obtained from the recasted FermiLAT~\cite{Ackermann:2015zua} as well as H.E.S.S.~\cite{Abdallah:2016ygi} are excluded.}}
	\label{fig:IDVectDM}  
\end{figure}

We observe that  the shape profile of the axial-vector induced contribution to the thermal averaged DM pair annihilation cross-section with varying DM mass is governed by the term $ \left( 1 + \frac{1}{3}\frac{m_\chi^2}{m_f^2}\, \vert \vec v_\chi\vert^2 +\cdots\right)$  in     equation \eqref{FDMIDQAV}, which is approximately $\sim$ 1 for the range of DM masses of phenomenological interest. This is in contrast to the case of relic density computation, where  the Wilson coefficient decreases with increasing  DM mass  to satisfy the relic density constraint. The use of these decreasing constrained  couplings  with increasing DM masses in figures \ref{fig:IDFb} and \ref{fig:IDFt} is responsible for the negative slope of the $\left\langle \sigma \left\vert\vec v_\chi\right\vert\right \rangle$.

\par In case  of the twist-2 type-1 operator the suppression due to  increment in the cosmological upper bound on $\Lambda$ is compensated with the increment in the DM mass alone. The negligible contributions from  the heavy-quark-philic scalar $\mathcal{O}^q_{\chi_S}$ operator given in equation \eqref{FDMIDQS} is attributed to the chiral suppression along with DM velocity dependence while  gluon-philic scalar $\mathcal{O}^g_{\chi_S}$, pseudo-scalar $\mathcal{O}^g_{\chi_{\rm PS}}$  and  twist-2 type-1  $\mathcal{O}^g_{\chi_{T_1}}$ operators given in equations \eqref{FDMIDGS}, \eqref{FDMIDGPS} and \eqref{FDMIDGT} respectively are $p$-wave suppressed ($\ll 10^{-29}$ cm$^3$ s$^{-1}$) and hence  not shown in the graph.  

\par Similarly, we plot the variation of the cosmological  bound of the thermally averaged DM annihilation cross-section ($\phi^0\,\phi^0\rightarrow b\,\bar{b}/\,\,t\,\bar{t}/\,\,g\,g$) with scalar DM mass   $m_{\phi^0}$ in figure \ref{fig:IDScalDM}.  The  $\left\langle \sigma \left\vert\vec v_{\phi^0}\right\vert\right \rangle$ induced by the  heavy-quark-philic twist-2 type-2 operator $\mathcal{O}^q_{\phi^0_{T_2}}$ given in  \eqref{SDMIDeqTwist} is chirally suppressed and therefore falls sharply with an increasing DM mass as shown in figure \ref{fig:RelicScalDM}.  The sharp fall in the $t$-quark-philic case, on the other hand, is flattened for DM mass ranges of less than 2 TeV. The gluon-philic twist-2 type-2 operator is $d$-wave $\propto v^4$ suppressed  as shown in equation \eqref{SDMIDeqGluonT}. 
Figure \ref{fig:IDVectDM} show the thermally averaged vector DM  pair annihilation cross-sections of the third generation quarks and gluons with varying vector DM mass corresponding to the  respective cosmological  bound on the Wilson coefficient as shown in figure \ref{fig:RelicVectDM}. Unlike  the  chiral $p$-wave suppressed  pseudo-scalar and axial-vector operators and $d$-wave suppressed twist-2 type-2 operators, we observe an appreciable contribution to the $\left\langle \sigma \left\vert \vec v_{V^0}\right\vert \right\rangle$  from the  heavy-quark-philic and gluon-philic scalar  operators which are of the order of $\sim 10^{-26}\, {\rm cm}^3 {\rm s}^{-1}$.


\subsection{DM-nucleon scattering}\label{direct}
\begin{figure}[h!]
	\centering
	\includegraphics[scale=0.4]{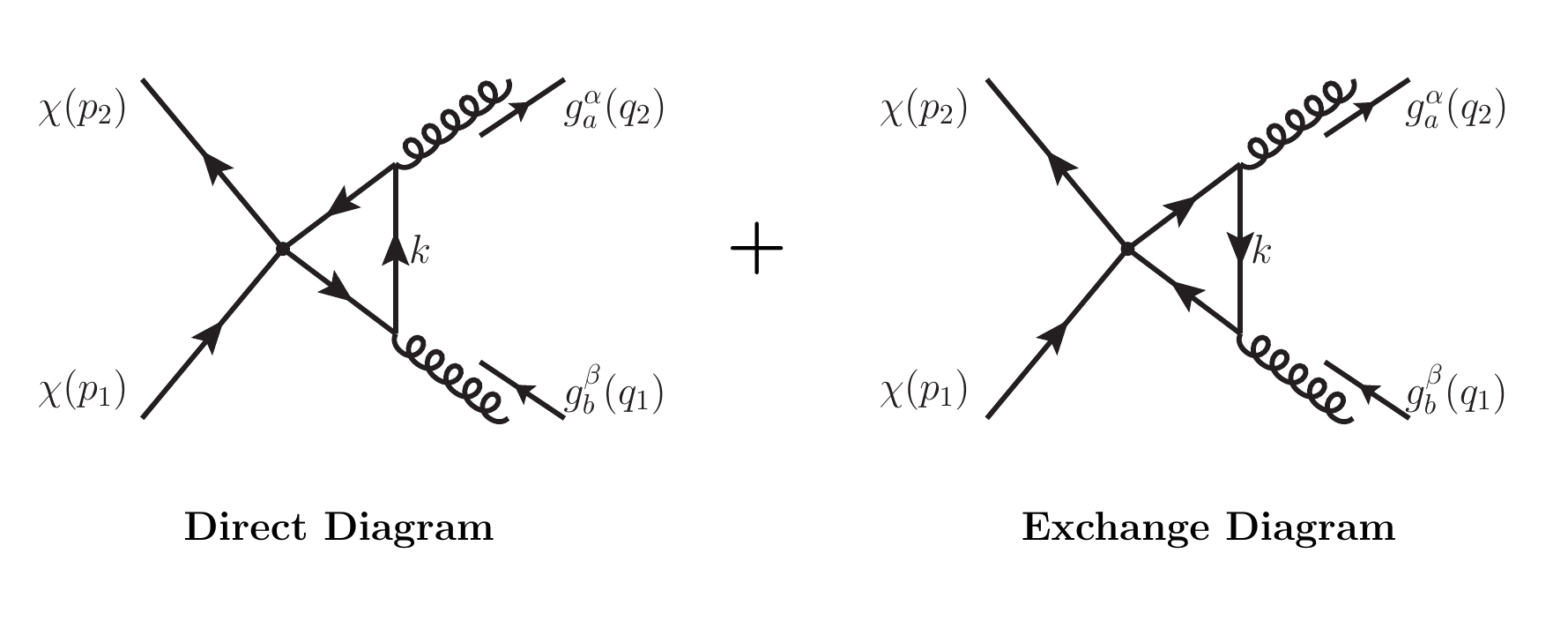}
	\caption[nooneline]{\justifying \em{One-loop Majorana DM-gluon scattering diagrams where the blob represents the four fermionic effective interactions induced by the scalar/axial-vector/twist-2 currents of heavy quarks.}}\label{DDfeyndia}
\end{figure}

In direct-detection experiments, the scattering of DM particles can be broadly classified as (a) DM-electron scattering, (b) DM-atom scattering, and (c) DM-nucleon scattering. In the absence of heavy sea quarks and anti-quarks inside nucleons at the direct-detection energy scale, the $b$-quark-philic and $t$-quark-philic DM interacts with the constituent gluons via a loop, as shown in figure  \ref{DDfeyndia}.
\par We compute the dominant DM-gluon one-loop elastic scattering amplitudes induced by the scalar, axial-vector, and twist-2 point interactions among DM and heavy quarks. The mass scale $m_Q$ of the heavy quarks running in the loop and the QCD coupling strength $\alpha_s$ characterise the loop-amplitudes. We derive the phenomenological effective DM-gluon interaction Lagrangians given in \eqref{LeffMDMgluon}, \eqref{LeffSDMgluon} and \eqref{LeffVDMgluon} which correspond to Majorana, real scalar, and real vector DM candidates, respectively.

\par Since the non-relativistic DM particles scatter the nucleons and not the free gluons, we perform the non-relativistic reduction of the interaction Lagrangian given in the Appendix \ref{DDappendix}. We connect the DM-gluon amplitudes induced by the scalar and twist-2 currents of heavy-quarks at one-loop order with their respective DM-effective nucleon interactions by evaluating the expectation values of the zero momentum scalar  and twist-2 gluon-philic operators between the initial and final nucleons in equations \eqref{fTGinfo} and \eqref{gluontwisNR} respectively. The Majorana, real scalar, and real vector DM-nucleon scattering cross-sections driven by the heavy-quark scalar, axial-vector, and twist-2 currents  are given as
\begin{subequations}
\bea 
\sigma^{q^{\chi N}}_{\rm S} 
&=&\frac{8}{81}\frac{1}{\pi} \left(C_{\chi_S}^q\right)^2\left[ \frac{1\, {\rm TeV}}{\Lambda}\right]^6\,\left[\frac{m_N}{1\, {\rm GeV}}\right]^2\left[\frac{\mu_{N\chi}}{1\,{\rm GeV}}\right]^2\,\left[\frac{\alpha_s\left(\Lambda\right)}{\alpha_s\left(\mu\right)}\right]^2\nn\\
&&\,\,\,\, \times   \left\vert I_S^{gg}\right\vert^2\, \left\vert f^N_{\rm TG}\right\vert^2 \left(3.9\times 10^{-46}\right)\, {\rm cm}^{2}\label{ScattScalMDM}\eea
\bea
\sigma^{q^{\chi N}}_{\rm AV} 
&=&\frac{16}{3}\frac{1}{\pi} \left(C_{\chi_{\rm AV}}^q\right)^2\left[ \frac{1\, {\rm TeV}}{\Lambda}\right]^4\,\left[\frac{\mu_{N\chi}}{1\,{\rm GeV}}\right]^2\,\left[\frac{\alpha_s\left(\Lambda\right)}{\alpha_s\left(\mu\right)}\right]^2\, \left[\frac{175\, {\rm GeV}}{m_Q}\right]^4\left\vert I_{\rm AV}^{gg}\right\vert^2\nn\\
&&\times \left[ \frac{\left\vert \vec q\right\vert}{10\, {\rm MeV}} \right]^4 \left[\displaystyle \sum_{q=u,\,d,\,s}\Delta_q^{(N)}\,\frac{\bar m}{m_q} \right]^2\left[\frac{{\cal J}+1}{\cal J}\right]S_N\,S_N^\prime \,  \left(4.13 \times 10^{-57}\right)\, {\rm cm}^{2}
\label{ScattAVMDM}\eea
\bea
\sigma^{q^{\chi N}}_{\rm T} 
&=& \frac{2}{\pi}\, \left(C^q_{\chi_T}\right)^2 \bigg[\frac{1\, {\rm TeV}}{\Lambda}\bigg]^8\, \bigg[\frac{\alpha_s(\Lambda)}{4 \pi}\bigg]^2\, \left[\frac{\mu_{\chi N}}{1\, {\rm GeV}}\right]^2\, \left[\frac{M_\chi}{100\, {\rm GeV}}\right]^2\, \left[\frac{m_N}{1\, {\rm GeV}}\right]^2\, \nn\\
&&\,\,\,\, \times \, \left[ \ln\left(\frac{\Lambda^2}{m_Q^2}\right) \right]^2\,\left\vert (g(2;\Lambda)\right\vert^2\, \left(1.98 \times 10^{-48}\right)\, {\rm cm^2} 
\label{ScattTwistMDM}
\eea
\end{subequations}
\begin{subequations}
\bea
\sigma^{q^{\phi^0 N}}_{ S} 
&=&\frac{4}{81}\,\frac{1}{\pi} \left(C_{\phi^0_S}^q\right)^2\left[ \frac{1\, {\rm TeV}}{\Lambda}\right]^4\,\left[ \frac{100\, {\rm GeV}}{m_{\phi^0}}\right]^2\left[\frac{m_N}{1\, {\rm GeV}}\right]^2\left[\frac{\mu_{N\chi}}{1\,{\rm GeV}}\right]^2\,\left[\frac{\alpha_s\left(\Lambda\right)}{\alpha_s\left(\mu\right)}\right]^2 \nn\\
&&\,\,\,\,\times  \left\vert I_S^{gg}\right\vert^2\, \left\vert f^N_{\rm TG}\right\vert^2 \left(3.9 \times 10^{-44}\right)\, {\rm cm}^{2}\label{ScattScalSDM}\eea
\bea
\sigma^{q^{\phi^0 N}}_{T_2} 
&=& \frac{1}{2\pi}\, \left(C^q_{\phi^0_{T_2}}\right)^2 \bigg[\frac{1\, {\rm TeV}}{\Lambda}\bigg]^8\, \bigg[\frac{\alpha_s(\Lambda)}{4 \pi}\bigg]^2\, \left[\frac{\mu_{\phi^0 N}}{1\, {\rm GeV}}\right]^2\, \left[\frac{M_{\phi^0}}{100\, {\rm GeV}}\right]^2\, \left[\frac{m_N}{1\, {\rm GeV}}\right]^2\, \nn\\
&&\,\,\,\, \times \, \left[ \ln\left(\frac{\Lambda^2}{m_Q^2}\right) \right]^2 \,\left\vert g(2;\Lambda)\right\vert^2\, \left(1.98 \times 10^{-48}\right)\, {\rm cm^2}  
\label{ScattTwistSDM}\eea
\end{subequations}
\begin{subequations}
\bea
\sigma^{q^{V^0 N}}_{S} 
&=&\frac{4}{243}\frac{1}{\pi} \left(C_{V^0_S}^q\right)^2\left[ \frac{1\, {\rm TeV}}{\Lambda}\right]^4\,\left[ \frac{100\, {\rm GeV}}{m_{V^0}}\right]^2\left[\frac{m_N}{1\, {\rm GeV}}\right]^2\left[\frac{\mu_{N\chi}}{1\,{\rm GeV}}\right]^2\,\left[\frac{\alpha_s\left(\Lambda\right)}{\alpha_s\left(\mu\right)}\right]^2\nn\\
&&\,\,\,\,\times   \left\vert I_S^{gg}\right\vert^2\, \left\vert f^N_{\rm TG}\right\vert^2 \left(3.9 \times 10^{-44}\right)\, {\rm cm}^{2}\label{ScattScalVDM}\eea
\bea
\sigma^{q^{V^0 N}}_{\rm AV} 
&=&\frac{32}{9}\frac{1}{\pi} \left(C_{V^0_{\rm AV}}^q\right)^2\left[ \frac{1\, {\rm TeV}}{\Lambda}\right]^4\,\left[\frac{\mu_{NV^0}}{1\,{\rm GeV}}\right]^2\,\left[\frac{\alpha_s\left(\Lambda\right)}{\alpha_s\left(\mu\right)}\right]^2\,  \left[\frac{175\, {\rm GeV}}{m_Q}\right]^4\left\vert I_{\rm AV}^{gg}\right\vert^2\nn\\
&&\times \left[ \frac{\left\vert \vec q\right\vert}{10\, {\rm MeV}} \right]^4 \left[\displaystyle \sum_{q=u,\,d,\,s}\Delta_q^{(N)}\,\frac{\bar m}{m_q} \right]^2\left[\frac{{\cal J}+1}{\cal J}\right]S_N\,S_N^\prime \,  \left(4.13 \times 10^{-57}\right)\, {\rm cm}^{2}\label{ScattAVVDM}\eea
\bea
\sigma^{q^{V^0 N}}_{T_2} 
&=& \frac{1}{6\pi}\, \left(C^q_{V^0_{T_2}}\right)^2 \left[\frac{1\, {\rm TeV}}{\Lambda}\right]^8\, \left[\frac{\alpha_s(\Lambda)}{4 \pi}\right]^2\, \left[\frac{\mu_{V^0 N}}{1\, {\rm GeV}}\right]^2\, \left[\frac{M_{V^0}}{100\, {\rm GeV}}\right]^2\, \left[\frac{m_N}{1\, {\rm GeV}}\right]^2\, \nn\\
&&\,\,\,\, \times \, \left[ \ln\left(\frac{\Lambda^2}{m_Q^2}\right) \right]^2\,   \left\vert g\left(2;\Lambda\right)\right\vert^2\, \times\left(1.98 \times 10^{-48}\right)\, {\rm cm^2}
\label{ScattTwistVDM}
\eea 
\end{subequations}
 
where $\mu_{N\,{\rm DM}}\equiv \left(m_N\,m_{\rm DM}\right) /\, \left(m_N+m_{\rm DM}\right)$ is the reduced mass for  the respective  DM-nucleon system and the scale $\mu$ is taken to be $Z^0$-boson mass. For the computation of scale dependent $\alpha_s\left(\mu\right), \alpha_s\left(\Lambda\right)$ and $g\left(x;\Lambda\right)$, we access CTEQ6l1~\cite{Pumplin:2002vw} PDF data set  from LHAPDF6 ~\cite{Buckley:2014ana} library. 
However, the scale dependence of the Wilson coefficients is found to be smaller than that of $\alpha_s$, as noted by the authors of the reference \cite{Belyaev:2018pqr}. It is to be noted that the observed logarithmically enhanced one-loop induced scattering cross-sections in equations \eqref{ScattTwistMDM}, \eqref{ScattTwistSDM} and \eqref{ScattTwistVDM} result from the explicit momentum dependence in the twist-2 operator interaction Lagrangian given in equations \eqref{MDMLag}, \eqref{SDMLag} and \eqref{VDMLag} respectively. The real vector DM-nucleon scattering cross-sections driven by the scalar and twist-2 type-2 currents of heavy quarks  in equations \eqref{ScattScalVDM} and \eqref{ScattTwistVDM} respectively   are found to be 1/3 of the scalar DM-nucleon scattering cross-sections corresponding to scalar and twist-2 type-2 currents of heavy-quarks in equations  \eqref{ScattScalSDM} and  \eqref{ScattTwistSDM} respectively. The velocity suppressed spin-dependent DM-nucleon scattering events induced by the pseudo-scalar operators are not analysed.  
\begin{figure}[h!]
	\centering
\hspace{-0.5cm}
  \begin{subfigure}{0.5\textwidth}
      \centering
	\includegraphics[scale=0.5]{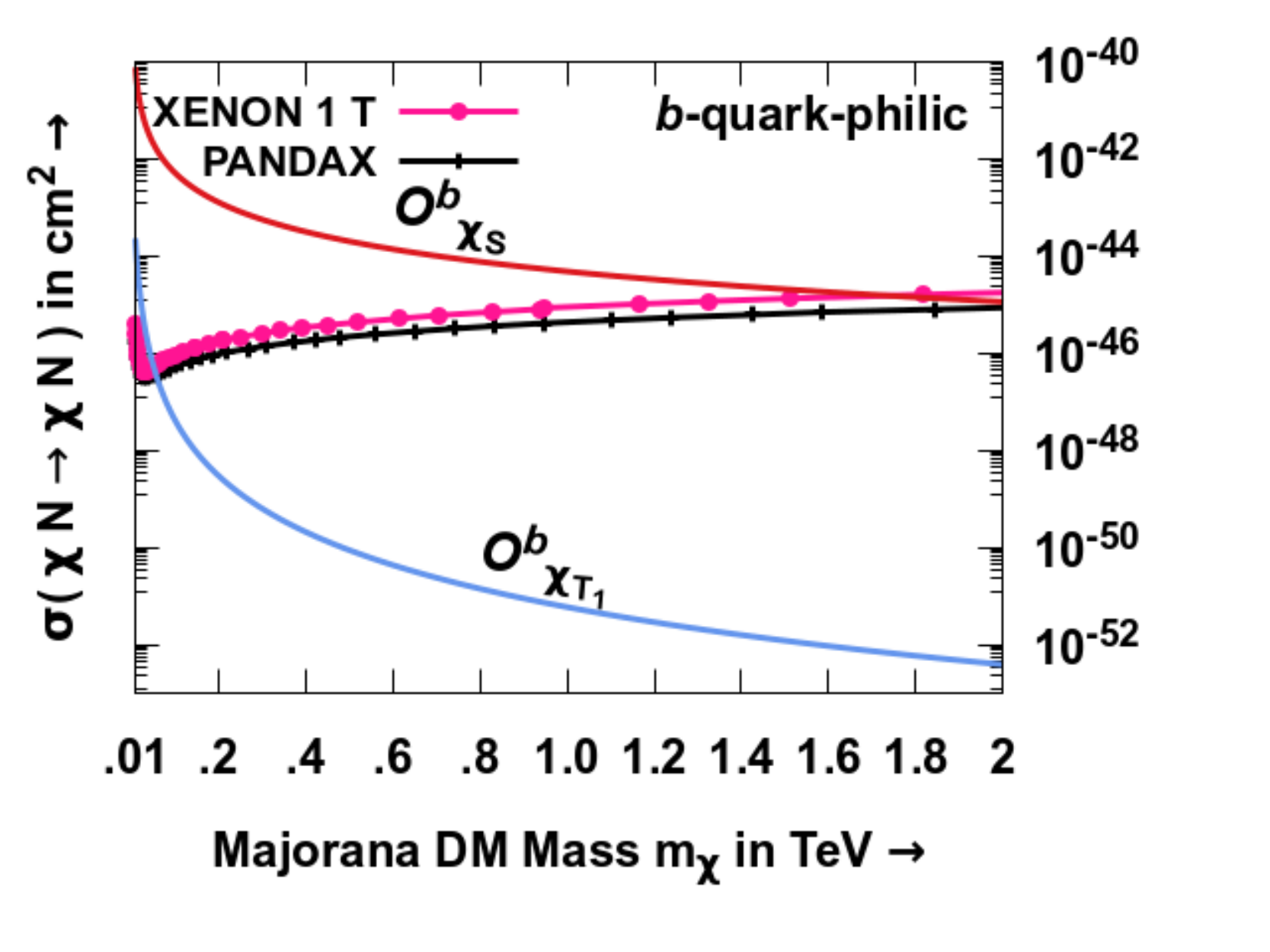}
      \caption{}	\label{fig:DDMDMb}
  \end{subfigure}%
\hspace{0.1cm}
  \begin{subfigure}{0.5\textwidth}
      \centering
	\includegraphics[scale=0.5]{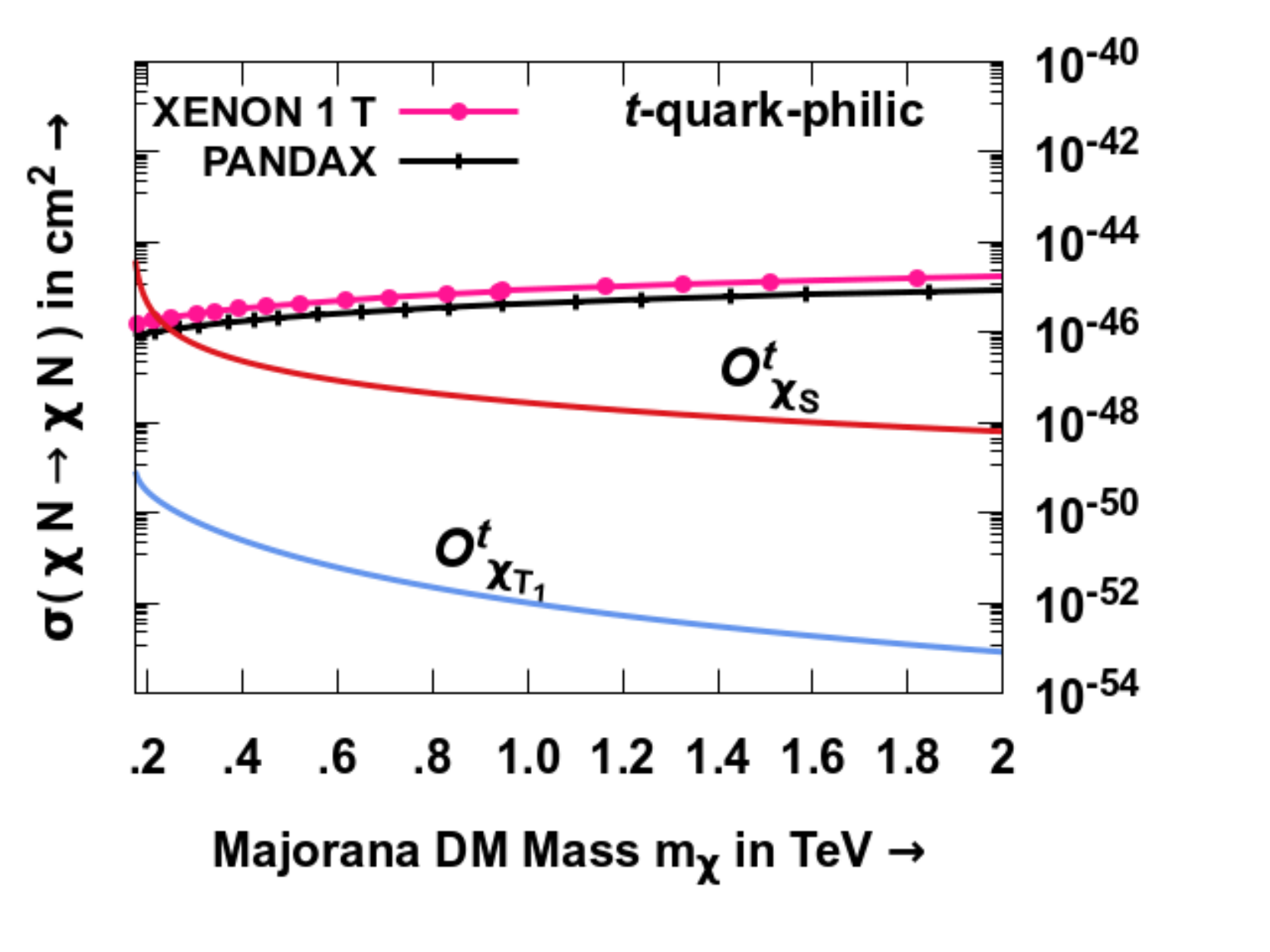}
      \caption{}	\label{fig:DDMDMt}
  \end{subfigure}%
  \\
    \begin{subfigure}{0.5\textwidth}
      \centering
	\includegraphics[scale=0.5]{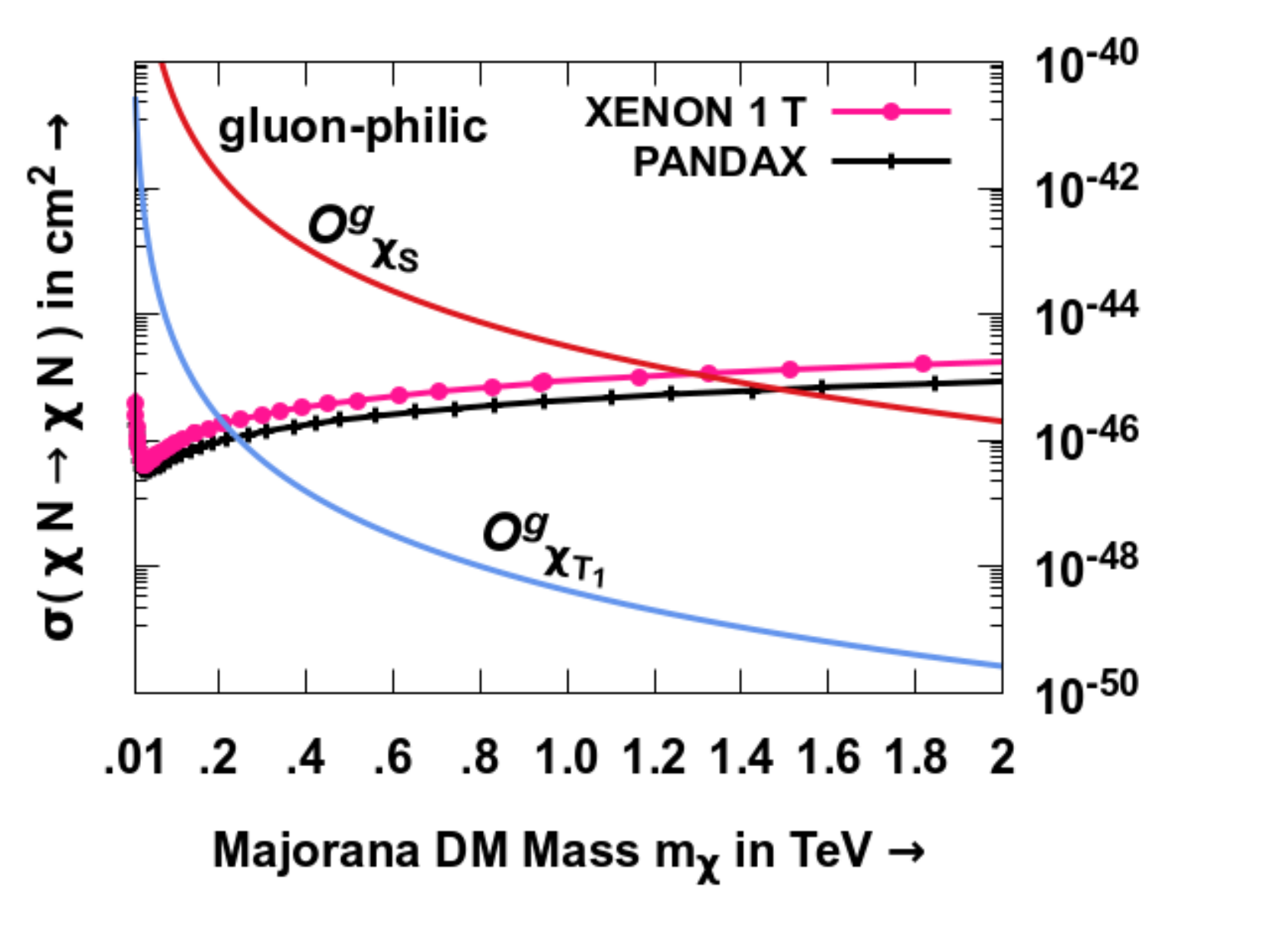}
      \caption{}	\label{fig:DDMDMg}
  \end{subfigure}%
	\caption[nooneline]{\justifying \em{ Figures \ref{fig:DDMDMb}, \ref{fig:DDMDMt} and \ref{fig:DDMDMg} depict the  spin-independent $b$-quark, $t$-quark and gluon-philic Majorana DM-nucleon    scattering cross-sections respectively. The scalar and twist-2 type-1 contributions in all panels  are evaluated using their respective  values for  $  \left\vert C^{q,\,g}_{\chi_{S,\,T_1}}/\, \Lambda^n\right\vert $ satisfying  $\Omega^{\chi} h^2$ = $0.1198\pm 0.0012$  \cite{Aghanim:2018eyx} as shown in figures \ref{fig:RelicFermDM}  and hence, regions above the solid curves  are cosmologically allowed. Regions above the experimental limits obtained from XENON-1T~\cite{XENON:2018voc} and PandaX-4T\cite{PandaX-4T:2021bab} are excluded.}}
	\label{fig:DDMDMX}  
\end{figure}

\begin{figure}[h!]
	\centering
\hspace{-0.5cm}
  \begin{subfigure}{0.5\textwidth}
      \centering
	\includegraphics[scale=0.5]{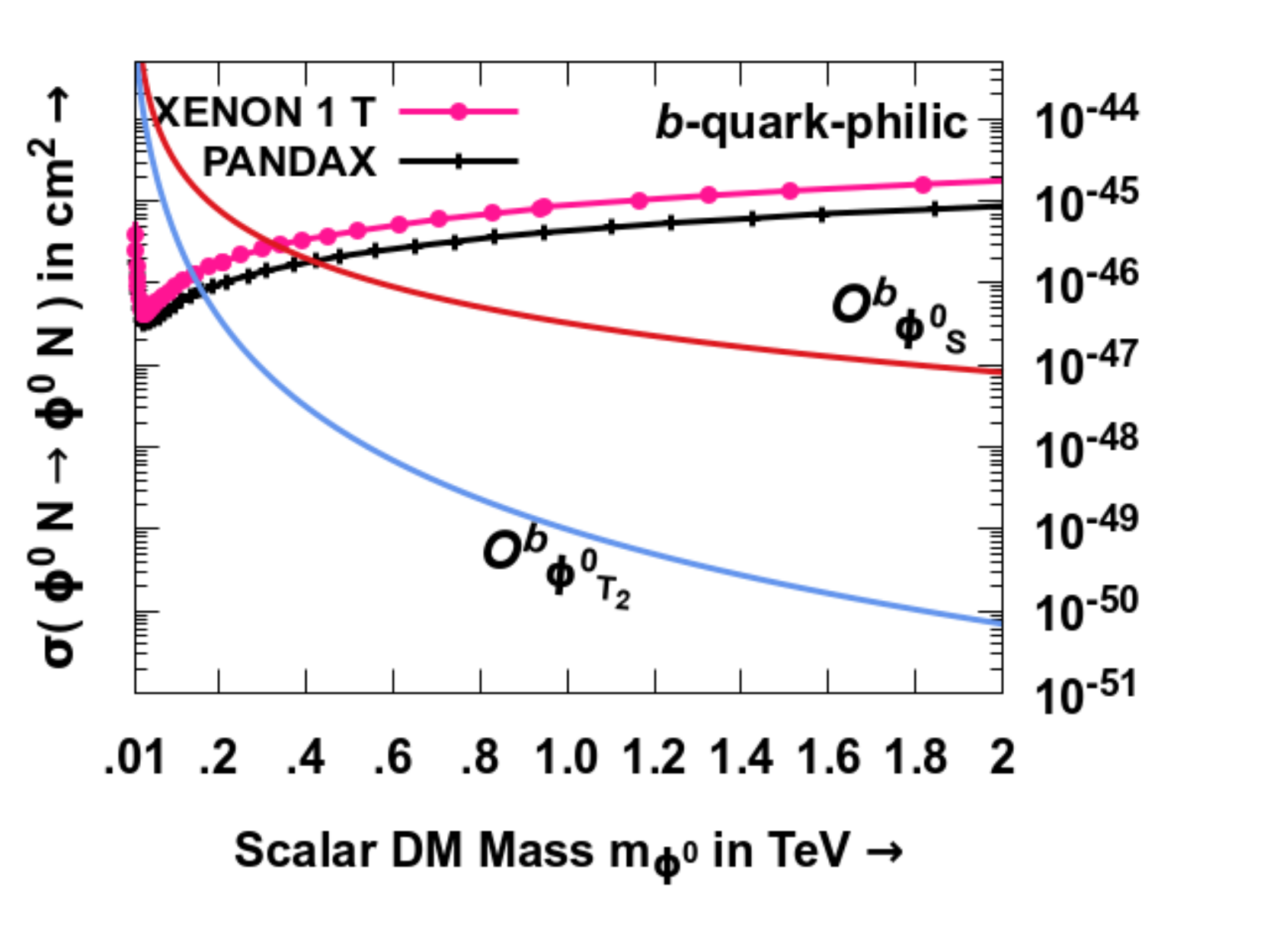}
      \caption{}	\label{fig:DDSDMb}
  \end{subfigure}%
\hspace{0.1cm}
  \begin{subfigure}{0.5\textwidth}
      \centering
	\includegraphics[scale=0.5]{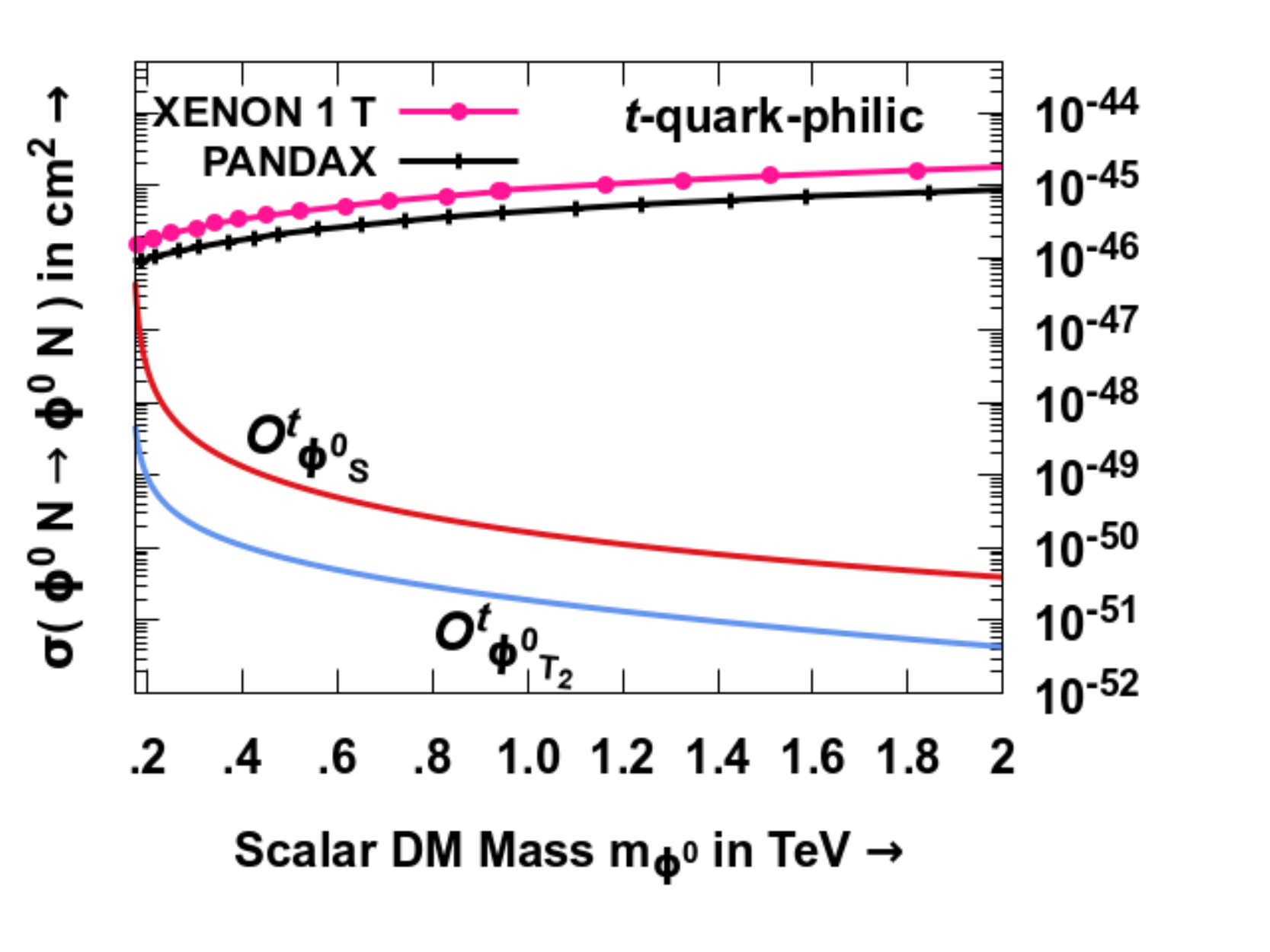}
      \caption{}	\label{fig:DDSDMt}
  \end{subfigure}%
  \\
    \begin{subfigure}{0.5\textwidth}
      \centering
	\includegraphics[scale=0.5]{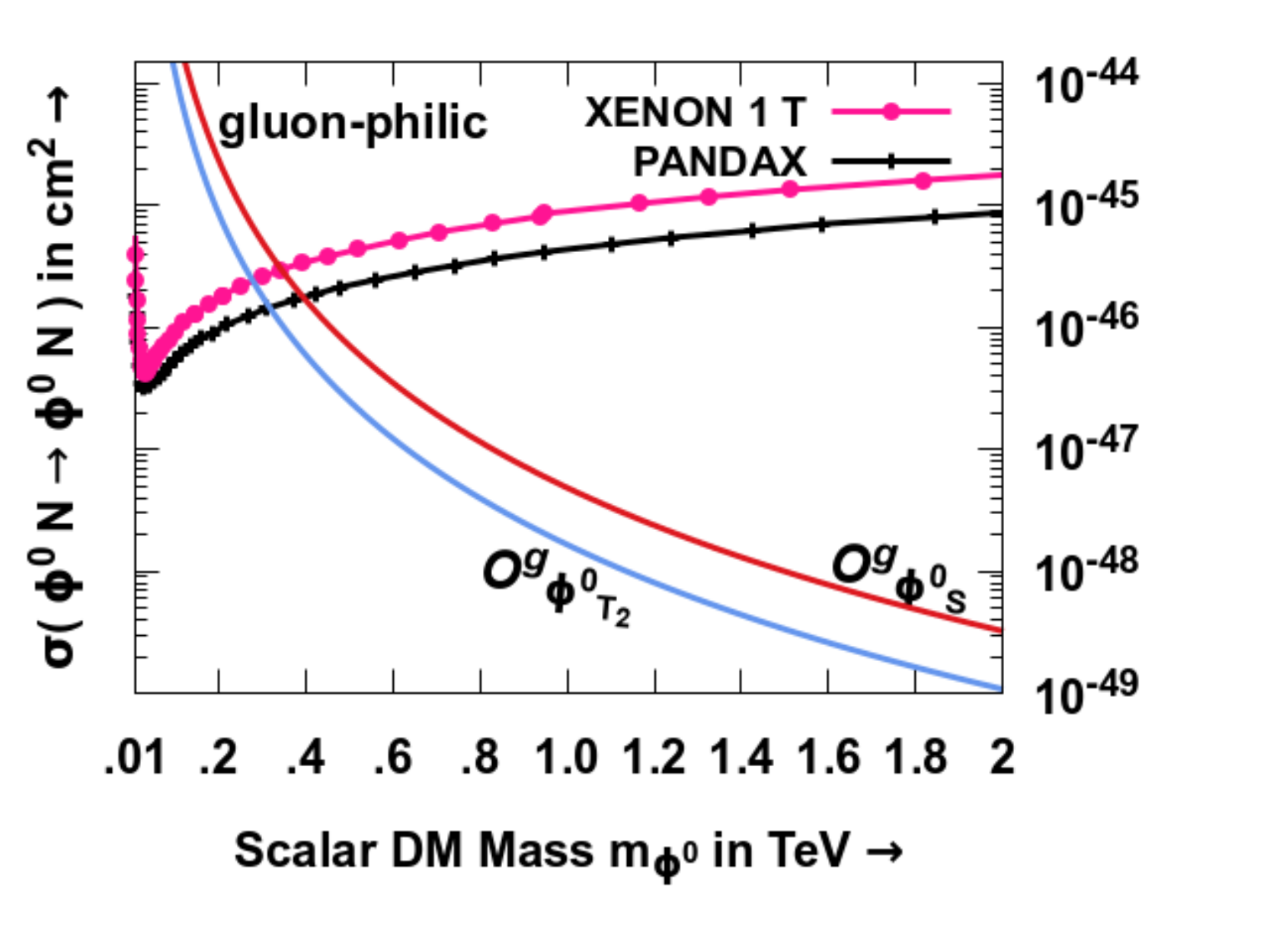}
      \caption{}	\label{fig:DDSDMg}
  \end{subfigure}%
	\caption[nooneline]{\justifying \em{Figures \ref{fig:DDSDMb}, \ref{fig:DDSDMt} and \ref{fig:DDSDMg} depict the  spin-independent $b$-quark, $t$-quark and gluon-philic scalar DM-nucleon  scattering cross-sections respectively. The scalar and twist-2 type-2 contributions in all the panels  are evaluated using their respective values for   $  \left\vert C^{q,\,g}_{\phi^0_{S,\,T_2}}/\, \Lambda^n\right\vert$ satisfying  $\Omega^{\phi^0} h^2$ = $0.1198\pm 0.0012$  \cite{Aghanim:2018eyx} as shown in figures \ref{fig:RelicScalDM}  and hence, regions above the solid curves  are cosmologically allowed. Regions above the experimental limits obtained from XENON-1T~\cite{XENON:2018voc} and PandaX-4T\cite{PandaX-4T:2021bab} are excluded.}}
	\label{fig:DDSDMX}  
\end{figure}
\begin{figure}[h!]
	\centering
\hspace{-0.5cm}
  \begin{subfigure}{0.5\textwidth}
      \centering
	\includegraphics[scale=0.5]{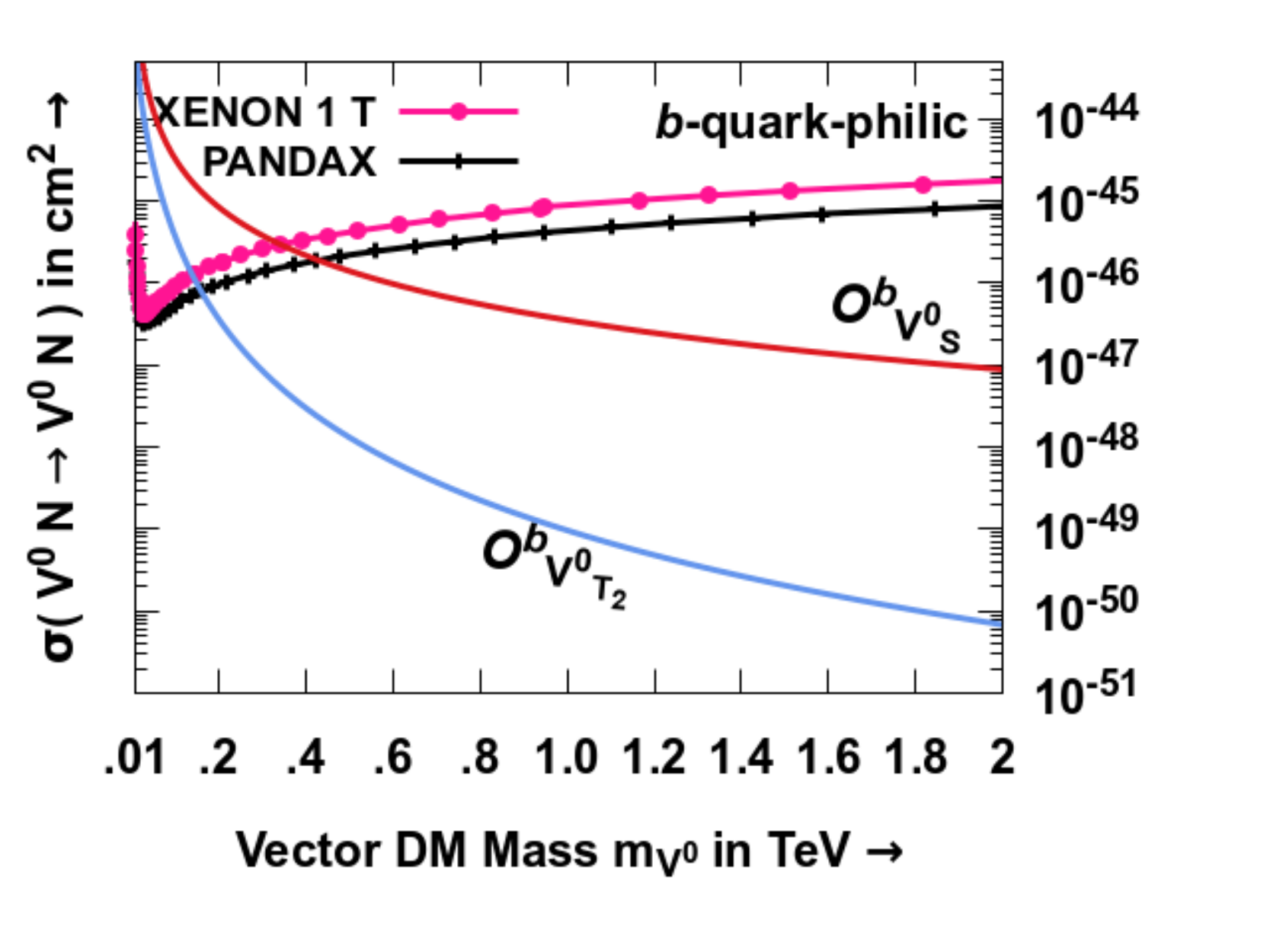}
      \caption{}	\label{fig:DDVDMb}
  \end{subfigure}%
\hspace{0.1cm}
  \begin{subfigure}{0.5\textwidth}
      \centering
	\includegraphics[scale=0.5]{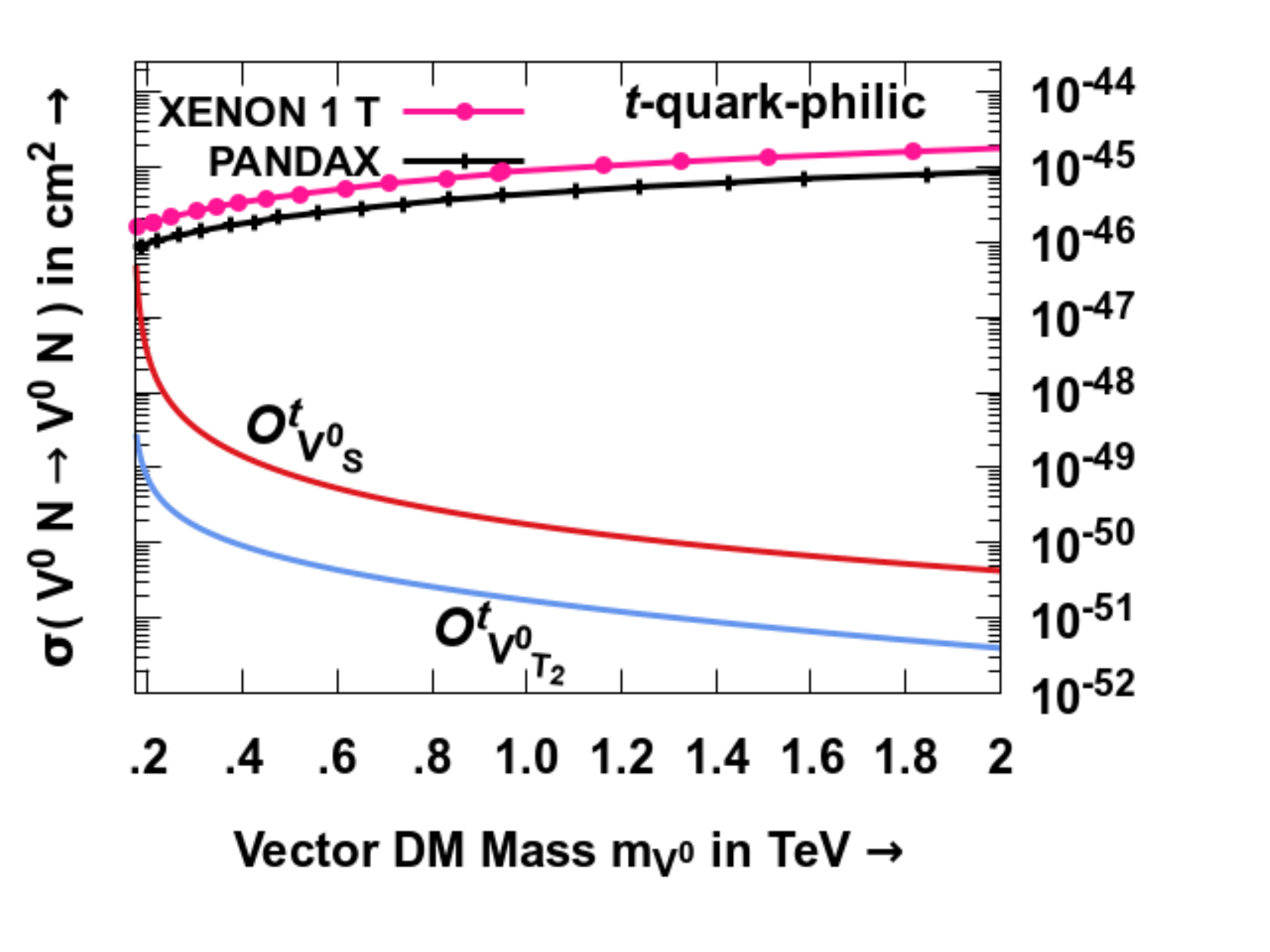}
      \caption{}	\label{fig:DDVDMt}
  \end{subfigure}%
  \\
    \begin{subfigure}{0.5\textwidth}
      \centering
	\includegraphics[scale=0.5]{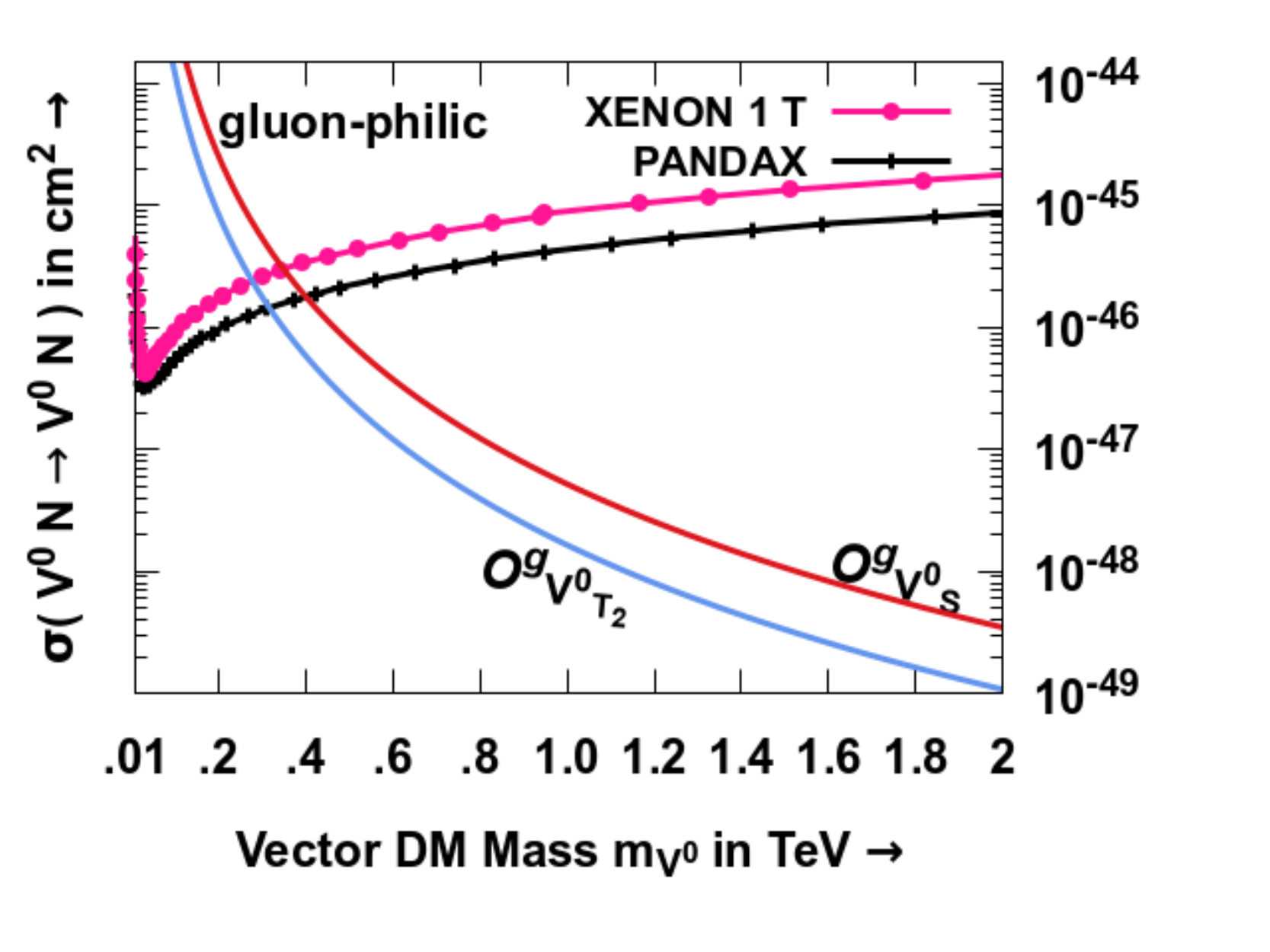}
      \caption{}	\label{fig:DDVDMg}
  \end{subfigure}%
	\caption[nooneline]{\justifying \em{Figures \ref{fig:DDVDMb}, \ref{fig:DDVDMt} and \ref{fig:DDVDMg} depict the  spin-independent $b$-quark, $t$-quark and gluon-philic vector DM-nucleon  scattering cross-sections respectively. The scalar and twist-2 type-2 contributions in all the panels  are evaluated using their respective  values for  $  \left\vert C^{q,\,g}_{V^0_{S,\,T_2}}/\, \Lambda^n\right\vert$ satisfying  $\Omega^{V^0} h^2$ = $0.1198\pm 0.0012$  \cite{Aghanim:2018eyx} as shown in figures \ref{fig:RelicVectDM}  and hence, regions above the solid curves  are cosmologically allowed. Regions above the experimental limits obtained from XENON-1T~\cite{XENON:2018voc} and PandaX-4T\cite{PandaX-4T:2021bab} are excluded.}}
	\label{fig:DDVDMX}  
\end{figure}

\par For completion, we compute and display the tree-level spin-independent Majorana DM-nucleon, scalar DM-nucleon, and vector DM-nucleon scattering cross-sections induced by the scalar and twist-2 current of gluons in equations \eqref{gScattScalMDM}-\eqref{gScattTwistMDM}, \eqref{gScattScalSDM}-\eqref{gScattTwistSDM} and \eqref{gScattScalVDM}-\eqref{gScattTwistVDM}   respectively as
\begin{subequations}
\begin{eqnarray}
\sigma^{g^{\chi N}}_{S} 
&=&\frac{128}{81}\frac{1}{\pi} \left(C_{\chi_S}^g\right)^2\left[ \frac{1\, {\rm TeV}}{\Lambda}\right]^8\,\left[\frac{m_\chi}{100\, {\rm GeV}}\right]^2\left[\frac{m_N}{1\, {\rm GeV}}\right]^2\left[\frac{\mu_{N\chi}}{1\,{\rm GeV}}\right]^2\,\nn\\
&&\,\,\,\times\,\,\left[\frac{\alpha_s\left(\Lambda\right)}{\alpha_s\left(\mu\right)}\right]^2   \left\vert f^N_{\rm TG}\right\vert^2 \times\left(3.9\times 10^{-48}\right)\, {\rm cm}^{2}
\label{gScattScalMDM}
\end{eqnarray}
\begin{eqnarray}
\sigma^{g^{\chi N}}_{T_1} 
&=&\frac{9}{8}\frac{1}{\pi} \left(C_{\chi_{T_1}}^g\right)^2\left[ \frac{1\, {\rm TeV}}{\Lambda}\right]^8\,\left[\frac{m_\chi}{100\, {\rm GeV}}\right]^2\left[\frac{m_N}{1\, {\rm GeV}}\right]^2\left[\frac{\mu_{N\chi}}{1\,{\rm GeV}}\right]^2\,\nn\\
&&\,\,\,\,\times\,\,\left\vert g\left(2;\,\mu_F\right)\right\vert^2 \times \left(3.9\times 10^{-48}\right)\, {\rm cm}^{2}\label{gScattTwistMDM}
\end{eqnarray}
\end{subequations}

\begin{subequations}
\begin{eqnarray}
\sigma^{g^{\phi^0 N}}_{S} 
&=&\frac{64}{81}\,\frac{1}{\pi} \left(C_{\phi^0_S}^g\right)^2\left[ \frac{1\, {\rm TeV}}{\Lambda}\right]^4\,\left[ \frac{100\, {\rm GeV}}{m_{\phi^0}}\right]^2\left[\frac{m_N}{1\, {\rm GeV}}\right]^2\left[\frac{\mu_{N\phi^0}}{1\,{\rm GeV}}\right]^2\,\left[\frac{\alpha_s\left(\Lambda\right)}{\alpha_s\left(\mu\right)}\right]^2 \nn\\
&&\,\,\,\,\times  \left\vert f^N_{\rm TG}\right\vert^2 \left(3.9 \times 10^{-44}\right)\, {\rm cm}^{2}\label{gScattScalSDM}
\end{eqnarray}
\begin{eqnarray}
\sigma^{g^{\phi^0 N}}_{T_2} 
&=& \frac{9}{32\pi}\, \left(C^g_{\phi^0_{T_2}}\right)^2 \bigg[\frac{1\, {\rm TeV}}{\Lambda}\bigg]^8\, \bigg[\frac{\alpha_s(\Lambda)}{4 \pi}\bigg]^2\, \left[\frac{\mu_{N\phi^0}}{1\, {\rm GeV}}\right]^2\, \left[\frac{m_{\phi^0}}{100\, {\rm GeV}}\right]^2\, \left[\frac{m_N}{1\, {\rm GeV}}\right]^2\, \nn\\
&&\,\,\,\, \times \left\vert g(2;\mu_F)\right\vert^2\, \left(3.9 \times 10^{-48}\right)\, {\rm cm^2}  \label{gScattTwistSDM}
\end{eqnarray}
\end{subequations}
\begin{subequations}
\begin{eqnarray}
\sigma^{g^{V^0 N}}_{S} 
&=&\frac{64}{243}\,\frac{1}{\pi} \left(C_{V^0_S}^g\right)^2\left[ \frac{1\, {\rm TeV}}{\Lambda}\right]^4\,\left[ \frac{100\, {\rm GeV}}{m_{V^0}}\right]^2\left[\frac{m_N}{1\, {\rm GeV}}\right]^2\left[\frac{\mu_{NV^0}}{1\,{\rm GeV}}\right]^2\,\left[\frac{\alpha_s\left(\Lambda\right)}{\alpha_s\left(\mu\right)}\right]^2 \nn\\
&&\,\,\,\,\times  \left\vert f^N_{\rm TG}\right\vert^2 \left(3.9 \times 10^{-44}\right)\, {\rm cm}^{2}\label{gScattScalVDM}
\end{eqnarray}
\begin{eqnarray}
\sigma^{g^{V^0 N}}_{T_2}
&=& \frac{3}{32\pi}\, \left(C^g_{V^0_{T_2}}\right)^2 \bigg[\frac{1\, {\rm TeV}}{\Lambda}\bigg]^8\, \bigg[\frac{\alpha_s(\Lambda)}{4 \pi}\bigg]^2\, \left[\frac{\mu_{NV^0}}{1\, {\rm GeV}}\right]^2\, \left[\frac{m_{V^0}}{100\, {\rm GeV}}\right]^2\, \left[\frac{m_N}{1\, {\rm GeV}}\right]^2\, \nn\\
&&\,\,\,\, \times \left\vert g(2;\mu_F)\right\vert^2\, \left(3.9 \times 10^{-48}\right)\, {\rm cm^2}  \label{gScattTwistVDM}
\end{eqnarray}
\end{subequations}
The analytical expressions for the gluon-philic DM-nucleon scattering cross-sections are in agreement with those given in reference \cite{Hisano:2015bma}. 

\par We display the spin independent Majorana, real scalar and real vector  DM-nucleon scattering cross-sections {\it w.r.t.} DM mass in  figures \ref{fig:DDMDMX}, \ref{fig:DDSDMX} and \ref{fig:DDVDMX} respectively. 
The scalar and twist-2 currents induced by $b$-quark-philic and $t$-quark-philic DM interactions are depicted in the left and right panels of all the three figures. Since, the cross-sections are evaluated using  the Wilson coefficients obtained from the relic density contours satisfying  $\Omega^{\rm DM} h^2$ = $0.1198$ in figures \ref{fig:RelicFermDM}, \ref{fig:RelicScalDM} and \ref{fig:RelicVectDM} for the Majorana, scalar and vector DM  respectively, the solid curves represent the cosmological lower-limits of the scattering cross-sections.
\par Each panel in figures \ref{fig:DDMDMX}, \ref{fig:DDSDMX} and \ref{fig:DDVDMX} also display the central values of the spin-independent cross-sections for a given DM mass from the interpolation of the observed data  in XENON-1T \cite{XENON:2018voc} and PandaX-4T data \cite{PandaX-4T:2021bab} experiments. The data for these cross-sections is derived from the lack of any excess  in the aforementioned experiments using statistical analysis of the recoil energy spectrum in the binned likelihood approach. These curves from the experiments determine the upper limit of direct-detection spin-independent cross-section. As a result, the experimental upper bound validates the region between the experimental upper bound and the specific cosmological lower limit curve, where the relic density constraint was satisfied using the corresponding fixed Wilson coefficient.

\par The profile of the spin-dependent DM-nucleon scattering cross-sections induced by the axial-vector coupling of heavy quarks with Majorana and Vector DM is characterised by their respective DM-gluon effective pseudo-scalar interaction lagrangians given in equations \eqref{LeffMDMgluon} and \eqref{LeffVDMgluon} respectively. Due to the entangled momentum dependency of the cross-section inside the velocity integral in the event rate calculation, an exclusive numerical estimation of the scattering cross-section corresponding to such operator is beyond the scope of our analysis. The upper limits on Wilson coefficients corresponding to the effective spin and momentum dependent pseudo-scalar interaction of gluons with fermionic DM  have been extracted by interpolating the data for event rates from the experimental data in references \cite{Kang:2018rad} and \cite{Kang:2018odb} for $m_{\rm DM}$ less than 1 TeV.

\subsubsection{Nuclear Recoil Spectrum}
 \label{recoil}
 For   a fixed target detector exposure  $\epsilon_T$ and target  nucleus  mass $m_{\rm nuc.}$, the differential nuclear recoil event rate {\it w.r.t.} nuclear recoil energy $dR_{\rm nuc.}/ dE_r$  corresponding to the DM-nucleon scattering cross-section $\sigma_N$ and DM velocity $\left\vert \vec v\right\vert$   is given as
\begin{eqnarray}
	\frac{dR_{\rm nuc.}}{dE_{\rm r}} &=&  \frac{\epsilon_T\,\rho_0}{m_{\rm nuc.}\ m_{\rm DM}}\ \int_{\left\vert\vec v\right\vert_{\rm min}}^{\left\vert \vec v_{\rm esc}\right\vert }\,\left\vert \vec v\right\vert f\left(\left\vert \vec v\right\vert\right)\ \frac{d\sigma}{dE_{\rm r}}\ d\left\vert \vec v\right\vert \nn\\
	&&\,\,\,\, =   \frac{\epsilon_T\,\rho_{\rm DM}}{2\, m_{\rm DM}}\, \frac{\sigma_N\, A^2}{\mu^2_N}\ \left\vert F\left(\left\vert\vec q\right\vert\,r_n\right)\right\vert^2\,\displaystyle \int_{ \left\vert \vec v\right\vert_{\rm min}}^{\left\vert \vec v\right\vert_{\rm max}} \frac{1}{\left\vert \vec v_{\rm DM}\right\vert}\,f\left(\left\vert \vec v\right\vert\right)\,\, d^3 \vec v
\label{Rate}
\end{eqnarray}
where $F\left(\left\vert\vec q\right\vert\,r_n\right)$  is the Helm form factor taking account of the non-vanishing finite size of the nucleus \cite{Helm:1956zz}. Here $r_n \equiv 1.2\times A^{1/3}$  is the effective radius of the nucleus with atomic mass $A$  and $\left\vert\vec q\right\vert$ is the momentum transfer corresponding to the recoil energy $E_r$.
Following reference \cite{Lewin:1995rx}, the  velocity integral for normalized  Maxwellian DM velocity distribution is solved as 
\begin{subequations}
\begin{eqnarray}
	&&\displaystyle\int_{\left\vert \vec v\right\vert_{\rm min}}^{\left\vert \vec v\right\vert_{\rm max}} \frac{f\left(\left\vert \vec v\right\vert\right)}{\left\vert \vec v\right\vert}\, d^3 \vec v =
	\displaystyle\int_{\sqrt{m_{\rm nuc.}\, E_r/\left(2\, \mu^2_{\rm nuc.}\right)}}^{\left\vert \vec v_{\rm esc}\right\vert} \frac{1}{\left\vert \vec v\right\vert}e^{-\left[  (\left\vert \vec v + \vec v_E\right\vert)/\,\left\vert \vec v_0\right\vert \right]^2} d^3\vec v \nn\\
	&&\,\,\,\,= \frac{1}{2\,  \left\vert\vec v_E\right\vert}\, {\rm erf}(\eta_-,\eta_+) - \frac{1}{\pi\, \left\vert \vec v_E\right\vert}\, (\eta_+ - \eta_-)\, e^{-\left[\left\vert \vec v_{\rm esc}\right\vert/ \left\vert\vec v_0\right\vert\right]^2}\hskip 0.5 cm {\rm where}\\
	&& \eta_\pm = \min\left[ \frac{\sqrt{m_{\rm nuc.}\, E_r/\left(2 \,\mu^2_{\rm nuc.}\right)}\pm \left\vert \vec v_E\right\vert}{\left\vert \vec v_0\right\vert}, \frac{\left\vert \vec v_{\rm esc}\right\vert }{\left\vert \vec v_0\right\vert} \right]\hskip 0.1 cm {\rm and} \hskip 0.1 cm \mu_{\rm nuc.}\equiv \frac{m_{\rm DM}\,m_{\rm nuc.}}{m_{\rm DM}+m_{\rm nuc.}}
\end{eqnarray}
\end{subequations}

\begin{figure}[h!]
	\centering
\hspace{-0.5cm}
  \begin{subfigure}{0.5\textwidth}
      \centering
	\includegraphics[scale=0.5]{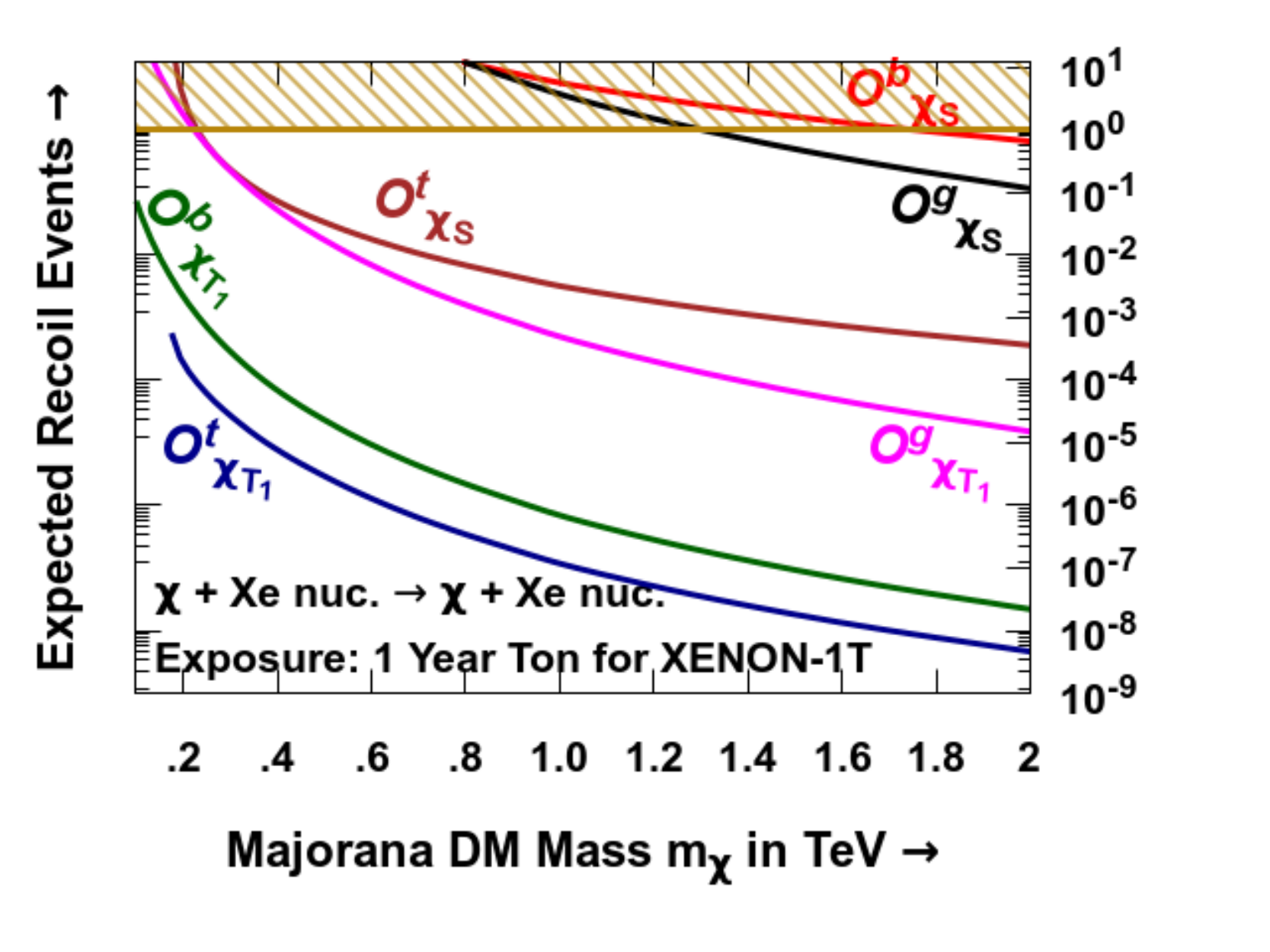}
      \caption{}	\label{fig:RecoilDDMDM}
  \end{subfigure}%
\hspace{0.1cm}
  \begin{subfigure}{0.5\textwidth}
      \centering
	\includegraphics[scale=0.5]{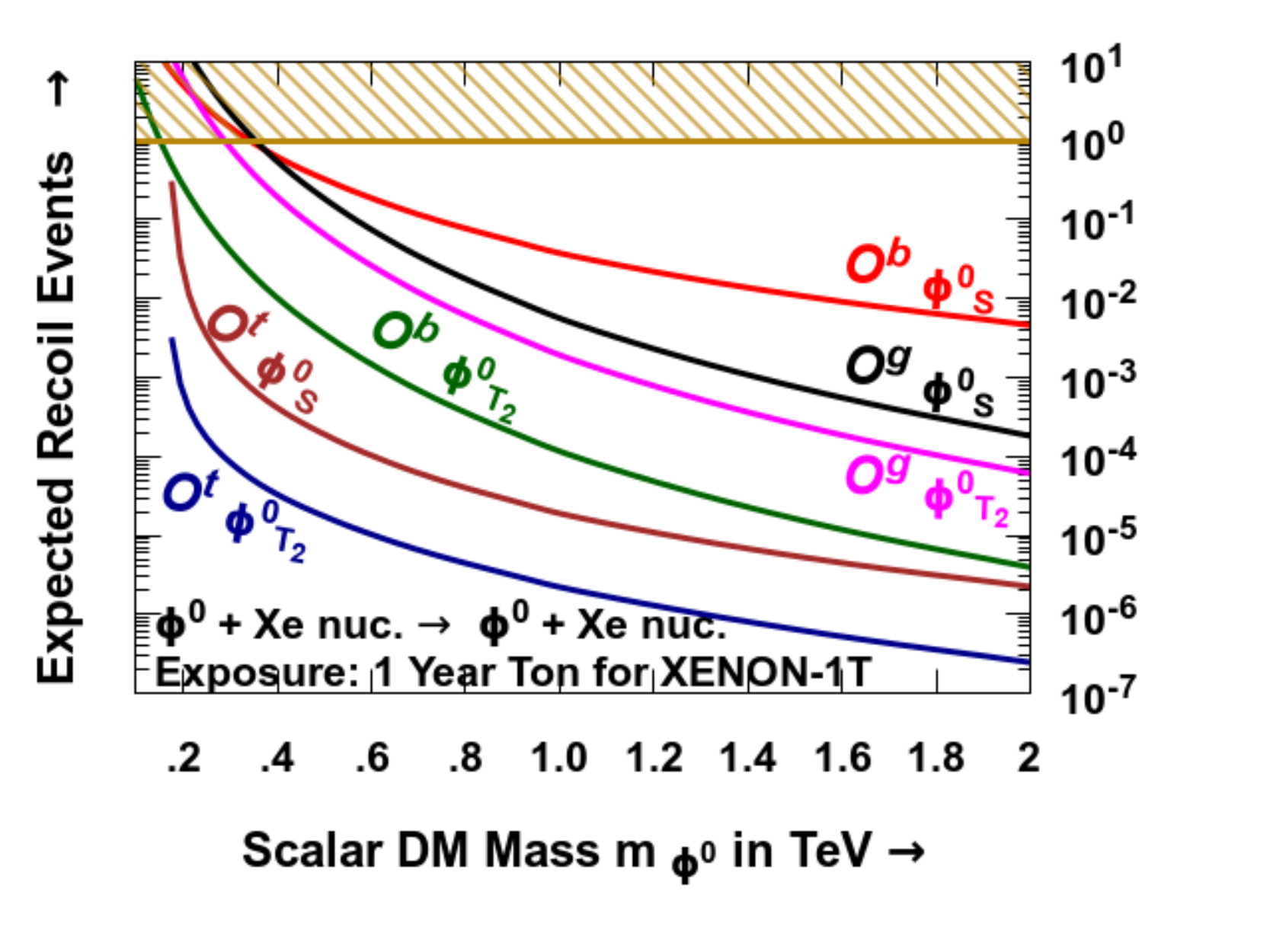}
      \caption{}	\label{fig:RecoilDDSDM}
  \end{subfigure}%
  \\
  \begin{subfigure}{0.5\textwidth}
      \centering
	\includegraphics[scale=0.5]{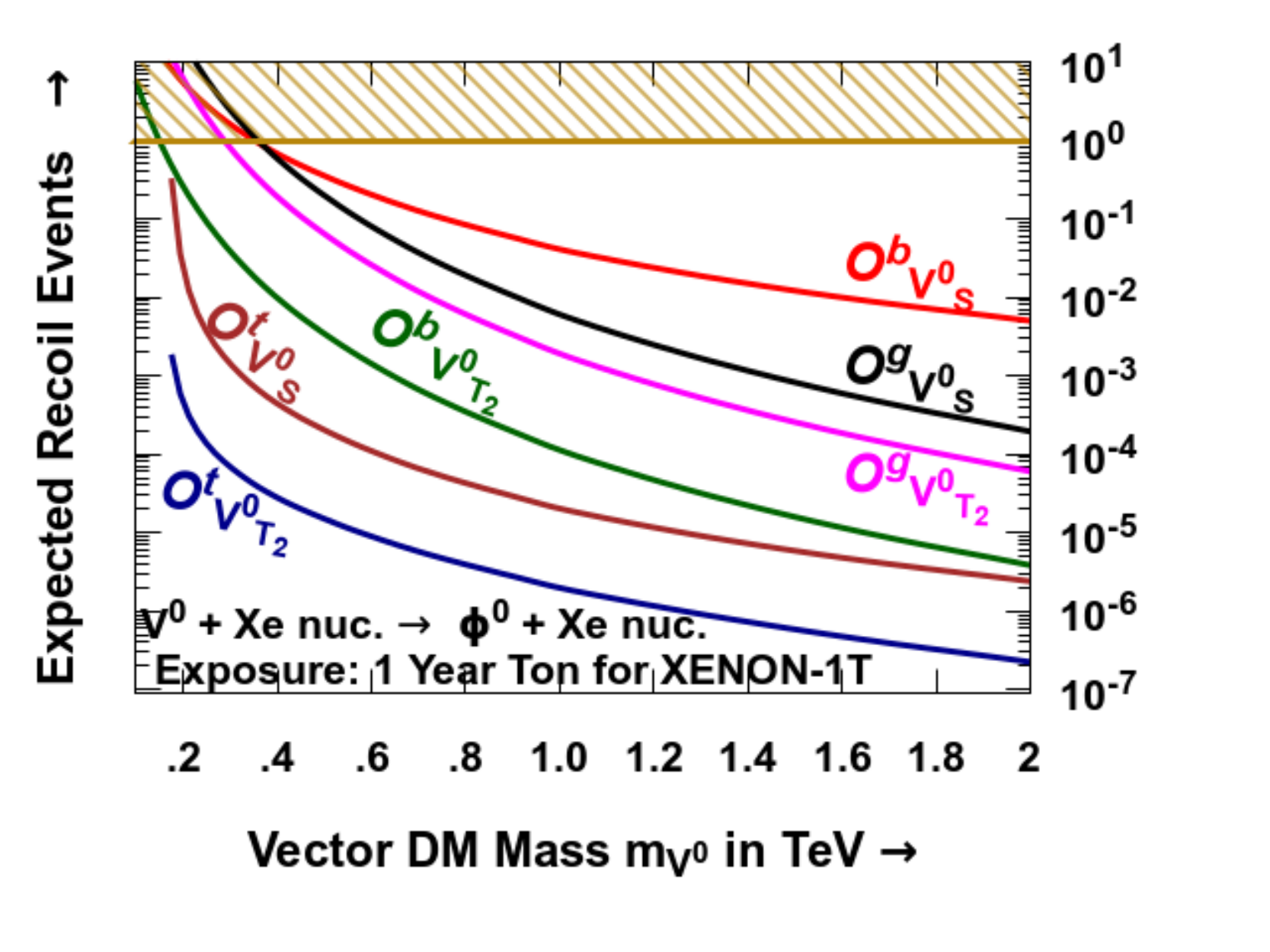}
      \caption{}	\label{fig:RecoilDDVDM}
  \end{subfigure}%
	\caption[nooneline]{\justifying \em{ Figures \ref{fig:RecoilDDMDM}, \ref{fig:RecoilDDSDM} and \ref{fig:RecoilDDVDM}  depict the  spin-independent recoiled nucleus event  due to respective scattering of Majorana, scalar and vector DM with Xe nucleus.  The scalar and twist-2   interactions  of $b$-quark, $t$-quark and gluon  in all the three panels  are evaluated using their respective  values for $  \left\vert C^{q,\,g}_{\chi_{S, \,T_1}}/\, \Lambda^n\right\vert$, $\left\vert C^{q,\,g}_{\phi^0_{S, \,T_2}}/\, \Lambda^n\right\vert$ and $\left\vert C^{q,\,g}_{V^0_{S, \,T_2}}/\, \Lambda^n\right\vert$ satisfying  $\Omega^{\rm DM} h^2$ = $0.1198\pm 0.0012$  \cite{Aghanim:2018eyx} as shown in figures \ref{fig:RelicFermDM}, \ref{fig:RelicScalDM} and \ref{fig:RelicVectDM} respectively and hence, regions above the solid curves  are cosmologically allowed. Shaded region is excluded from XENON-1T data.}}
	\label{fig:DDRecoilEvents}  
\end{figure}

\par For numerical computation we have taken   Earth's velocity relative to galactic frame to be $\left\vert \vec v_E\right\vert$ = 232 km/s, $\left\vert\vec v_0\right\vert $ = 220 km/s and the escape velocity $\left\vert\vec v_{\rm esc}\right\vert$ = 544 km/s. Further, in order to incorporate the detector based effects, the nucleus recoil event rate $dR_{\rm nuc.}/\,dE_{r}$ is convoluted with the detector efficiency \cite{XENON:2018voc}. Integrating  over the recoil energy from $E_{\rm th} \sim$ 4.9 KeV to the maximum $E_{\rm max}$ for a fixed duration and size of the detector, we can estimate the expected  recoil nucleus events.

\par As an illustration, we predict and plot the probable number of Xe nuclear recoil events {\it w.r.t.} varying DM mass in a XENON-1T \cite{XENON:2018voc} setup where 1.3 tonnes of Xe target are exposed for a duration of 278.8 days, which is equivalent to one Ton-year of net target exposure.

\par The expected number of recoil nucleus events due to Majorana DM-Xe nucleus scattering, scalar DM-Xe nucleus scattering, and vector DM-Xe nucleus scattering, respectively, are depicted in figures \ref{fig:RecoilDDMDM}, \ref{fig:RecoilDDSDM}, and \ref{fig:RecoilDDVDM}, which are induced by scalar and twist-2 currents of heavy quarks at one loop level and of gluons at tree level. Since the solid curves in the figures are created with the corresponding lower bound on the Wilson coefficients satisfying the relic density restriction $\Omega^{\rm DM} h^2$ = $0.1198$, they correspond to the fixed number of events. This results in lower-limits on the estimated number of recoil nucleus events for a particular DM mass. The regions above the respective curves are cosmologically permissible. 

\par The upper bound on the spin-independent scattering events corresponding to the central value of the upper bound obtained in the direct-detection cross-section for the XENON-1T experiment \cite{XENON:2018voc} is then compared with the theoretical predictions for the event rate with fixed Wilson coefficients in each panel of figures \ref{fig:RecoilDDMDM}, \ref{fig:RecoilDDSDM}, and \ref{fig:RecoilDDVDM}. We find that the XENON-1T experiment rules out the contributions of $b$-quark-philic and gluon-philic Majorana DM scalar interactions for $m_\chi \le $ 1.8 and 1.3 TeV, respectively.  Barring the said two operators, we find that contributions from all other operators pertaining to Majorana, scalar and vector DM for $m_{\rm DM}\ge $ 200 GeV may be probed in the ongoing and future direct-detection experiments with enhanced target exposure.

 
\section{Summary and conclusions} \label{summary}

In this article we assess the viability of the $b$-quark-philic, $t$-quark-philic, and gluon-philic self-conjugated spin 1/2, 0 and 1 DM  candidates in the EFT approach. We have formulated the generalised effective interaction Lagrangian induced by the scalar, pseudo-scalar, axial-vector, and twist-2 operators for the real particles in section \ref{model}. 

\par For a given DM mass, the relic abundance of Majorana, scalar, and vector DM is computed in section \ref{relic} using the thermally averaged cross-sections in appendix \ref{TherAvgCS}. Figures  \ref{fig:RelicFermDM}, \ref{fig:RelicScalDM} and \ref{fig:RelicVectDM}  show twenty-eight interaction strengths in the form of Wilson coefficients $  \left\vert C^{q,\,g}_{{\rm DM}_i\,O_j}/\, \Lambda^n\right\vert$ in TeV$^{-n}$ (eleven for Majorana DM, six for scalar DM and eleven for vector DM) satisfying the relic density constraint  $\Omega^{\rm DM} h^2$ $\approx$ 0.1198 \cite{Aghanim:2018eyx}.

\par Using the constrained Wilson coefficients, we  study the thermal averaged DM pair annihilation cross-sections for the indirect detection of varying DM masses ($0.01$ - 2 TeV)  in figures \ref{fig:IDFermDM}, \ref{fig:IDScalDM} and \ref{fig:IDVectDM}. The contributions driven by  the lower-limit of the effective couplings to the annihilation cross-sections in $b\, \bar b$, $t\,\bar t$ and $g\,g$ channels are found to be consistent when compared   with  the upper limits of the $b\,\bar b$ annihilation cross-sections obtained from  FermiLAT~\cite{Ackermann:2015zua} and H.E.S.S.~\cite{Abdallah:2016ygi} indirect experiments.

\par The scattering of the incident heavy-quark-philic DM particle  off the static nucleon  induced by the effective one loop-interactions of the Majorana / scalar / vector DM with gluons in the direct-detection experiments are studied. The contributions of gluon-philic Majorana, scalar and vector DM are revisited and found to be in agreement with results in the literature \cite{Hisano:2015bma}.  Lower bounds on the dominant spin-independent scalar and twist currents induced scattering cross-sections are derived by switching the respective cosmological lower-limits on Wilson coefficients one by one for Majorana, scalar, and vector DM, as illustrated in figures \ref{fig:DDMDMX}, \ref{fig:DDSDMX} and \ref{fig:DDVDMX}. They are compared with the available results  from XENON-1T~\cite{XENON:2018voc} and PandaX-4T \cite{PandaX-4T:2021bab}. 
Furthermore, with an identical target exposure of 1 tonne-year in the XENON-1T experiment, we compute the lower-bound on the predicted number of recoil nucleus events due to Majorana, scalar and vector DM-nucleon scattering and are shown in figures  \ref{fig:RecoilDDMDM}, \ref{fig:RecoilDDSDM}  and \ref{fig:RecoilDDVDM}. 
\begin{figure}[h!]
	\centering
\hspace{-0.5cm}
  \begin{subfigure}{0.5\textwidth}
      \centering
	\includegraphics[scale=0.5]{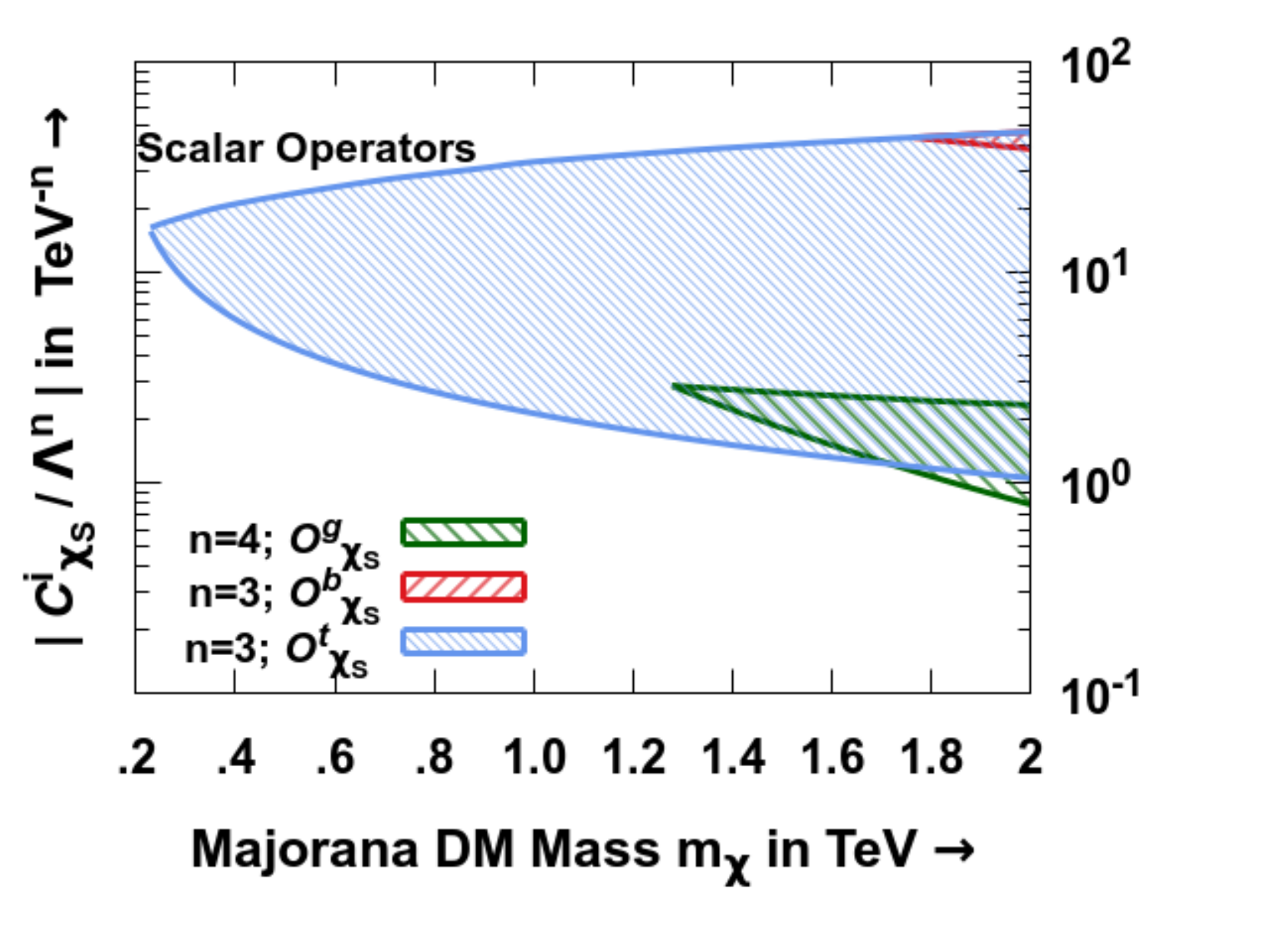}
      \caption{}	\label{fig:ComboMDMScal}
  \end{subfigure}%
\hspace{0.1cm}
  \begin{subfigure}{0.5\textwidth}
      \centering
	\includegraphics[scale=0.5]{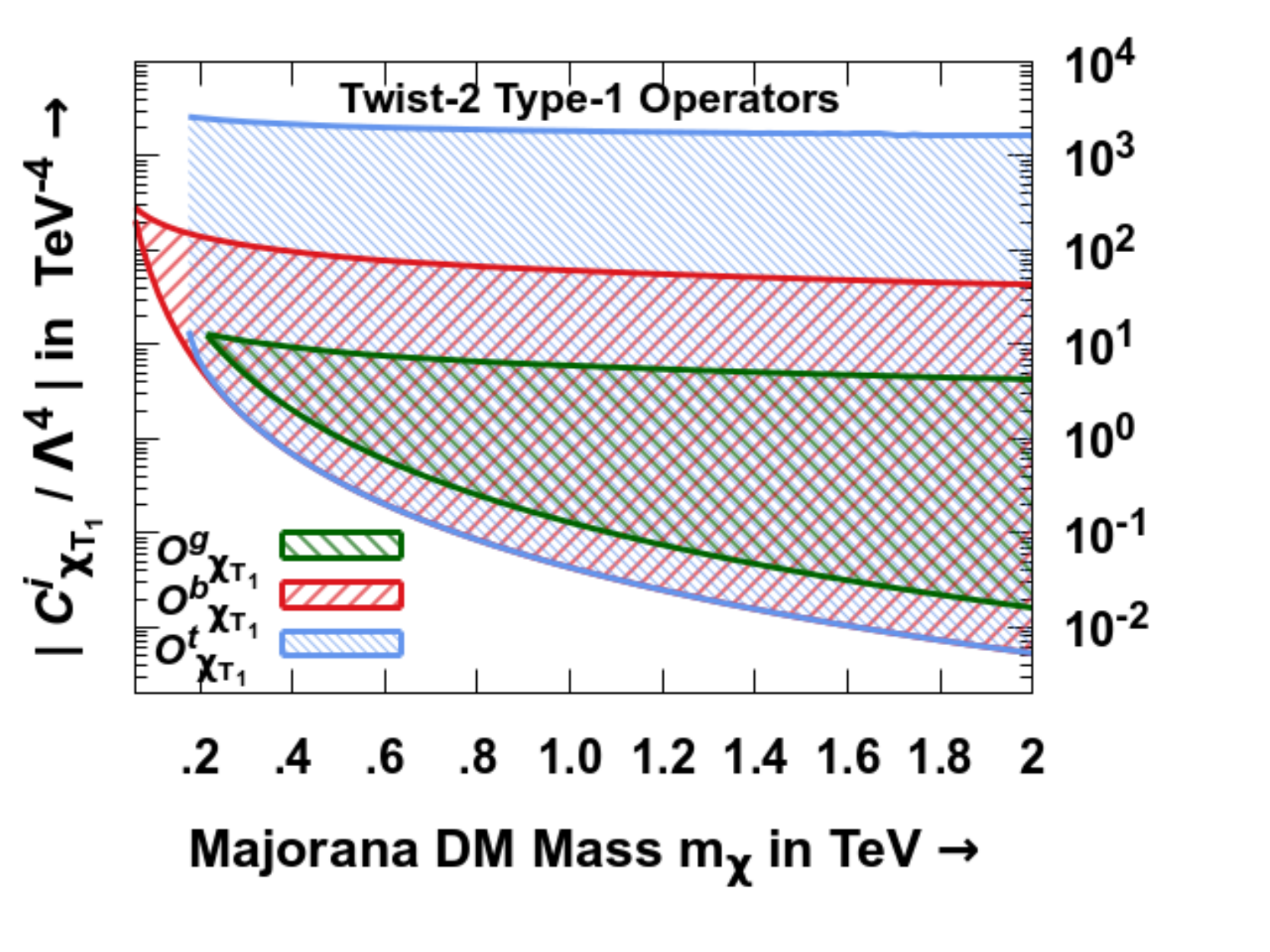}
      \caption{}	\label{fig:ComboMDMTwist}
  \end{subfigure}%
	\caption[nooneline]{\justifying \em{The shaded region in figures \ref{fig:ComboMDMScal} and \ref{fig:ComboMDMTwist} depict the allowed range of the Wilson coefficients corresponding to scalar and twist-2 type-2 operators from the relic density constraint~\cite{Aghanim:2018eyx} and direct detection limits from XENON-1T experiment ~\cite{XENON:2018voc}, shown in red, blue and green for the $b$-quark, $t$-quark and gluon-philic Majorana DM respectively.}}
	\label{fig:MDMCombined}  
\end{figure}

\begin{figure}[h!]
    \begin{subfigure}{0.5\textwidth}
	\includegraphics[scale=0.49]{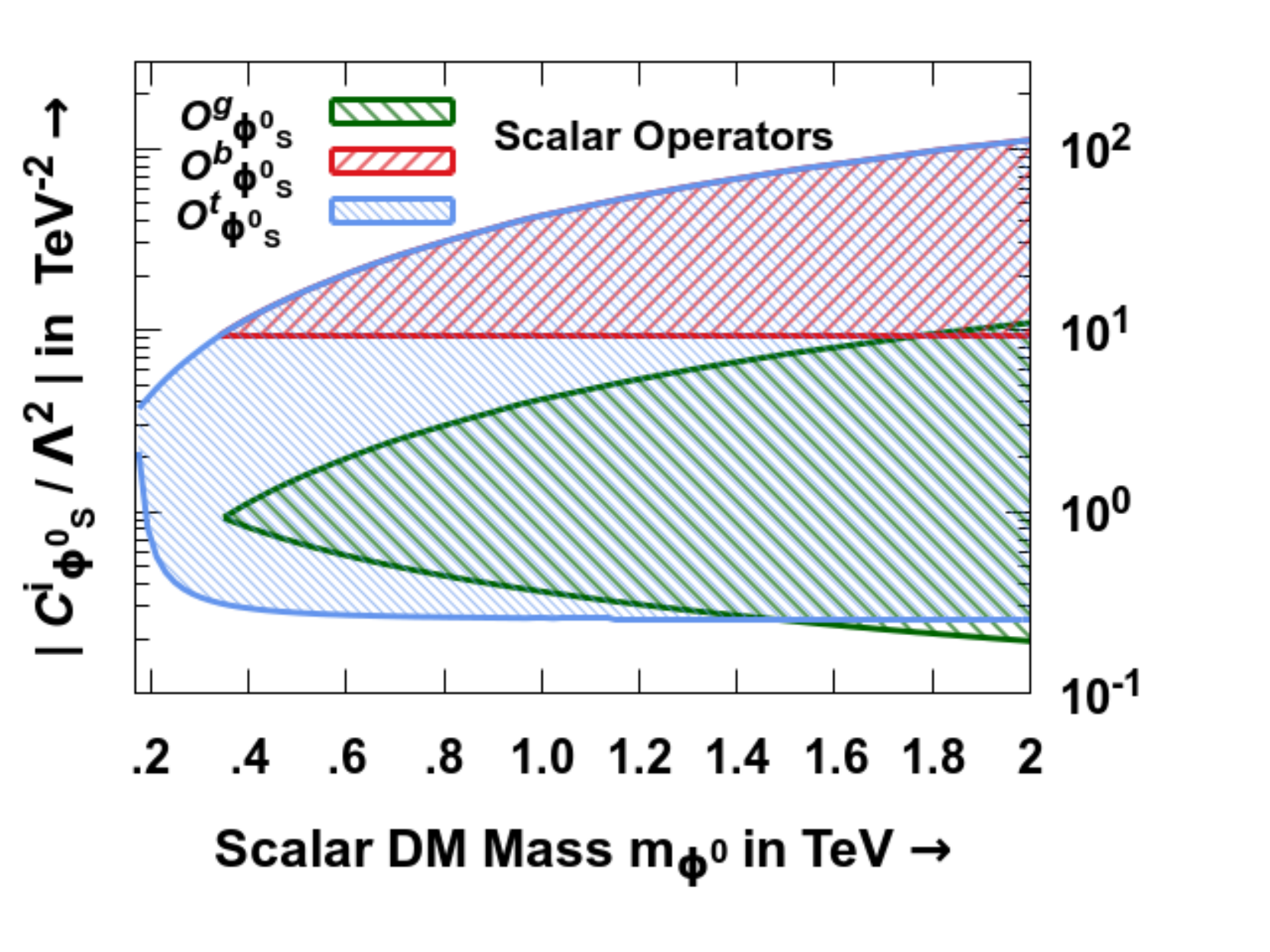}
      \caption{}	\label{fig:ComboSDMScal}
  \end{subfigure}%
 \begin{subfigure}{0.5\textwidth}
	\includegraphics[scale=0.49]{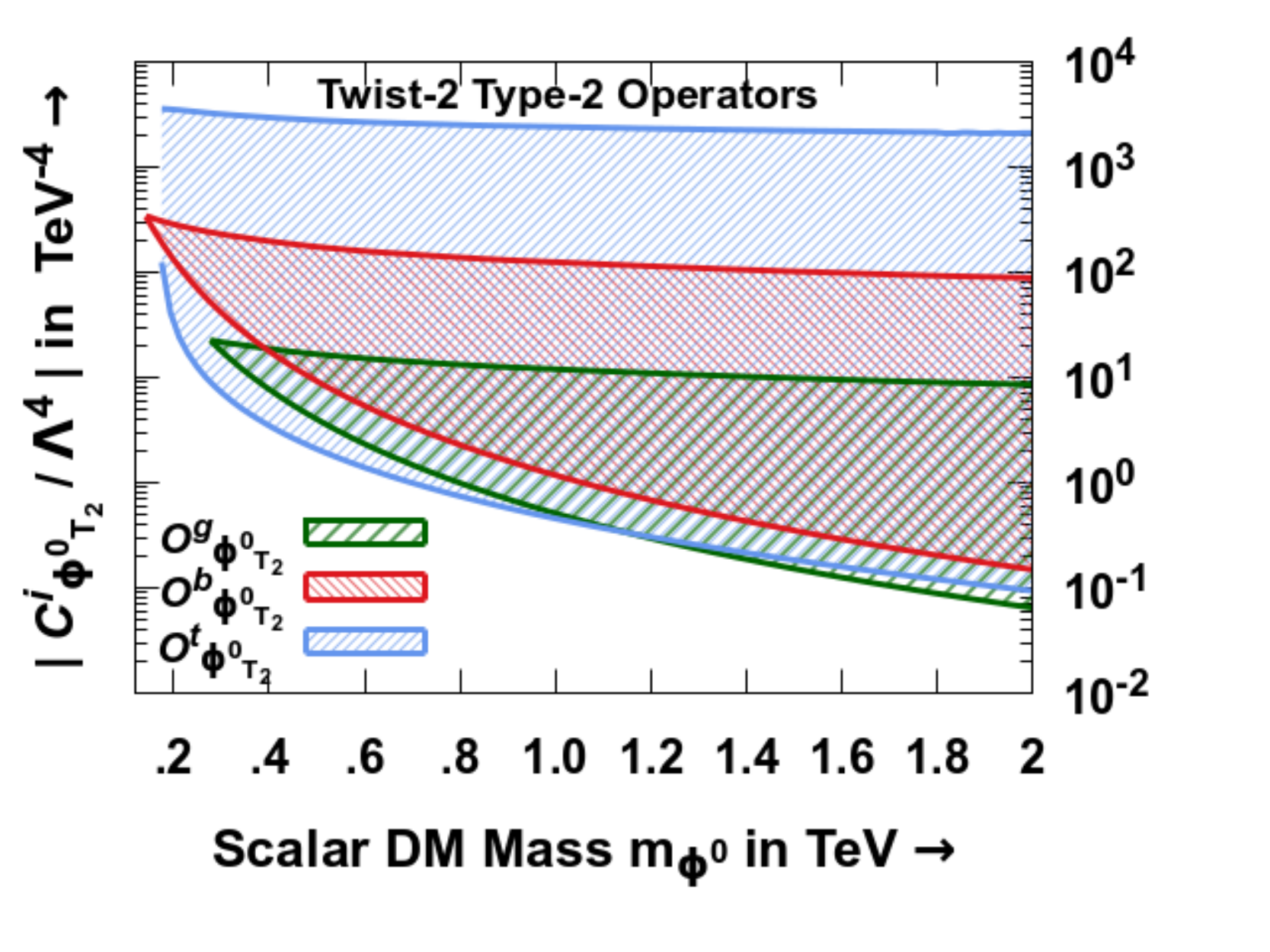}
      \caption{}	\label{fig:ComboSDMTwist}
  \end{subfigure}%
	\caption[nooneline]{\justifying \em{The shaded region in figures \ref{fig:ComboSDMScal} and \ref{fig:ComboSDMTwist} depict the allowed range of the Wilson coefficients corresponding to scalar and twist-2 type-2 operators from the relic density constraint~\cite{Aghanim:2018eyx} and direct detection limits from XENON-1T experiment ~\cite{XENON:2018voc}, shown in red, blue and green for the $b$-quark, $t$-quark and gluon-philic scalar DM respectively.}}
	\label{fig:SDMCombined}  
\end{figure}

\par Finally,  figures \ref{fig:MDMCombined}, \ref{fig:SDMCombined} and \ref{fig:VDMCombined}  corresponding to real spin 1/2, 0 and 1 DM candidates respectively encapsulate the predicted range  for the eighteen Wilson coefficients associated with scalar and twist-2 operators when activated individually. The cosmological relic density~\cite{Aghanim:2018eyx} puts the lower bounds on the Wilson coefficients for varying DM mass $\sim 0.01 \le m_{\rm DM}\le$ 2 TeV, while the upper limits on spin-independent scattering cross-sections obtained in the XENON-1T direct-detection experiment \cite{XENON:2018voc} puts the upper bounds on the Wilson coefficients. However, with the simultaneous switching of these operators the  shaded allowed band of the Wilson coefficients  shown in figures \ref{fig:MDMCombined}, \ref{fig:SDMCombined} and \ref{fig:VDMCombined} may shift below.  Findings of our analysis are summarised as follows:
\begin{figure}[h!]
	\centering
\hspace{-0.5cm}
    \begin{subfigure}{0.5\textwidth}
      \centering
	\includegraphics[scale=0.5]{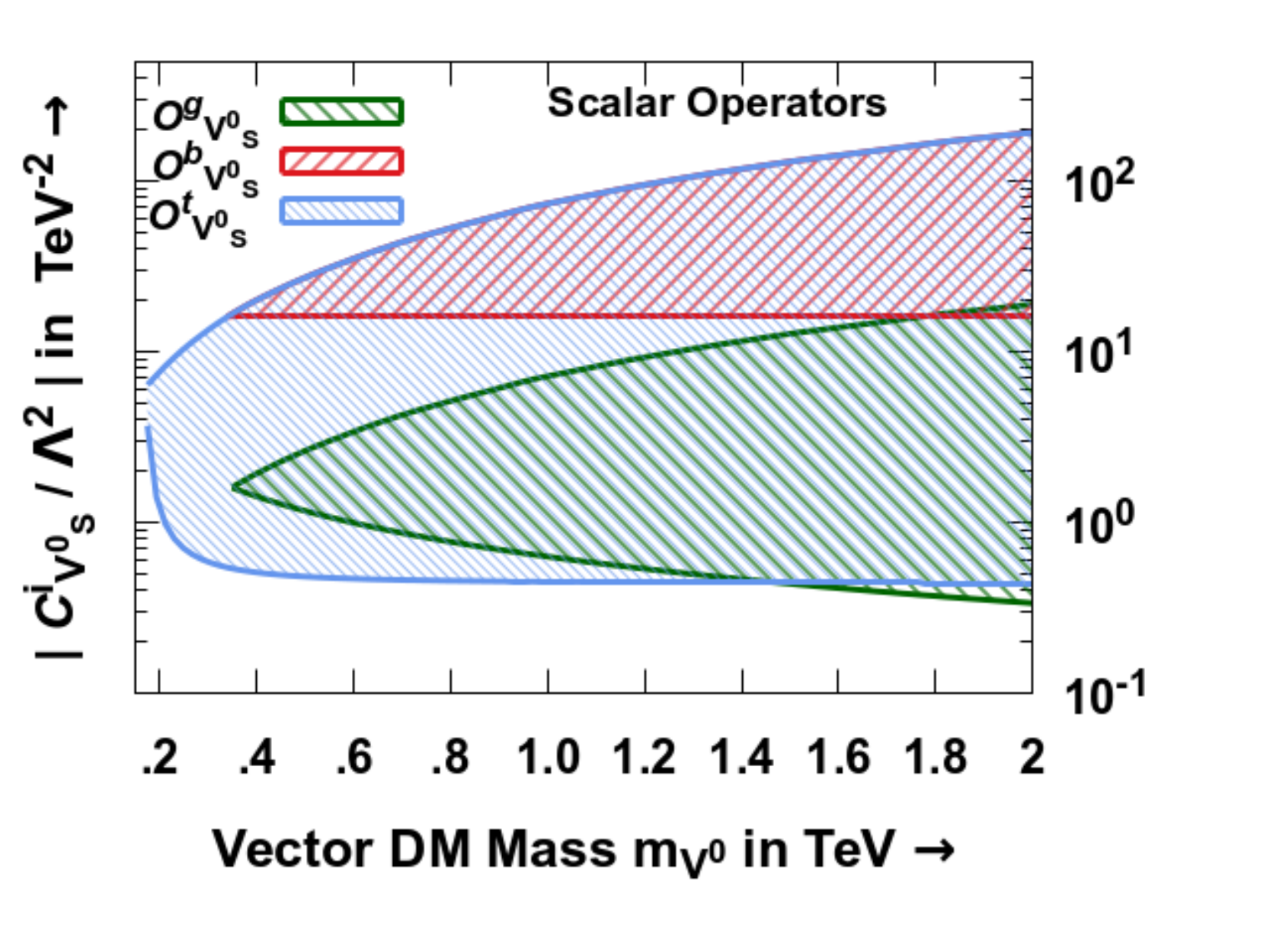}
      \caption{}	\label{fig:ComboVDMScal}
  \end{subfigure}%
\hspace{0.1cm}
 \begin{subfigure}{0.5\textwidth}
      \centering
	\includegraphics[scale=0.5]{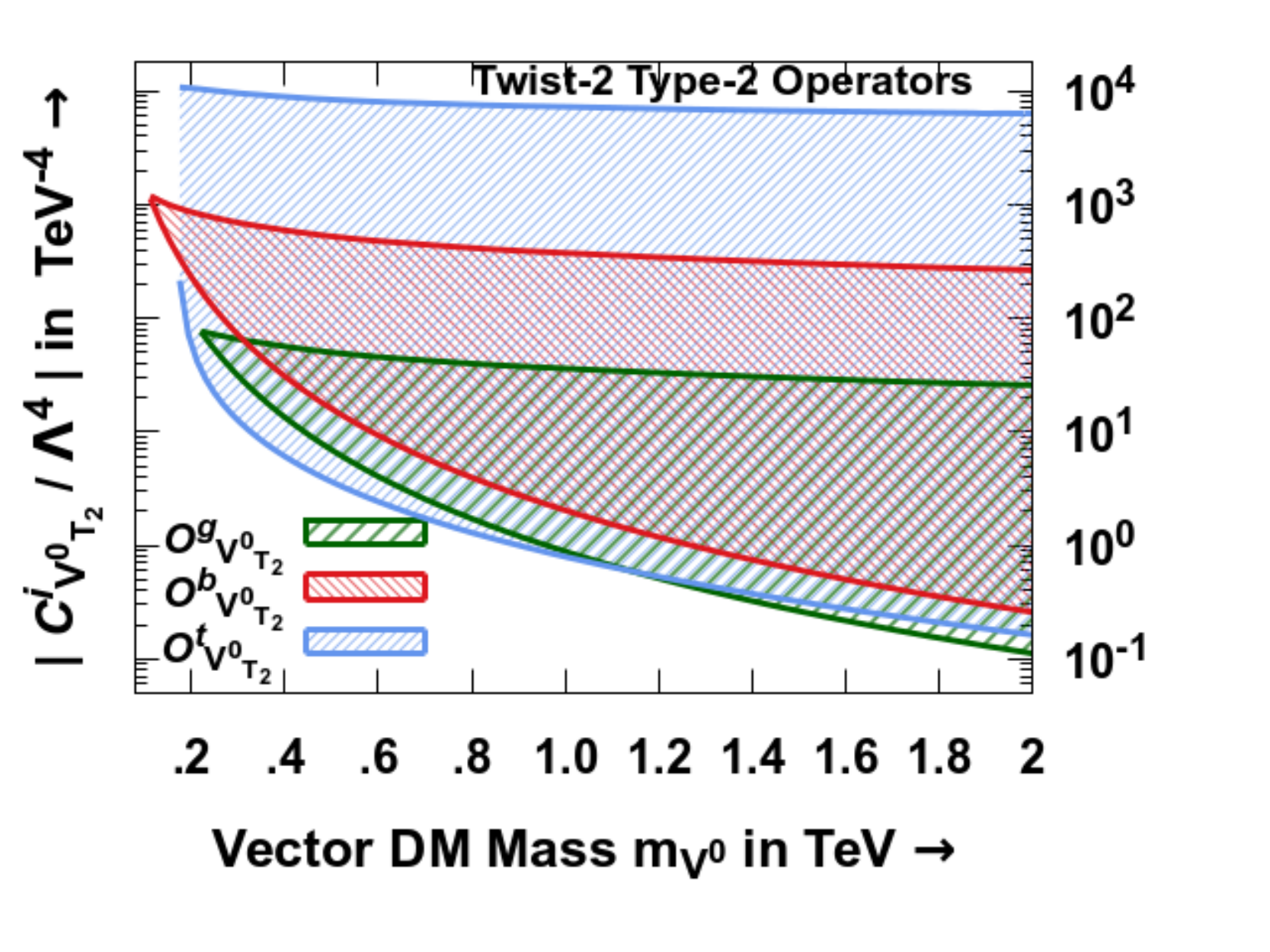}
      \caption{}	\label{fig:ComboVDMTwist}
  \end{subfigure}%
	\caption[nooneline]{\justifying \em{The shaded region in figures \ref{fig:ComboVDMScal} and \ref{fig:ComboVDMTwist} depict the allowed range of the Wilson coefficients corresponding to scalar and twist-2 type-2 operators from the relic density constraint~\cite{Aghanim:2018eyx} and direct detection limits from XENON-1T experiment~\cite{XENON:2018voc}, shown in red, blue and green for the $b$-quark, $t$-quark and gluon-philic vector DM respectively.}}
	\label{fig:VDMCombined}  
\end{figure}
\begin{itemize}
	\item Contributions from $b$-quark-philic, $t$-quark-philic, and gluon-philic Majorana DM scalar operators are found to be viable solutions for $m_\chi\ge $ 1.8 TeV, 200 GeV, and 1.25 TeV, respectively, as shown in figure \ref{fig:ComboMDMScal}. The contributions of $b$-quark-philic, $t$-quark-philic, and gluon-philic twist-2 type-1 Majorana DM operators shown in figure \ref{fig:ComboMDMTwist} have been validated for a much wider spectrum of $m_\chi\ge$ 60 GeV, 200 GeV, and 200 GeV, respectively.
\item In figure \ref{fig:ComboSDMScal} contributions from $b$-quark-philic, $t$-quark-philic and gluon-philic scalar DM scalar operators are found to be consistent with current experimental data for $m_{\phi^0}\ge $ 350 GeV, 200 GeV and 350 GeV, respectively. The corresponding  contributions  induced by twist-2 type-2  DM operators are  validated for $m_{\phi^0}\ge $ 150 GeV, 200 GeV and 280 GeV, respectively as shown in the  figure \ref{fig:ComboSDMTwist} . 
	\item In figure \ref{fig:ComboVDMScal} contributions from $b$-quark-philic, $t$-quark-philic and gluon-philic vector DM scalar operators are validated for $m_{V^0}\ge$ 350 GeV, 180 GeV and 350 GeV, respectively. The corresponding contributions induced by twist-2 type-2 DM operators are  validated for $m_{V^0}\ge $ 100 GeV, 180 GeV and 200 GeV, respectively as shown in figure \ref{fig:ComboVDMTwist}.
	
\end{itemize}

\par This analysis shows that the third-generation heavy-quark-philic and gluon-philic real DM are promising for $m_{\rm DM}\ge $ 200 GeV and are likely to be probed and falsified otherwise in the ongoing and future upgraded direct detection experiments. This study also opens up the scope    for future collider based DM search analysis through additional channels induced by the cosmologically  constrained operators.


\bigskip
\begin{acknowledgments}
We thank Abhaya Kumar Swain, Debajyoti Choudhury  and Manvinder Pal Singh for  fruitful discussions. We acknowledge the partial financial support from core research grant CRG/2018/004889 SERB, Government  of India. LKS acknowledges the UGC SRF fellowship for the partial financial support.
\end{acknowledgments}


\appendix
\begin{center}
{\bf \Large Appendix}
\end{center}

\section{DM Pair Annihilation Cross-section}\label{annihilationCS}
The annihilation cross-sections of a pair of Majorana DM $\chi$ to a pair of third generation heavy quarks  induced by the  scalar, pseudo-scalar, axial-vector and twist-2 type-1 operators are given as:
\begin{widetext}
\begin{subequations}
\begin{eqnarray}
	\sigma_{S}(\chi \bar{\chi} \rightarrow f \bar{f}) &=& {\cal C}_a\left[\frac{C^f_{\chi_S}}{\Lambda^3}\right]
	^2\ \frac{1}{16\, \pi}\, m_f^2\,\, \beta_f^3\, \beta_\chi\, s \label{hqSigMDMScal}\\
	\sigma_{\rm PS}(\chi \bar{\chi} \rightarrow f \bar{f}) &=& {\cal C}_a\left[\frac{C^f_{\chi_{\rm PS}}}{\Lambda^3}\right]
	^2\ \frac{1}{16\, \pi}\, m_f^2\,\, \frac{\beta_f}{\beta_\chi}\, s \label{hqSigMDMPS}\\
	\sigma_{\rm AV}(\chi \bar{\chi} \rightarrow f \bar{f}) &=& {\cal C}_a\left[\frac{C^f_{\chi_{\rm AV}}}{\Lambda^2}\right]^2\ \frac{1}{12\, \pi}\,\,\Big[1 - 4\, (x_\chi + x_f) + 28\, x_\chi\, x_f \Big]\,\, \frac{\beta_f}{\beta_\chi}\, s\, \label{hqSigMDMAxial}
\end{eqnarray}
\begin{eqnarray}
	\sigma_{T_1}(\chi \bar{\chi} \rightarrow f \bar{f}) &=& {\cal C}_a\left[\frac{C^f_{\chi_{T_1}}}{\Lambda^4}\right]^2 \frac{1}{960\, \pi}\,\,  \Big[4 x_\chi^2\, \left(92\, x_f^2 + 9\, x_f - 8 \right) +\, 3\, x_\chi \left(12 \,x_f^2 + 9\, x_f + 2 \right) \nn \\
	&& \hspace{4cm} - 32 x_f^2 + 6 x_f + 8\Big] \frac{\beta_f}{\beta_\chi} \,\, s^3\hspace{1cm} \label{hqSigMDMTwist} 
\end{eqnarray}
\end{subequations}
\end{widetext}
	where ${\cal C}_a$ = $3$,  $x_i$ = $m_i^2/\,s$ and $\beta_i$ = $\sqrt{1 - 4 x_i}$. 
\par As mentioned in the text, we include the study of gluon-philic Majorana DM interactions induced by the scalar, pseudo-scalar and twist-2 type-1 operators. The annihilation cross-sections of the Majorana DM pair to a pair of gluons  are given as 	
\begin{subequations}
\begin{eqnarray}
	\sigma_{S}(\chi \bar{\chi} \rightarrow g \,g) &=& {\cal C}_a\, {\cal C}_f \left[\frac{C^g_{\chi_S}}{\Lambda^4}\right]^2\ \frac{1}{4\, \pi}\, \, \left(\frac{\alpha_s}{\pi}\right)^2 \,\, m_{\chi}^2\,\,\beta_\chi\, \, s^2 \label{gSigMDMScal} \\
	\sigma_{\rm PS}(\chi \bar{\chi} \rightarrow g \,g) &=& {\cal C}_a\, {\cal C}_f\left[\frac{C^g_{\chi_{\rm PS}}}{\Lambda^4}\right]^2\ \frac{1}{2\, \pi}\, \, \left(\frac{\alpha_s}{\pi}\right)^2\, m_{\chi}^2\,\, \beta_\chi\, \,  s^2 \label{gSigMDMPS} \\
	\sigma_{T_1}(\chi \bar{\chi} \rightarrow g \,g) &=& {\cal C}_a\, {\cal C}_f\left[\frac{C^g_{\chi_{T_1}}}{\Lambda^4}\right]^2\ \frac{1}{80\, \pi}\,\, \left(1 + \frac{8}{3}\, x_\chi\right) \,\,  \beta_\chi\, s^3\label{gSigMDMTwist}
\end{eqnarray}
\end{subequations}
where ${\cal C}_f$ = $4/3$.
\par The annihilation cross-sections of a pair of real spin 0 DM $\phi^0$ to a pair of third generation heavy quarks  induced by the  scalar and twist-2 type-2 operators are given as:
\begin{subequations}
\begin{eqnarray}
	\sigma_{S}(\phi^0 \phi^0 \rightarrow f \bar{f}) &=& {\cal C}_a\left[\frac{C^f_{\phi^0_S}}{\Lambda^2}\right]^2\ \frac{1}{2\, \pi}\,m_f^2\,\, \frac{\beta_f^3}{\beta_\phi} \label{hqSigSDMScal}\\
	\sigma_{T_2}(\phi^0 \phi^0 \rightarrow f \bar{f}) &=&{\cal C}_a \left[\frac{C^f_{\phi^0_{T_2}}}{\Lambda^4}\right] \frac{1}{240 \pi}\frac{\beta_f^3}{\beta_\phi} s^3 \Bigg[ 2 x_\phi^2 (23 x_f + 8)- 4 x_\phi\, (7 x_f + 2) + 6 x_f + 1 \Bigg]\nn\\
	\label{hqSigSDMTwist} \end{eqnarray}
	\end{subequations}
The cross-sections for the pair of scalar DM annihilation into a pair of gluons induced by scalar and twist-2 type-2 operators are given as
	\begin{subequations}
	\begin{eqnarray}
	\sigma_{S}(\phi^0 \phi^0 \rightarrow g \,g) &=&{\cal C}_a\, {\cal C}_f \left[\frac{C^g_{\phi^0_S}}{\Lambda^2}\right]^2\ \left( \frac{\alpha_s}{\pi} \right)^2\ \frac{2}{\pi}\,\, \frac{1}{\beta_{\phi^0}}\,\,s\label{gSigSDMScal} \\
	\sigma_{T_2}(\phi^0 \phi^0 \rightarrow g \,g) &=&{\cal C}_a\, {\cal C}_f \left[\frac{C^g_{\phi^0_{T_2}}}{\Lambda^4}\right]^2\  \frac{1}{60\,\pi}\,\,  \beta_{\phi^0}^3\,\, s^3 \label{gSigSDMTwist}
\end{eqnarray}
\end{subequations}
\par Similarly, the production of a pair of third generation heavy quarks   as a result of the annihilation of a pair of real vectors DM   induced by the scalar, pseudo-scalar, axial-vector, and twist-2 type-2 operators is given as
\begin{widetext}
\begin{subequations}
\begin{eqnarray}
	\sigma_{S}(V^0\, V^0 \rightarrow f \bar{f}) &=& {\cal C}_a\left[\frac{C^f_{V^0_S}}{\Lambda^2}\right]^2\ \frac{1}{72\, \pi}\, \, m_f^2\, \frac{1}{x_{V^0}^2}\,\Big[ 1 - 4 x_{V^0} + 12 x_{V^0}^2\Big]\,\,\frac{\beta_f^3}{\beta_{V^0}}\,  \label{hqSigVDMScal}\\
	\sigma_{\rm PS}(V^0\, V^0 \rightarrow f \bar{f}) &=& {\cal C}_a\left[\frac{C^f_{V^0_{\rm PS}}}{\Lambda^4}\right]^2\,\frac{2}{9\,\pi}\,\,m_f^2\,\,\beta_f\, \beta_{V^0} \,\, s^2 \label{hqSigVDMPS}\\	
	\sigma_{\rm AV}(V^0 \,V^0 \rightarrow f \bar{f}) &=& {\cal C}_a\left[\frac{C^f_{V^0_{\rm AV}}}{\Lambda^2}\right]^2 \frac{1}{27\, \pi}\,\, \frac{1} {x_{V^0}}\,\Big[1 - 4\, (x_{V^0} + x_f) + 28\, x_{V^0}\, x_f \Big]\,\,  \beta_f\, \beta_{V^0}\, s\label{hqSigVDMAxial}\\
	\sigma_{T_2}(V^0 \,V^0 \rightarrow f \bar{f}) &=& {\cal C}_a\left[\frac{C^f_{V^0_{T_2}}}{\Lambda^4}\right]^2 \frac{1}{8640\, \pi}\,\, \frac{1}{x_{V^0}^2}\,\Big[\Big(1 + 6\, x_{V^0}\Big)\, \Big(1 - 4 x_{V^0} + 12 x_{V^0}^2\Big)\Big]  \,\beta_f^3\, \beta_{V^0}^3\,  s^3  \label{hqSigVDMTwist}\nn\\ 
\end{eqnarray}
\end{subequations}
\end{widetext}
The annihilation cross-sections of a pair of vector DM induced by the scalar and twist-2 type-2 gluon currents are given as
\begin{widetext}
\begin{subequations}
\begin{eqnarray}
	\sigma_{S}(V^0\, V^0 \rightarrow g\, g) &=&{\cal C}_a\, {\cal C}_f \left[\frac{C^g_{V^0_S}}{\Lambda^2}\right]^2 \frac{1}{18\, \pi}\,\,\left( \frac{\alpha_s}{\pi} \right)^2\,\frac{1}{x_{V^0}^2} \,\Big[1 - 4 x_{V^0} + 12 x_{V^0}^2\Big]\,\, \frac{1}{\beta_{V^0}}\,\, s \label{gSigVDMScal}\\
	\sigma_{\rm PS}(V^0\, V^0 \rightarrow g g) &=&  {\cal C}_a\, {\cal C}_f \left[\frac{C^g_{V^0_{\rm PS}}}{\Lambda^4}\right]^2\,\frac{1}{9\,\pi}\,\, \beta_{V^0}\, s^3 \label{gSigVDMPS}\\
	\sigma_{T_2}(V^0V^0 \rightarrow g g) &=& {\cal C}_a\, {\cal C}_f\left[\frac{C^g_{V^0_{T_2}}}{\Lambda^4}\right]^2 \frac{1}{2160\, \pi}\,\,\frac{1}{x_{V^0}^2} \,\Big[1 - 4 x_{V^0} + 12 x_{V^0}^2\Big]\,\, \beta_{V^0}^3\,\, s   \label{gSigVDMTwist}
\end{eqnarray}
\end{subequations}
\end{widetext}

\section{Thermally-averaged annihilation cross-section}\label{TherAvgCS}
In order to compute the probability of a DM particle
being annihilated by another one, the aforementioned annihilation cross-sections are re-written in terms of the magnitude of relative velocity $\left\vert \vec v_{{\rm DM}_1}-\vec v_{{\rm DM}_2}\right\vert\equiv \left\vert\vec v_{\rm DM}\right\vert$ and the dimensionless ratio $\xi_f\equiv m_f^2/\, m^2_{\rm DM}$. 
\par The thermal average of the heavy-quark-philic Majorana DM pair annihilation cross-sections given in equations \eqref{hqSigMDMScal}, \eqref{hqSigMDMPS}, \eqref{hqSigMDMAxial}  and \eqref{hqSigMDMTwist} corresponding to scalar and  pseudo-scalar operators respectively  are given as
\begin{widetext}
\begin{subequations}
\begin{eqnarray}
\left\langle \sigma_{S} (\chi \chi \rightarrow f \bar{f})  \left\vert\vec v_{\chi}\right\vert  \right\rangle&=&  {\cal C}_a\left[\frac{C^f_{\chi_S}}{\Lambda^3}\right]^2 \frac{1}{8\,\pi} \,m_f^2\, m_\chi^2\,\,\left(1 - \xi_f\right)^{3/2}\,\,\left\vert\vec v_{\chi}\right\vert^2   \label{FDMIDQS}\\
\left\langle \sigma_{\rm PS} (\chi \chi \rightarrow f \bar{f})  \left\vert\vec v_{\chi}\right\vert  \right\rangle&=&  {\cal C}_a\left[\frac{C^f_{\chi_{\rm PS}}}{\Lambda^3}\right]^2 
\frac{1}{8\,\pi} \,m_f^2\, m_\chi^2\,\,\left(1 - \xi_f\right)^{1/2}\,\, \left[4+\frac{1}{2}\,\frac{\xi_f}{1-\xi_f}\,\left\vert\vec v_{\chi}\right\vert^2\right]\label{FDMIDQPS}
\end{eqnarray}
The axial-vector contribution is analytically expanded in terms of two independently converging series expansions in terms of $\left\vert\vec v_{\chi}\right\vert^2$ and given as
\begin{eqnarray}
\left\langle \sigma_{\rm AV} (\chi \chi \rightarrow f \bar{f}) \left\vert\vec v_{\chi}\right\vert  \right\rangle &= &{\cal C}_a\left[\frac{C^f_{\chi_{\rm AV}}}{\Lambda^2}\right]^2 \frac{1 }{2 \pi} \left(1-\xi_f\right)^{1/2} \Bigg[  m_\chi^2\bigg\{ \frac{1}{3}\frac{1}{1-\xi_f}\left\vert\vec v_{\chi}\right\vert^2 + \frac{1}{6} \frac{1}{(1-\xi_f)^2}\left\vert\vec v_{\chi}\right\vert^4 + \cdots\bigg\} \nn \\
&&\, + m_f^2\,\bigg\{1 +  \frac{1}{24} \frac{(- 28 + 23 \xi_f)}{1-\xi_f} \left\vert\vec v_{\chi}\right\vert^2 + \frac{\left(-72 -48 \xi_f + 53 \xi_f ^2\right)}{384 (1-\xi_f)^2}\left\vert\vec v_{\chi}\right\vert^4 + \cdots\bigg\}\Bigg] \nn\\
& = & {\cal C}_a\left[\frac{C^f_{\chi_{\rm AV}}}{\Lambda^2}\right]^2 \frac{1 }{2\, \pi} \,\, m_f^2\,\,\left(1-\xi_f\right)^{1/2} \left[ 1 +  \frac{1}{24} \,\frac{8\xi_f^{-1} - 28 + 23 \xi_f}{1-\xi_f}\, \left\vert\vec v_{\chi}\right\vert^2+\cdots\right] 
\label{FDMIDQAV}\nn\\
\end{eqnarray}

And finally the contribution from the twist-2 type-1 operator induced by the heavy quark-philic Majorana DM interaction is given as
\begin{eqnarray}
\left\langle \sigma_{T_1} (\chi \chi \rightarrow f \bar{f}) \left\vert\vec v_{\chi}\right\vert  \right\rangle &=& {\cal C}_a\left[\frac{C^f_{\chi_{T_1}}}{\Lambda^4}\right]^2 \, \frac{1}{2\,\pi}\,\, m_\chi^6\,\, \left(1-\xi_f\right)^{1/2} \,\left(2 + \xi_f\right)\nn \\
&& \hspace{1cm} \times \left[ 1 + \frac{1}{48} \,\, \frac{56 - 41 \xi - 8 \xi_f^2 + 11 \xi_f^3}{\left(1-\xi_f\right)\,\left(2+\xi_f\right)}\,\,\left\vert\vec v_{\chi}\right\vert^2  \right]\,\  \label{FDMIDQT1}
\end{eqnarray}
\end{subequations}
\end{widetext}
Using annihilation cross-sections in \eqref{gSigMDMScal}, \eqref{gSigMDMPS}  and \eqref{gSigMDMTwist} the thermal averaged annihilation cross-sections for the gluon-philic Majorana DM are given  as
\begin{subequations}
\begin{eqnarray}
	\left\langle \sigma_{S} (\chi \chi \rightarrow g \,g) \left\vert\vec v_{\chi}\right\vert  \right\rangle &=& {\cal C}_a\, {\cal C}_f\left[\frac{C^g_{\chi_S}}{\Lambda^4}\right]^2 \left(\frac{\alpha_s}{\pi}\right)^2 \frac{2}{\pi}\, \, m_\chi^6 \,\,\left\vert\vec v_{\chi}\right\vert^2 \label{FDMIDGS} \\
	\left\langle \sigma_{\rm PS} (\chi \chi \rightarrow g \,g) \left\vert\vec v_{\chi}\right\vert  \right\rangle &=& {\cal C}_a\, {\cal C}_f\left[\frac{C^g_{\chi_{\rm {\rm PS}}}}{\Lambda^4}\right]^2 \left(\frac{\alpha_s}{\pi}\right)^2 \frac{4}{\pi}\,  m_\chi^6 \,\, \left\vert\vec v_{\chi}\right\vert^2 \label{FDMIDGPS} \\
	\left\langle \sigma_{T_1} (\chi \chi \rightarrow g\, g) \left\vert\vec v_{\chi}\right\vert  \right\rangle &=& {\cal C}_a \,{\cal C}_f\left[\frac{C^g_{\chi_{T_1}}}{\Lambda^4}\right]^2\ \frac{2}{3\,\pi}\,\, m_\chi^6\,\,  \left\vert\vec v_{\chi}\right\vert^2	\label{FDMIDGT}
\end{eqnarray}
\end{subequations}

\par The thermal average of heavy-quark-philic scalar DM pair annihilation cross-sections dipalyed in equations \eqref{hqSigSDMScal} and \eqref{hqSigSDMTwist} are given as  
\begin{widetext}
\begin{subequations}
\begin{eqnarray}
	\left\langle \sigma_S(\phi^0 \phi^0 \rightarrow f \bar{f}) \left\vert\vec v_{\phi^0}\right\vert  \right\rangle  &=&{\cal C}_a \left[\frac{C^f_{\phi^0_S}}{\Lambda^2}\right]^2 \frac{1}{\pi}\,\, m_f^2\,\,\left(1 - \xi_f\right)^{3/2}\,  \left[ 1 - \frac{1}{8}\, \frac{2 -  5 \xi_f}{1 - \xi_f} \,\,\left\vert\vec v_{\phi^0}\right\vert^2\right] \label{SDMIDeqScal} \\
	\left\langle \sigma_{T_2} (\phi^0 \phi^0 \rightarrow f \bar{f}) \left\vert\vec v_{\phi^0}\right\vert  \right\rangle &=&{\cal C}_a \left[\frac{C^f_{\phi^0_{T_2}}}{\Lambda^4}\right]^2\, \frac{1}{4\,\pi} \,\, m_f^2 \, m_{\phi^0}^4\,\,\left(1-\xi_f\right)^{3/2} \left[1 + \frac{1}{24}\,\frac{10 -\xi_f }{1-\xi_f}\, \left\vert\vec v_{\phi^0}\right\vert^2\right] \nn\\&&\label{SDMIDeqTwist}
\end{eqnarray}
\end{subequations}
\end{widetext}
Similarly, the thermal average of heavy-quark-philic scalar DM pair annihilation cross-sections displayed in equations \eqref{gSigSDMScal} and \eqref{gSigSDMTwist} are given as  
\begin{subequations}
\begin{eqnarray}
	\left\langle \sigma_{S} (\phi^0 \phi^0 \rightarrow g\, g) \left\vert\vec v_{\phi^0}\right\vert  \right\rangle &=&{\cal C}_a\, {\cal C}_f \left[\frac{C^g_{\phi^0_S}}{\Lambda^2}\right]^2 \left(\frac{\alpha_s}{\pi}\right)^2\,\, \frac{16}{\pi}\,\, m_{\phi^0}^2 \label{SDMIDeqGluon} \\
	\left\langle \sigma_{T_2} (\phi^0 \phi^0 \rightarrow g\,g) \left\vert\vec v_{\phi^0}\right\vert  \right\rangle &=&{\cal C}_a\, {\cal C}_f \left[\frac{C^g_{\phi^0_{T_2}}}{\Lambda^4}\right]^2\, \frac{2}{15\, \pi} \,\,\, m_{\phi^0}^6\,\, \,\left\vert\vec v_{\phi^0}\right\vert^4 \label{SDMIDeqGluonT} 
\end{eqnarray}
\end{subequations}
\par The thermal average of the vector DM pair annihilation cross-sections given in equations \eqref{hqSigVDMScal}, \eqref{hqSigVDMPS}, \eqref{hqSigVDMAxial} and \eqref{hqSigVDMTwist}   corresponding to the scalar, pseudo-scalar, axial-vector and twist-2 type-2 operators, respectively, are given as
\begin{widetext}
\begin{subequations}
\begin{eqnarray}
\left\langle \sigma_{S} (V^0\,V^0\rightarrow f \bar{f}) \left\vert\vec v_{V^0}\right\vert  \right\rangle &=&{\cal C}_a \left[\frac{C^f_{V^0_S}}{\Lambda^2}\right]^2\, \frac{1}{3\, \pi}\, m_f^2\, \left(1-\xi_f\right)^{3/2} \left[1 + \frac{1}{24}  \frac{2 + 7 \xi_f}{1-\xi_f} \left\vert\vec v_{V^0}\right\vert^2\right] \label{VDMIDeqScal} \\
\left\langle \sigma_{\rm PS} (V^0\,V^0\rightarrow f \bar{f}) \left\vert\vec v_{V^0}\right\vert  \right\rangle &=&  {\cal C}_a \left[\frac{C^f_{V^0_{\rm PS}}}{\Lambda^4}\right]^2\, \frac{16}{9\, \pi}\,\, m_f^2\,\, m_{V^0}^4\,\, \left\vert\vec v_{V^0}\right\vert^2 \label{VDMIDeqPS} \\
	\left\langle \sigma_{\rm AV} (V^0\,V^0 \rightarrow f \bar{f}) \left\vert\vec v_{V^0}\right\vert  \right\rangle &=&{\cal C}_a \left[\frac{C^f_{V^0_{\rm AV}}}{\Lambda^2}\right]^2\, \frac{2}{9\,\pi}\,  \,m_f^2\,\,  \left(1-\xi_f\right)^{1/2}\,\,  \left\vert\vec v_{V^0}\right\vert^2 \label{VDMIDeqAxial} \\	
	\left\langle \sigma_{T_2} (V^0\,V^0 \rightarrow f \bar{f}) \left\vert\vec v_{V^0}\right\vert  \right\rangle &=& {\cal C}_a \left[\frac{C^f_{V^0_{T_2}}}{\Lambda^4}\right]^2\, \frac{1}{90\, \pi}\,\,  m_{V^0}^6\,\,   \left(1-\xi_f\right)^{3/2} \left[1 + \frac{3}{2}\, \xi_f\right] \,\, \left\vert\vec v_{V^0}\right\vert^4\label{VDMIDeqTwist} 
\end{eqnarray}
\end{subequations}
\end{widetext}
	Similarly, the thermal average of the  vector DM pair annihilation to gluon channels as displayed in equations  \eqref{gSigVDMScal}, \eqref{gSigVDMPS} and \eqref{gSigVDMTwist} corresponding to the scalar, pseudo-scalar and twist-2  currents respectively are given as
\begin{widetext}
\begin{subequations}
\begin{eqnarray}
	\left\langle \sigma_{S} (V^0\,V^0 \rightarrow g\, g) \left\vert\vec v_{V^0}\right\vert  \right\rangle &=&{\cal C}_a\, {\cal C}_f \left[\frac{C^g_{V^0_S}}{\Lambda^2}\right]^2\, \left(\frac{\alpha_s}{\pi}\right)^2\, \frac{16}{3\,\pi} \,\,  m_{V^0}^2\,\left[ 1 + \frac{1}{3}\,\,\left\vert\vec v_{V^0}\right\vert^2\right] \label{VDMIDeqGluonS} \\	
	\left\langle \sigma_{\rm PS} (V^0\,V^0 \rightarrow g\, g) \left\vert\vec v_{V^0}\right\vert  \right\rangle &=&  {\cal C}_a\, {\cal C}_f \left[\frac{C^g_{V^0_{\rm PS}}}{\Lambda^4}\right]^2\, \left(\frac{\alpha_s}{\pi}\right)^2\, \frac{64}{9\,\pi} \,\,  m_{V^0}^6\, \vert\vec v_{V^0}\vert^2 \label{VDMIDeqGluonPS} \\	
	\left\langle \sigma_{T_2} (V^0\,V^0 \rightarrow g\,g) \left\vert\vec v_{V^0}\right\vert  \right\rangle &=&{\cal C}_a\, {\cal C}_f \bigg[\frac{C^g_{V^0_{T_2}}}{\Lambda^4}\bigg]^2\, \frac{2}{45\,\pi}\,\, m_{V^0}^6\, \,\,\left\vert\vec v_{V^0}\right\vert^4  \label{VDMIDeqGluonT}
\end{eqnarray}
\end{subequations}
\end{widetext}


\section{Effective DM-nucleon Interactions from  1-Loop DM-Gluon amplitudes}
\label{DDappendix}
The DM-gluon scattering  arising due  to the one loop Feynman-diagrams  in figure \ref{DDfeyndia}   is induced by the scalar, axial-vector and twist-2 currents of the  heavy quarks. We do not consider the  contributions of the pseudo-scalar operators as they are spin and velocity suppressed. These one-loop amplitudes characterise the  effective point interaction Lagrangian for the gluon with Majorana, real scalar   and real vector DM candidates respectively and are given as
\begin{widetext}
\begin{subequations}
\begin{eqnarray}
	{\cal L}^{\chi\chi gg}_{\rm eff.} &=& \frac{C_{\chi_S}^q}{\Lambda^3}\,\frac{\alpha_s}{4\pi}\,\left(\bar\chi\,\chi\right) \left(G^a\right)_{\alpha\beta}\, \left(G^a\right)^{\alpha\beta}\, I_S^{gg} + \frac{C_{\chi_{\rm AV}}^q}{\Lambda^2}\,\frac{\alpha_s}{4\pi}\,m_\chi\left(\bar\chi\,i\,\gamma^5\,\chi\right) \widetilde{\left(G^a\right)_{\alpha\beta}}\, \left(G^a\right)^{\alpha\beta}\,4\, I_{\rm AV}^{gg}\nn\\
	&&\,\,\,\, 
	+  \frac{C_{\chi_{T_{1}}}^q}{\Lambda^4}\,\, \frac{\alpha_s}{4\pi}\,\left[\frac{4}{3}\, \ln\left(\frac{\Lambda^2}{m_Q^2}\right)\right]\, \left(\bar\chi\,i\,\partial^\mu\,\gamma^\nu\,\chi\right) {\cal O}^g_{\mu\nu}\label{LeffMDMgluon}\\
{\cal L}^{\phi^0\phi^0 gg}_{\rm eff.} &=& \frac{C_{\phi^0_S}^q}{\Lambda^2}\,\frac{\alpha_s}{4\pi}\,\left(\phi^0\,\phi^0\right) \left(G^a\right)_{\alpha\beta}\, \left(G^a\right)^{\alpha\beta}\, I_S^{gg} + \frac{C_{\phi^0_{T_{2}}}^q}{\Lambda^4}\,\,\frac{\alpha_s}{4 \pi}\, \left[\frac{4}{3}\, \ln\left(\frac{\Lambda^2}{m_Q^2}\right)\right]\,\left(\phi^0\,i\,\partial^\mu\,i\,\partial^\nu\,\phi^0\right) {\cal O}^g_{\mu\nu}\label{LeffSDMgluon}\,\,\,\nn \\ \\
{\cal L}^{V^0V^0 gg}_{\rm eff.} &=& \frac{C_{V^0_S}^q}{\Lambda^2}\frac{\alpha_s}{4\pi} \left(V^0\right)^\mu \left(V^0\right)_\mu\left(G^a\right)^{\alpha\beta} \widetilde{\left(G^a\right)_{\alpha\beta}} I_S^{gg} + \frac{C^q_{V^0_{\rm AV}}}{\Lambda^2} \frac{\alpha_s}{4\pi} \left(V^0\right)^{\mu\nu} \widetilde{\left(V^0\right)_{\mu\nu}} \left(G^a\right)^{\alpha\beta} \widetilde{\left(G^a\right)_{\alpha\beta}} 4 I_{\rm AV}^{gg}\nn\\
&&\ +\ \frac{C_{V^0_{T_{2}}}^q}{\Lambda^4}\,\frac{\alpha_s}{4\pi}\,\left[\frac{4}{3}\, \ln\left(\frac{\Lambda^2}{m_Q^2}\right)\right]\,\left(V^0\right)^\rho\,i\,\partial^\mu\,i\,\partial^\nu\,\left(V^0\right)_\rho\, {\cal O}^g_{\mu\nu}\label{LeffVDMgluon}
\end{eqnarray}
\end{subequations}
\end{widetext}
The dimensionless one-loop integrals $I_S^{gg}$, $I_{\rm AV}^{gg}$, and $I_{T}^{gg}$ are defined as 
\begin{subequations}
\bea
I_S^{gg} &=& \displaystyle \int_0^1 dx \displaystyle \int_0^{1-x} dy \frac{1-4\,x\,y}{1- \frac{q^2}{m_Q^2}\,x\,y} \label{DDloopint1}\\
I_{\rm AV}^{gg} &=& \displaystyle \int_0^1 dx \displaystyle \int_0^{1-x} dy \frac{x^2 - x -x\,y}{1- \frac{q^2}{m_Q^2}\,x\,y} \label{DDloopint2}
\eea
\end{subequations}
where the square of the four momentum transferred $q^2\approx 2\, m_N\, E_r$ ($E_r\lesssim$ 100 KeV  is the recoil energy of the nucleon). $m_Q$ and $m_N$ are   the concerned heavy quark mass ($m_b/\,m_t$) running in the loop and nucleon mass respectively.

\par We observe that the one-loop effective contact interactions generated from quark scalar, twist  and axial vector currents contain the scalar, pseudo-scalar and  twist gluon operators respectively. We perform the non-relativistic reduction of these operators for  zero momentum partonic gluons and  evaluate the gluonic operators between the nucleon states. The zero momentum gluon contribution to the hadronic matrix element $f^N_{\rm TG}\approx .923$  is
extracted in terms of light quark as \cite{Cirelli:2013ufw,Belanger:2013oya,Cheng:2012qr}
\begin{widetext}
\bea
f_{TG}^N \equiv -\frac{1}{m_N}\left\langle N\left \vert \frac{9}{8}\, \frac{\alpha_s}{\pi}\,  G_{\alpha\beta}^a G^{\alpha\beta}_a \right\vert N\right \rangle= 1 - \displaystyle\sum_{q=u,\,d,\,s} f_{Tq}^N \equiv 1-  \sum_{q=u,\,d,\,s} \left\langle N \left  \vert  \frac{m_q}{m_N}\,\left(\bar qq\right) \right\vert N \right\rangle\nn\\ \label{fTGinfo}
\eea
\end{widetext}
According to \cite{Cheng:2012qr} and \cite{Cheng:1988im}, the pseudoscalar gluon operator between nucleon states is computed as
\bea
\left\langle N\left \vert  G_{\alpha\beta}^a \widetilde{G^{\alpha\beta}_a} \right\vert N\right \rangle = m_N\, \bar m\, \displaystyle\sum_{q=u,\,d,\,s} \frac{1}{m_q}\,\,\Delta^{(N)}_q; \nn \\
\hskip 0.5cm {\rm where}\,\, \bar m =\left[\frac{1}{m_u}+\frac{1}{m_d} + \frac{1}{m_s}\right]^{-1} \label{coeffdeltainfo}
\eea
The axial vector current $\left\langle N\left\vert \bar q\gamma^\mu\gamma^5q\right\vert N\right\rangle = 2\Delta^{(N)}_q\,s^\mu$  specifies the coefficient $\Delta^{(N)}_q$ as the spin content   of the nucleon's quark $q$. The coefficients  for light quarks satisfy
$\Delta^{(p)}_u=\Delta^{(n)}_d$, $\Delta^{(n)}_u=\Delta^{(p)}_d$ and $\Delta^{(p)}_s=\Delta^{(n)}_s$, while the contribution from heavy quarks is found to be vanishingly small \cite{Polyakov:1998rb}. It is important to observe that a quark axial-vector current is related to the gluonic pseudoscalar current by PCAC \cite{Cheng:1988im}. The zero momentum nucleonic matrix element for  gluon twist-2 operators is defined as
\bea
\left\langle N\left\vert {\cal O}^g_{\mu\nu} \right\vert N\right\rangle &=& -\,\frac{1}{m_N} \left[k^{\mu}\,k^{\nu}-\frac{1}{4} g^{\mu\nu}\, m^2_N\right]\, g\left(2;\mu_{\rm R} \right); \nn\\
{\rm where} &&\,\,\, \,\,\,\,\,\, g\left(2;\mu_{\rm R}\right) = \displaystyle\int_0^1  x\, g\left(x;\mu_{\rm R} \right)\,dx\label{gluontwisNR}
\eea


\end{document}